\documentclass[]{aa}
\usepackage{graphicx}
\usepackage{txfonts}
\usepackage{multirow}  
\usepackage{mathtools} 
\usepackage[vlined]{algorithm2e} 
\SetAlgorithmName{Alg.}{Alg.}{List of Algs.} 
\SetAlgoCaptionSeparator{.}
\usepackage[breaklinks = true,
            colorlinks = true,
            urlcolor   = blue,
            citecolor  = blue,
            linkcolor  = blue]{hyperref}

\bibpunct{(}{)}{;}{a}{}{,} 

\graphicspath{ {figs/} }

\newcommand*{ \dd   }{ \ensuremath{ \mathrm{d}    } } 
\newcommand*{ \red  }{ \ensuremath{ \mathrm{red}  } } 
\newcommand*{ \blue }{ \ensuremath{ \mathrm{blue} } } 
\newcommand*{ \dex  }{ \ensuremath{ \mathrm{dex}  } }
\newcommand*{ \iter }{ \ensuremath{ \mathrm{iter} } }

\begin{document}

\title{
  Partial redistribution in 3D non-LTE radiative transfer in solar atmosphere
  models
}

\author{
  Andrii V.~Sukhorukov\inst{\ref{ISP},\ref{MAO}}
  \and
  Jorrit Leenaarts\inst{\ref{ISP}}
}

\institute{
  Institute for Solar Physics, Department of Astronomy, Stockholm University,
  AlbaNova University Centre, SE-106 91 Stockholm, Sweden,
  \email{andrii.sukhorukov@astro.su.se, jorrit.leenaarts@astro.su.se}
  \label{ISP}
  \and
  Main Astronomical Observatory, National Academy of Sciences of Ukraine,
  27 Akademika Zabolotnoho str., 03680 Kyiv, Ukraine,
  \label{MAO}
}

\date{Received ; accepted }

\abstract
  { Resonance spectral lines such as \ion{H}{I} Ly\,$\alpha$, \ion{Mg}{II} h\&k,
    and \ion{Ca}{II} H\&K that form in the solar chromosphere are influenced by
    the effects of 3D radiative transfer as well as partial redistribution
    (PRD).  So far no one has modeled these lines including both effects
    simultaneously owing to the high computing demands of existing algorithms.
    Such modeling is however indispensable for accurate diagnostics of the
    chromosphere. }
  { We present a computationally tractable method to treat PRD scattering in
    3D~model atmospheres using a 3D non-LTE radiative transfer code. }
  { To make the method memory-friendly, we use the hybrid approximation of
    \citet{2012A&A...543A.109L} for the redistribution integral.  To make it
    fast, we use linear interpolation on equidistant frequency grids.
    We verify our algorithm against computations with the RH code and
    analyze it for stability, convergence, and usefulness of acceleration
    using model atoms of \ion{Mg}{II} with the h\&k lines and \ion{H}{I} with
    the Ly\,$\alpha$ line treated in PRD. }
  { A typical 3D~PRD solution can be obtained in a model atmosphere with
    $ 252 \times 252 \times 496 $ coordinate points in 50\,000--200\,000~CPU
    hours, which is a factor ten slower than computations assuming complete
    redistribution.  We illustrate the importance of the joint action of PRD and
    3D effects for the \ion{Mg}{II} h\&k lines for disk-center intensities as
    well as the center-to-limb variation. }
  { The proposed method allows simulating PRD lines in time series of
    radiation-MHD models in order to interpret observations of chromospheric
    lines at high spatial resolution. }


\maketitle
\section{Introduction}

At the beginning of the development of radiative transfer theory,
astrophysicists assumed that scattering in spectral lines is coherent.
This assumption is however unable to reproduce observed intensity profiles and
their center-to-limb variation in detail.
In the 1940s, a better approximation called complete frequency redistribution
(CRD) was introduced.
\citet{1942PhDT........12H} 
validated it using systematic observations of various line profiles in the solar
spectrum.

The CRD approximation means that during line scattering there is no correlation
between absorbed and emitted photons.
CRD requires atomic levels of the line transition to be strongly perturbed by
elastic collisions with other atoms.
This is true in dense layers of stellar atmospheres such as the photosphere of
the Sun.
With a few exceptions, all lines visible in the solar spectrum are formed in
CRD.
The CRD approximation makes the line source function constant, which strongly
simplifies analytical and numerical solutions of the radiative transfer problem
in such lines.

In the 1970s, as reviewed by
\citet{1985ASIC..152....1L} 
and
\citet{1987Ap.....26...90N}, 
it became clear that strong chromospheric lines actually demonstrate
partially-coherent scattering (nowadays more commonly referred to as partial
frequency redistribution, PRD).
Although the theory of PRD was developed earlier, only then it became possible
to model PRD lines due to high computational demands for the redistribution
functions.

Contrary to CRD, PRD includes the fact that photon scattering is not arbitrary
and could even be fully correlated in the most extreme case of coherent
scattering.
Line scattering becomes partially or fully coherent only if several conditions
are met together
\citep[see p.\,519]{2014tsa..book.....H}. 
The upper atomic level of the line transition must be weakly perturbed by
elastic collisions to be able to retain radiatively-excited sublevels during the
finite radiative lifetime of the level.
This is true in low-density layers of stellar atmospheres such as the
chromosphere of the Sun.
The chemical element of the line has to be abundant and exist in a dominant
ionization stage so that the line extinction dominate over the continuum
extinction by several orders of magnitude.
The line has to be a resonance transition, or a transition whose lower level is
metastable

Only a small number of lines in the solar spectrum are affected by PRD but they
are indispensable as diagnostics of the outer atmosphere of the Sun.
Among them are: the strongest lines of \ion{H}{I} Lyman series such as the
Ly\,$\beta$ 102.6\,nm and the Ly\,$\alpha$ 121.6\,nm lines
\citep{1973ApJ...185..709M,
       1995ApJ...455..376H}; 
the strongest chromospheric \ion{Mg}{II} k~279.6\,nm and h~280.4\,nm
\citep{1974ApJ...192..769M} 
as well as the \ion{Ca}{II} K~393.4\,nm and H~396.8\,nm lines
\citep{1974SoPh...38..367V,
       1975ApJ...199..724S}; 
strong resonance UV lines of abundant neutrals such as the \ion{Mg}{I} 285.2\,nm
line
\citep{1977ApJ...216..654C} 
or the \ion{O}{I} resonance triplet at 130\,nm
\citep{2002ApJ...566..500M}; 
and resonance lines of other alkali and alkaline-earth metals such as the
\ion{Na}{I} D~589\,nm doublet or the \ion{Ba}{II} 455.4\,nm line, which are
mostly formed in the photosphere but show PRD effects towards the extreme limb
\citep{1992A&A...265..268U,
       1993KPCB....9...52B,
       1979ApJ...231..277R}. 

In the chromosphere, where PRD lines are formed, the 3D spatial transport of
radiation becomes essential
\citep{2009ApJ...694L.128L,
       2010ApJ...709.1362L,
       2012ApJ...749..136L,
       2015ApJ...803...65S,
       2015SoPh..290..979J}. 
%
So far no one has modeled PRD lines including the effects of 3D non-LTE
radiative transfer due to the large computational effort that is required.

In this paper, we present a method to perform 3D non-LTE radiative transfer with
PRD effects, which was implemented in the Multi3D code
\citep{2009ASPC..415...87L},
and investigate whether such modeling is significant.
We are motivated to make up for a lack of accurate physical models needed to
interpret observations in the \ion{Mg}{II} h\&k lines from the IRIS satellite
\citep{2014SoPh..289.2733D}, 
measurements in the \ion{H}{I} Ly\,$\alpha$ line obtained by the CLASP rocket
experiment
\citep{2014ASPC..489..307K}, 
and observations in the \ion{Ca}{II} H\&K lines using the new imaging
spectrometer CHROMIS
at the Swedish 1-m Solar Telescope.

The structure of the paper is as follows.
Section~\ref{sec:PRDinRH} briefly explains the method to solve the 3D~non-LTE
radiative transfer problem in PRD lines.
Section~\ref{sec:setup} describes the computational setup: model atmospheres,
model atoms, and the code.
Section~\ref{sec:numerical-considerations} lays out the most important
computational specifics.
Section~\ref{sec:results} presents the results for our calculations, which we
verify, analyze for stability, convergence speed, and applicability of
acceleration.
Using intensities computed in the \ion{Mg}{II} h\&k lines, we illustrate the
importance of 3D non-LTE radiative transfer with PRD effects.
In Section~\ref{sec:discussion-conclusions}, we give some conclusions.
Technical details on the algorithm are given in Appendix~\ref{sec:appendix}.

\section{Method}
\label{sec:PRDinRH}

The iterative solution algorithm of Multi3D employs preconditioning of the rate
equations as formulated by
\citet{1991A&A...245..171R,
       1992A&A...262..209R}. 
This method was extended by
\citet{2001ApJ...557..389U} 
to include the effects of PRD and implemented in the RH code%
\footnote{\url{http://www4.nso.edu/staff/uitenbr/rh.html}},
the de facto standard for non-LTE radiative transfer calculations in
plane-parallel geometry.
The RH code has been extended to allow for parallel computation of a large
number of plane-parallel atmospheres (1.5D~approximation) by
\citet{2015A&A...574A...3P}. 
Below we briefly review the algorithm and the extension by
\citet{2012A&A...543A.109L}, 
who devised a method to compute a fast approximate solution of the full
angle-dependent PRD problem in moving atmospheres.

\subsection{The PRD algorithm in the Rybicki-Hummer framework}
\label{subsec:PRDdesc}

The statistical equilibrium non-LTE radiative transfer problem for an atom with
$ N_\mathrm{L} $ levels consists of solving the rate equations for the atomic
level populations:
\begin{equation} \label{eq:RateEq}
  \dfrac{\dd n_i}{\dd t} =
  \sum_{ j,\,j \ne i }^{ N_\mathrm{L} }
    n_{\!j}
    P_{\!ji}
  - n_i
  \sum_{ j,\,j \ne i }^{ N_\mathrm{L} }
    P_{i\!j}
  = 0,
\end{equation}
with $ n $ the atomic level populations, and $ P_{i\!j} = C_{i\!j} + R_{i\!j} $
the total rates consisting of collisional rates $ C_{i\!j} $ and radiative rates
$ R_{i\!j} $.
The radiative rates depend on the specific intensity $ I( \vec{n}, \nu ) $,
which in turn depends on the direction $ \vec{n} $ and the frequency $\nu$, and
is computed from the transfer equation:
\begin{equation}
  \frac{ \dd I( \vec{n}, \nu ) }{ \dd s }
    =
    j( \vec{n}, \nu ) - \alpha( \vec{n}, \nu )\, I( \vec{n}, \nu ).
\end{equation}
The transfer equation couples different locations in the atmosphere and makes
the problem non-linear and non-local.

The assumption of complete redistribution (CRD) is that the frequency and
direction of the absorbed and emitted photon in a scattering event are
uncorrelated, so that the normalized emission profile
$ \psi_{\!ji}(\vec{n}, \nu) $ and absorption profile
$ \varphi_{\!ji}(\vec{n}, \nu) $ in a transition $i {\to} j$ are equal:
\begin{equation} \label{eq:psi=phi}
  \psi_{\!ji}( \vec{n}, \nu ) =
    \varphi_{\!ji}( \vec{n}, \nu ),
\end{equation}
for all frequencies $ \nu $ and directions $ \vec{n} $.
If PRD is important then this equality is no longer valid, and following
\citet{1989A&A...213..360U} 
we define:
\begin{equation}
  \rho_{i\!j}( \vec{n}, \nu )
  \equiv
    \frac{ \psi_{\!ji}( \vec{n}, \nu ) }
         { \varphi_{i\!j}( \vec{n}, \nu ) }.
\end{equation}
The profile ratio $ \rho_{i\!j}(\vec{n}, \nu) $ describes for the transition
$ i {\to} j $ how strongly the line emissivity correlates with the line opacity
for the direction $ \vec{n} $ and the frequency $ \nu $.
It depends on the local particle densities, temperature and the local radiation
field $ I( \vec{n}, \nu ) $.
It is discussed in more detail in Section~\ref{subsec:profile-ratio}.
The contributions into the opacity and emissivity from the PRD transition
$ i {\to} j $ are
\begin{align}
  \alpha_{i\!j} (\vec{n}, \nu) &
    =
    \frac{ h\nu }{ 4\pi }
    B_{i\!j}
    \varphi_{i\!j}( \vec{n}, \nu )
    \biggl[
      n_i - n_{\!j}
        \frac{ g_i }{ g_{\!j} }
        \rho_{i\!j}( \vec{n}, \nu )
    \biggr]
    \quad\text{and}
    \label{eq:opacity}
  \\
  j_{i\!j} (\vec{n}, \nu) &
    =
    \frac{ h\nu }{ 4\pi }
    A_{\!ji}
    \varphi_{i\!j}( \vec{n}, \nu )
    \Bigl[
      n_{\!j}
      \rho_{i\!j}( \vec{n}, \nu )
    \Bigr],
    \label{eq:emissivity}
\end{align}
with $ n $ and $ g $ the corresponding level populations and statistical
weights, and $ A_{\!ji} $, $ B_{\!ji} $, and $ B_{i\!j} $ the Einstein
coefficients.
This means that the line source function
\begin{equation} \label{eq:Sij}
  S_{\!i\!j} (\vec{n}, \nu) =
    \frac{ n_{\!j} A_{\!ji} \rho_{i\!j}( \vec{n}, \nu ) }
         { n_i B_{i\!j} - n_{\!j} B_{\!ji} \rho_{i\!j}( \vec{n}, \nu ) } =
    \frac{ 2h\nu^3 }
         { c^2     }
    \biggl[
      \frac{ 1       }{ \rho_{i\!j}( \vec{n}, \nu )  }
      \frac{ n_i     }{ n_{\!j} }
      \frac{ g_{\!j} }{ g_i     }
      - 1
    \biggr]^{-1}
\end{equation}
depends on frequency and direction, in contrast to CRD where it is constant.

The radiative rates in the PRD transition $ i {\to} j $ are
\begin{align}
  R_{i\!j} &
    =
    B_{i\!j} \bar{J}_{i\!j}^\varphi
    \quad\text{and}
    \label{eq:Rij}
  \\
  R_{\!ji} &
    =
    A_{\!ji} + B_{\!ji} \bar{J}_{\!ji}^\psi,
    \label{eq:Rji}
\end{align}
where the mean angle-averaged intensities integrated with the absorption and
emission profiles are denoted by
\begin{align}
  \bar{J}_{\!i\!j}^\varphi &
    \equiv
    \oint\mspace{-11mu}\int\!
    I( \vec{n}, \nu )\,
    \varphi_{i\!j}( \vec{n}, \nu )\,
    \dd\nu\,
    \frac{\dd\Omega}{4\pi}
    \quad\text{and}
    \label{eq:Jij}
  \\
  \bar{J}_{\!\!ji}^\psi    &
    \equiv
    \oint\mspace{-11mu}\int\!
    I( \vec{n}, \nu )\,
    \psi_{i\!j}( \vec{n}, \nu )\,
    \dd\nu\,
    \frac{\dd\Omega}{4\pi}.
    \label{eq:Jji}
\end{align}
The Rybicki-Hummer method of solving the non-LTE problem in PRD consists of the
following conceptual steps:
\begin{enumerate}
  \item The level populations are initialized with a guess solution.
        Popular choices are LTE populations, or populations based on assuming a
        zero radiation field, or a previously computed solution assuming CRD.
  \item The profile ratios $ \rho_{i\!j}( \vec{n}, \nu ) $ are initalized to
        unity for each PRD transition $ i {\to} j $.
  \item The formal solution of the radiative transfer equation is performed for
        all directions and frequencies to obtain intensities.
  \item Using the resulting intensities a preconditioned set of rate equations
        is formulated.
  \item The solution of the preconditioned equations gives an improved estimate
        of the true solution for the level populations.
  \item A number of extra PRD sub-iterations are performed, where the level
        populations are kept fixed and only intensity is redistributed:
  \begin{enumerate}
    \item The formal solution is computed again only for the frequencies in the
          PRD transitions.
    \item Using the new intensities, the redistribution integral is computed in
          each PRD transition to update the profile ratio
          $ \rho_{i\!j}( \vec{n}, \nu ) $.
    \item Using the new profile ratio, line opacities and emissivities are
          updated.
    \item Steps (a)--(c) are repeated until the changes to the radiation field
          are smaller than a desired value.
  \end{enumerate}
  \item From $ \rho_{i\!j}( \vec{n}, \nu ) $ updated opacities and emissivities
        are computed.
  \item Steps 3-7 are repeated until convergence.
\end{enumerate}
For more details we refer the reader to
\citet{2001ApJ...557..389U} 
and
\citet{1991A&A...245..171R,
       1992A&A...262..209R}. 
%

\subsection{The profile ratio $ \rho_{i\!j}( \vec{n}, \nu ) $}
\label{subsec:profile-ratio}

The core of the PRD scheme is the calculation of the redistribution integrals to
compute the profile ratio $ \rho_{i\!j}( \vec{n}, \nu ) $.
Following
\citet{1989A&A...213..360U}, 
we consider the PRD transition $ i {\to} j $ with all subordinate transitions
$ k {\to} j\colon k < j$, which share the same broadened upper level $ j $.
The redistribution integrals, computed in each subordinate transition
$ k {\to} j $ including the transition $ i {\to} j $ itself then contribute into
the profile ratio:
\begin{multline} \label{eq:rho-definition}
  \rho_{i\!j}( \vec{n}, \nu )
    =
    1 + \sum_{k < j} \dfrac{ n_k B_{k\!j} }{n_{\!j} P_{\!j}}
  \\
    {\times}\oint\mspace{-11mu}\int\!\!
    I\bigl( \vec{n}^\prime\!, \nu^\prime \bigr)
    \Biggl[
      \frac{ R_{k\!ji}\bigl( \vec{n}^\prime\!, \nu^\prime\!;
                             \vec{n},          \nu        \bigr) }
           { \varphi_{i\!j}( \vec{n}, \nu )                      }
      -
      \varphi_{k\!j}\bigl( \vec{n}^\prime\!, \nu^\prime \bigr)
    \Biggr]
    \dd\nu^\prime \frac{\dd\Omega^\prime\!}{4\pi},
\end{multline}
with $ R_{k\!ji} $ the inertial-frame redistribution function,
$ P_{\!j} = \sum_{k \neq j } P_{\!jk} $ the total depopulation rate out of the
upper level $ j $, where each double redistribution integral is taken over
photons with frequencies $ \nu^\prime\! $ coming along directions
$ \vec{n}^\prime $ within solid angles $ \Omega^\prime $ that are absorbed in
the subordinate lines $ k {\to} j $ including the resonance transition
$ i {\to} j $.

As the inertial frame redistribution function $ R_{k\!ji} $ we use a function
for transitions with a sharp lower level and a broadened upper level, which is
generalised for two possible cascades of a normal redistribution in the
$ i {\to} j {\to} i $ transition itself and resonance Raman scattering (also
called cross-redistribution) in $ k {\to} j {\to} i\colon k \neq i $ subordinate
transitions sharing the same upper level $ j $.
The function $ R_{k\!ji} $ consists of two weighted components
\citep{1982JQSRT..27..593H}: 
\begin{equation} \label{eq:Rkji}
  R_{k\!ji}
    \bigl(
      \vec{n}^\prime\!, \nu^\prime\!;
      \vec{n},          \nu
    \bigr)
    =
    \gamma
    R_{k\!ji}^\mathrm{II}
      \bigl(
        \vec{n}^\prime\!, \nu^\prime\!;
        \vec{n},          \nu
      \bigr)
    +
    (1 - \gamma)
    \varphi_{k\!j}\bigl( \vec{n}^\prime\!, \nu^\prime \bigr)
    \varphi_{i\!j}\bigl( \vec{n},          \nu        \bigr).
\end{equation}
Here $ R_{k\!ji}^\mathrm{II} $ is the inertial frame correlated redistribution
function for a transition with a sharp lower level and a broadened upper level,
representing coherent scattering.
The product $ \varphi_{k\!j} \varphi_{i\!j} $ approximates the inertial frame
non-correlated redistribution function $ R_{k\!ji}^\mathrm{III} $, representing
complete redistribution.
The quantity
$ \gamma \equiv P_{\!j} / \bigl( P_{\!j} + Q_{\!j}^\mathrm{E} \bigr) $
is the coherence fraction, i.e., the ratio of total rate $ P_{\!j} $ out of
level $ j $ to its sum with the rate of elastic collisions
$ Q_{\!j}^\mathrm{E} $.
Entering~\eqref{eq:Rkji} into~\eqref{eq:rho-definition} we finally obtain
\begin{multline} \label{eq:rho-angle-dependent}
  \rho_{i\!j}(\vec{n}, \nu)
    =
    \\
    1 + \gamma\!
    \sum_{k < j}
    \frac{ n_k    B_{k\!j} }
         { n_{\!j} P_{\!j} }
    \left[
      \oint\mspace{-11mu}\int\!\!
      I\bigl( \vec{n}^\prime\!, \nu^\prime \bigr)
      \frac{
        R_{k\!ji}^\mathrm{II}
        \bigl(
          \vec{n}^\prime\!, \nu^\prime\!;
          \vec{n},          \nu
        \bigr)
      }{
        \phi_{i\!j}( \vec{n}, \nu )
      }
      \dd\nu^\prime
      \frac{\dd\Omega^\prime\!}{4\pi}
      -
      \bar{J}_{\!k\!j}^\varphi
    \right]
\end{multline}
%

\subsection{The hybrid approximation}
\label{subsec:hybrid-approximation}

Computing the profile ratio using Eq.\,\eqref{eq:rho-angle-dependent} is
computationally expensive, as it involves evaluating the angle-dependent
redistribution function $ R_{k\!ji}^\mathrm{II} / \varphi_{i\!j} $ and computing
the double integral along each absorption frequency $ \nu^\prime $ and direction
$ \vec{n}^\prime $ for each emission frequency $ \nu $ and direction $ \vec{n}$.
\citet{2012A&A...543A.109L} 
demonstrated in plane-parallel computations using RH, that evaluating
Eq.\,\eqref{eq:rho-angle-dependent} takes about a factor 100 more time than
computing the formal solution for all angles and frequencies.

A common additional assumption to speed up the computations is to assume that
the radiation field $ I( \vec{n}, \nu ) $ is isotropic in the inertial frame%
\footnote{Here and below we distinguish between two reference frames.
          1) The \textit{inertial frame}, also called the
          \textit{laboratory frame} or the \textit{observer's frame}.
          2) The \textit{comoving frame}, where gas is locally at rest, in
          other texts also called the \textit{gas frame}, the
          \textit{fluid frame}, or the \textit{rest frame}.}.
In that case the angle integral in Eq.~\eqref{eq:rho-angle-dependent} can be
performed analytically to obtain
\begin{equation} \label{eq:rho-inertial-frame}
  \rho_{i\!j}( \nu ) =
    1 + \gamma\!
    \sum_{k < j}
    \frac{ n_k     B_{k\!j} }
         { n_{\!j} P_{\!j}  }
    \Biggl[
      \int\!\!
      J\bigl( \nu^\prime \bigr)\,
      \varg_{k\!ji}^\mathrm{II}\bigl( \nu^\prime\!, \nu \bigr)
      \,\dd\nu^\prime\!
      -
      \bar{J}_{k\!j}^\varphi
    \Biggr].
\end{equation}
with an angle-averaged redistribution function $ \varg_{k\!ji}^\mathrm{II} $
\citep[e.g.,][]{1986A&A...160..195G,
                1989A&A...213..360U}. 
The assumption of isotropy is not valid if the velocities in the model
atmosphere are larger than the Doppler width of the absorption profile, a common
situation in radiation-MHD simulations of the solar atmosphere
\citep{1974A&A....35..233M,
       1976ApJ...205..492M}. 

Therefore
\citet{2012A&A...543A.109L} 
proposed what they called the \textit{hybrid approximation}, which is a fast and
sufficiently accurate representation of Eq.\,\eqref{eq:rho-angle-dependent}
based on Eq.\,\eqref{eq:rho-inertial-frame}.
The idea behind the hybrid approximation is to take the redistribution integral
in the comoving frame assuming the radiation field to be isotropic there.

The angle-averaged intensity in the comoving frame
$ J^{\star\!}\bigl( \nu^{\star\!} \bigr) $
is computed from the intensity in the inertial frame $ I(\vec{n}, \nu) $, taking
into account Doppler shifts due to the local velocity $ \vec{\varv} $:
\begin{equation} \label{eq:J*}
  J^{\star\!}\bigl( \nu^{\star\!} \bigr)
    =
    \oint
    I
    \biggl(
      \vec{n},
      \nu^{\star\!}
      \biggl[
        1 + \frac{ \vec{n}\cdot\vec{\varv} }
                 { c                       }
      \biggr]
    \biggr)
      \frac{\dd \Omega}{4\pi}.
\end{equation}
The assumption of an isotropic specific intensity in the comoving frame
$ I^{\star\!}\bigl( \vec{n}, \nu^{\star\!} \bigr) =
  J^{\star\!}\bigl( \nu^{\star\!} \bigr) $
implies that we can use Eq.\,\eqref{eq:rho-inertial-frame} with the
angle-averaged redistribution function in the comoving frame to compute the
comoving-frame profile ratio $ \rho_{i\!j}^\star $ that depends only on
frequency:
\begin{equation} \label{eq:rho*}
  \rho_{i\!j}^{\star\!}\bigl( \nu^{\star\!} \bigr) =
    1 + \gamma\!
    \sum_{k < j}
    \frac{ n_k     B_{k\!j} }
         { n_{\!j} P_{\!j}  }
    \Biggl[
      \int\!\!
      J^{\star\!}\bigl( \nu^\prime \bigr)\,
      \varg_{k\!ji}^\mathrm{II}\bigl( \nu^\prime\!, \nu^{\star\!} \bigr)
      \dd \nu^\prime\!
      -
      \bar{J}_{k\!j}^\varphi
    \Biggr],
\end{equation}
where the integration is taken over the absorption frequency in the comoving frame.
Finally, the comoving profile ratio $ \rho_{i\!j}^{\star\!} $ is then
transformed into the inertial profile ratio $ \rho_{i\!j} $ using an inverse
Doppler transform so that $ \rho_{i\!j} $ again becomes dependent on both
frequency and angle:
\begin{equation} \label{eq:rho}
 \rho_{i\!j}( \vec{n}, \nu )
    =
    \rho_{i\!j}^{\star\!}
    \biggl(
      \nu
      \biggl[
        1 - \frac{ \vec{n}\cdot\vec{\varv} }
                 { c                       }
      \biggr]
    \biggr).
\end{equation}
Note that Eqs.\,(8)--(9) in
\citet{2012A&A...543A.109L} 
contain two mistakes.  First, both signs before $ \vec{n}\cdot\vec{u}/c $ have
to be inverted.  Second, the subscript `r' of $ \nu_\mathrm{r} $ in Eq.\,(8) has
to be dropped.  We give the correct transforms here in Eqs.\,\eqref{eq:J*} and
\eqref{eq:rho}.

There are several different possibilities for numerically computing the
interpolations in Eqs.\,\eqref{eq:J*} and \eqref{eq:rho}.
For Eq.\,\eqref{eq:J*}, storing $ I( \vec{n}, \nu ) $ and then performing the
interpolation and integration in one go is the most straightforward and rather
fast.
However, storing $ I( \vec{n}, \nu ) $ takes typically 1--4\,GiB of memory per
subdomain, which is too much on current-generation supercomputers, which
normally have 2\,GiB of memory per core (see the discussion in
Sect.\,\ref{sec:numerical-considerations}).

Another method is not to store the intensity, but instead to interpolate and
incrementally add it to $ J^{\star\!} $ during the formal solution.
This method does not require as much storage, and we implemented various
variants in Multi3D:
\begin{itemize}
  \item General interpolation -- interpolation indices and weights are computed
        on the fly.  This method is slow but requires no additional storage.
  \item Precomputed interpolation -- interpolation indices and weights are
        computed once, stored and then re-used.  This method is fast, but
        requires a large memory storage.
  \item Interpolation on an equidistant frequency grid -- this method is
        moderately fast, and requires only modest extra storage because the
        interpolation indices and weights depends only on the direction
        $ \vec{n} $ and not on frequency $ \nu $.
\end{itemize}
We describe each of these algorithms in detail in Appendix~\ref{sec:appendix}.

\section{Computation setup}
\label{sec:setup}
\subsection{Model atmospheres}
\label{subsec:atmospheres}
%
\begin{figure}
  \includegraphics[width=\columnwidth]{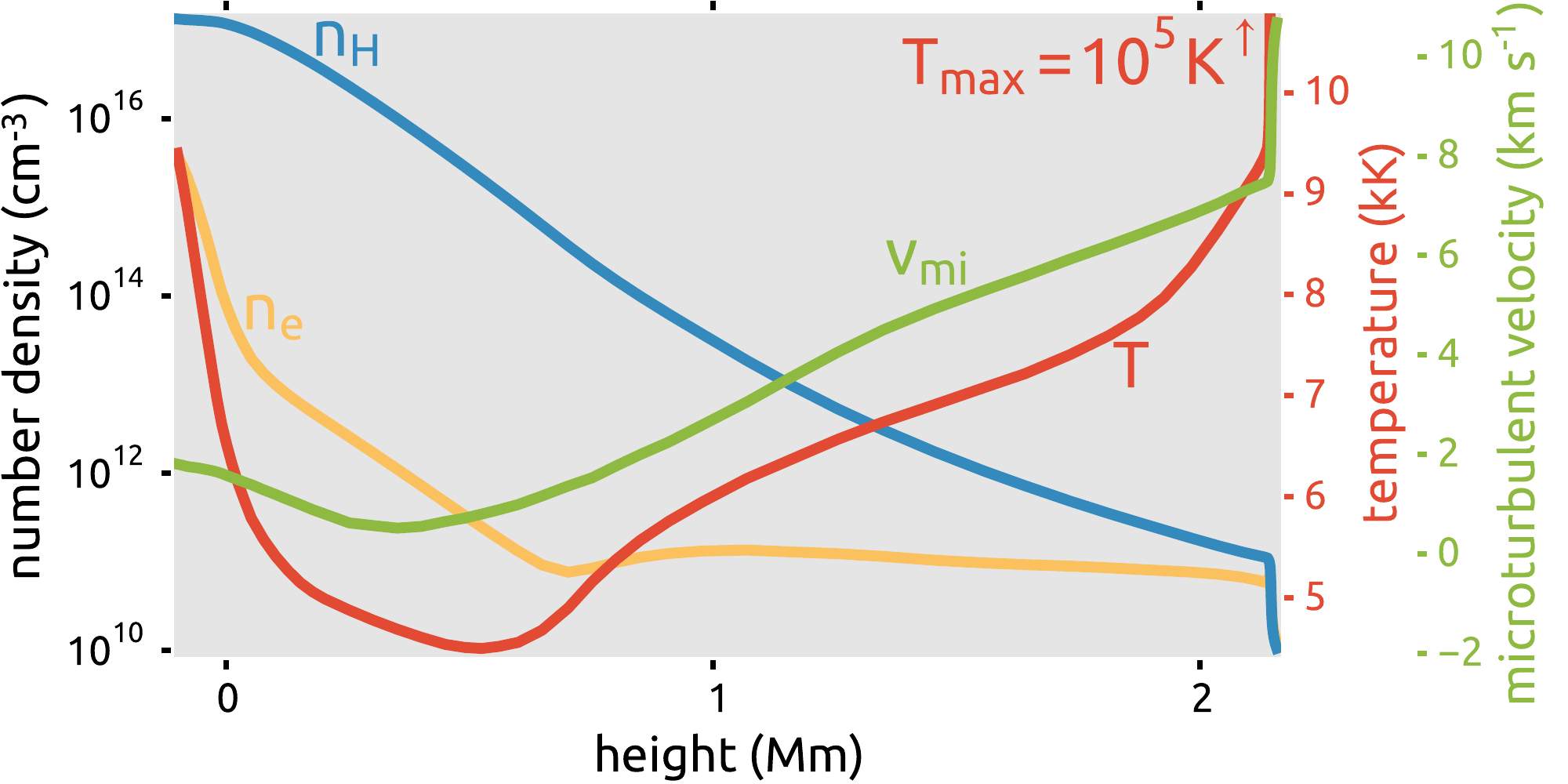}
  \\
  \includegraphics[width=\columnwidth]{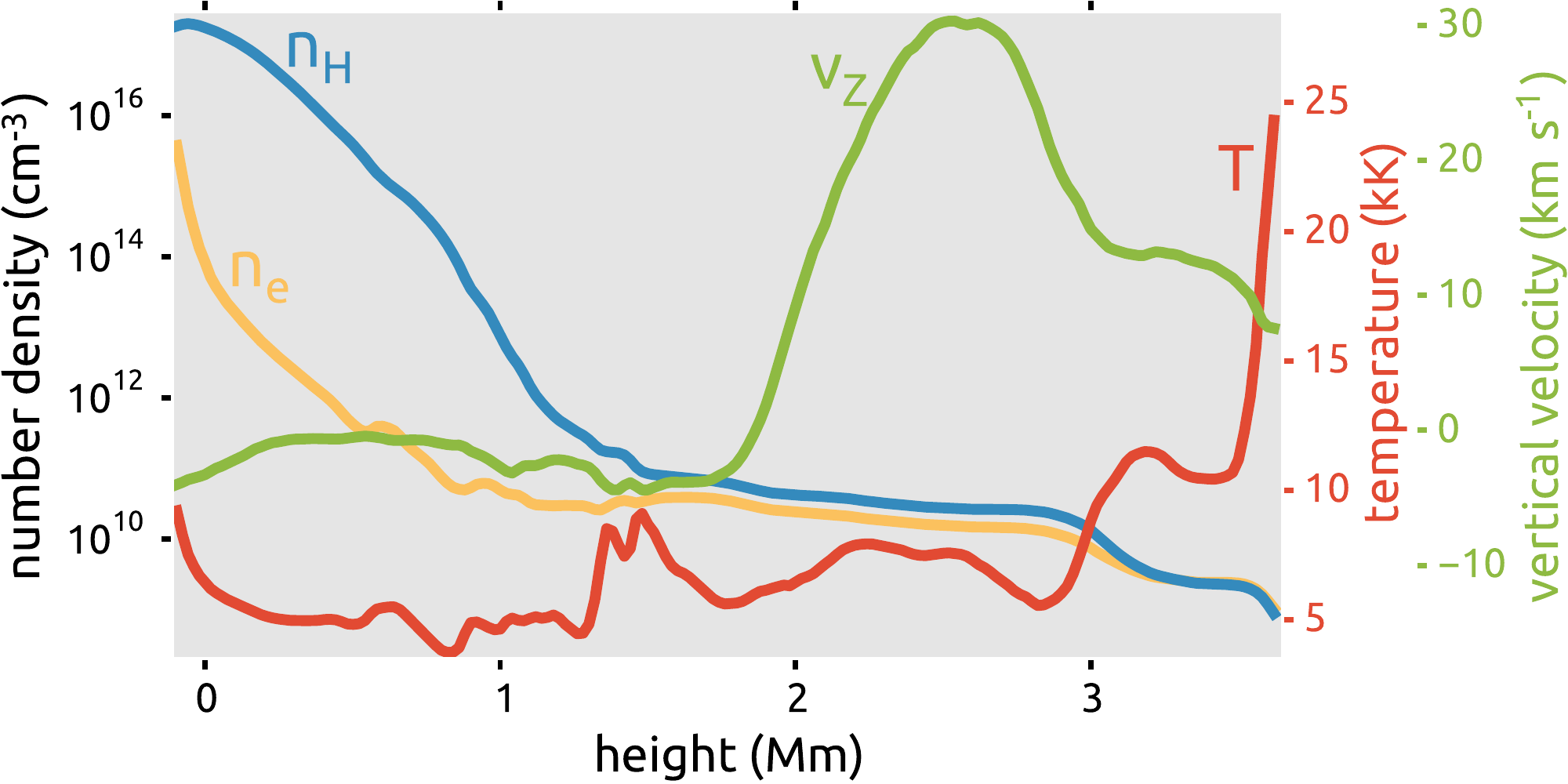}
  \caption{Height dependence of electron and hydrogen number densities,
    temperature, and velocities in the FAL-C (top) and the 1D Bifrost (bottom)
    model atmospheres.  The vertical velocity is zero in the FAL-C model
    and the microturbulent velocity is zero in the 1D Bifrost model.}
  \label{fig:1d-atmospheres}
\end{figure}
We tested our PRD algorithm using two different 1D model atmospheres with
different velocity distributions.  The standard FAL-C model atmosphere
\citep{1993ApJ...406..319F} 
shown in Fig.\,\ref{fig:1d-atmospheres}~(top) was modified to contain constant
vertical velocities of ${-}10,\,0,\,{+}10$\,km\,s$^{-1}$, two monotonically
increasing from ${-}10$ to ${+}10$\,km\,s$^{-1}$ and decreasing from ${+}10$ to
${-}10$\,km\,s$^{-1}$ velocity gradients, two discontinuous velocity jumps from
${+}10$ to ${-}10$\,km\,s$^{-1}$ and back representing toy shock waves, one
smooth velocity composition of two waves with 10 and 6\,km\,s$^{-1}$ amplitudes,
and one random normal velocity distribution with a zero mean and a standard
deviation of 10\,km\,s$^{-1}$.  Small insets in Fig.\,\ref{fig:m3d-vs-rh_falc}
illustrate some of these configurations.  For timing, convergence stability,
convergence acceleration, and other tests we employed a 1D column extracted from
a radiation-MHD simulation performed with the Bifrost code (details given
below).  This column corresponds to coordinates $ X = 0 $ and $ Y = 0 $ of the
original snapshot and is resampled into a vertical grid with 188~points in the
Z-direction.  Physical properties of this model atmosphere are illustrated in
Fig.\,\ref{fig:1d-atmospheres}~(bottom).  In this paper we simply refer to it as
the \textit{1D~Bifrost} model atmosphere.

To investigate how the algorithm performs in realistic situations, we used a
snapshot from the 3D radiation-MHD simulation by
\citet{2016A&A...585A...4C} 
corresponding to run time $ t = 3850 $\,s performed with the Bifrost code
\citep{2011A&A...531A.154G}. 
This snapshot contains a bipolar region of enhanced magnetic flux with unsigned
field strength of 50\,G in the photosphere.  The simulation box spans from the
bottom of the photosphere up to the corona having physical sizes of
$ 24 \times 24 \times 16.8 $\,Mm.  To save computational time, the horizontal
resolution of the snapshot was halved to 96\,km so that a new coarser grid with
$ 252 \times 252 \times 496 $ points is used.  This snapshot has, amongst other
studies, been used previously to study the formation of the hydrogen H$\alpha$
line
\citep{2012ApJ...749..136L}, 
the \ion{Mg}{II}~h\&k lines
\citep{2013ApJ...772...90L}, 
and the \ion{C}{II} lines
\citep{2015ApJ...811...81R}. 

\subsection{Model atoms} \label{subsec:atoms}

As a first test model atom we used a minimally sufficient model of \ion{Mg}{II}
with four levels and the continuum of \ion{Mg}{III}, where first two resonance
transitions represent the k~279.64\,nm and h~280.35\,nm doublet.
\citet{2013ApJ...772...89L} 
provides more details on this $ 4{+}1 $ level model atom.

As a second test model atom we used a model of \ion{H}{I} with five levels and
continuum of \ion{H}{II}, where the first two resonance transitions represent
the Ly\,$\alpha$ 121.6\,nm and the Ly\,$\beta$ 102.6\,nm lines.

For the timing experiments summarized in Table~\ref{tab:ndex-tdex}, we
initialized all corresponding model atom transitions on a frequency grid of
$ N_\nu^\mathrm{total} $ points with $ N_\nu^\mathrm{PRD} $ of them covered by
PRD lines.
PRD frequencies cover the two overlapping h\&k lines for the \ion{Mg}{II} model
atom and by the Ly\,$\alpha$ line for the \ion{H}{I} model atom.

\subsection{Numerical code} \label{subsec:code}

We perform numerical simulations using an extensively updated version of the
Multi3D code
\citep{2009ASPC..415...87L}.
The code solves the non-LTE radiative transfer problem for a given model atom in
a given 3D model atmosphere using the method of short characteristics to
integrate the formal solution of the transfer equation and the method of
multilevel accelerated $\Lambda$-iterations (M-ALI) with pre-conditioned rates
in the system of statistical equilibrium equations following
\citet{1991A&A...245..171R,1992A&A...262..209R}. 
The current version of the code allows MPI parallelization both for frequency
and 3D~Cartesian domains.
A typical geometrical subdomain size is $ 32 \times 32 \times 32 $ grid points.
The code also employs non-overshooting 3rd order Hermite interpolation for the
source function and the optical depth scale
\citep{2003ASPC..288....3A,2013A&A...549A.126I}.
We use the 24-angle quadrature (A4 set) from
\citet{carlson1963}.

To treat resonance lines in PRD, we implemented non-coherent scattering into the
code using the hybrid approximation with the three interpolation schemes
mentioned in Sects.\,\ref{subsec:PRDdesc}--\ref{subsec:profile-ratio}.
We validate our Multi3D implementation with RH computations.
In the RH code, the hybrid approximation for the PRD redistribution was
implemented by
\citet{2012A&A...543A.109L} 
for plane-parallel model atmospheres, and shown to give results consistent with
full angle-dependent PRD.

\section{Numerical considerations}
\label{sec:numerical-considerations}
\subsection{Computational expenses of interpolation methods}
\label{subsec:interpolation-timing}

We compared the time efficiency of each of the three interpolation methods
mentioned in Sect.\,\ref{subsec:hybrid-approximation} using system clock
routines in the computation of Eq.\,\eqref{eq:J*} (which we call forward
transform) and Eq.\,\eqref{eq:rho} (which we call backward transform).
Given a model atmosphere of $ N_\mathrm{X} N_\mathrm{Y} N_\mathrm{Z} $ spatial
grid points, using $ N_\mu $ angles, and a frequency grid with $ N_\nu $ knots,
we computed the mean time spent per spatial point per ray direction per
frequency for the forward transform ($t_\blacktriangleright$) and the backward
transform ($t_\blacktriangleleft$).
We also estimated how much memory each algorithm requires to store interpolation
indices and weights.
The numerical details of each interpolation method are given in
Appendix~\ref{sec:appendix}.

\subsubsection{General interpolation}
\label{subsubsec:general}

An advantage of the general interpolation algorithm is that it is applicable to
any type of frequency grid with arbitrary spacing between knots and requires
no memory storage for interpolation indices and weights.  Drawbacks are that it
is very time-consuming and it does not scale linearly with the number of grid
points.

To perform the forward transform, for each knot in the inertial frame two full
scans are done along knots in the comoving frame, resulting in $ 2N_\nu^2$
operations for the whole line profile. The mean time $ t_\blacktriangleright $
spent per knot therefore scales as $ O(N_\nu)$.

To perform the backward transform, for each knot in the inertial frame a
bisection search is done on the whole knot range in the comoving frame.
The search can be correlated%
\footnote{Usually this is called \textit{hunting} in textbooks on numerical
          methods.}
if the interpolation is performed along the frequency direction but this is
not the case in Multi3D.  The total number of operations is therefore around
$2 N_\nu \log_2 N_\nu$ and $t_\blacktriangleleft$ scales as $O(\log_2 N_\nu)$.

\subsubsection{Precomputed interpolation}
\label{subsubsec:precomputed}

The precomputed interpolation algorithm is applicable to any frequency grid,
just like the general interpolation method. It is both very fast and scales
well because interpolation indices and weights are computed only once based
on the atmosphere, the chosen angles, and the frequency grid.  The drawback is
that the algorithm is memory consuming.

For each knot in the inertial frame two operations are performed for the forward
transform, and thus $t_\blacktriangleright$ scales as $O(1)$.

In the backward transform, only one operation is done per inertial-frame
knot, resulting in $t_\blacktriangleleft = O(1)$.

If we have a single non-overlapping PRD line (such as Ly\,$\alpha$), the forward
transform requires storage of
$3 N_\nu N_\mu N_\mathrm{X} N_\mathrm{Y} N_\mathrm{Z} = 3 N_\Sigma$
numbers for interpolation indices and weights.  For the backward transform, we
need only $2 N_\Sigma$ numbers (see Appendix~\ref{sec:appendix} for details).
For single-precision (4-byte) integers and floating point numbers this results
in $20 N_\Sigma$ bytes.  For a typical run using the 3D model atmosphere we have
$ N_\nu \approx 200 $, $ N_\mu = 24 $, and a spatial subdomain size of $32^3$
grid points, so that storing the interpolation indices and weighs takes 2.9\,GiB
of memory per CPU.  The data can be stored more efficiently by using less
precise floating point numbers, so that the storage requirement is lowered to
$14 N_\Sigma$ bytes, corresponding to 2\,GiB for our typical use case.

\subsubsection{Precomputed interpolation on equidistant grid}
\label{subsubsec:equidistant}

The general interpolation algorithm is slow, and precomputed interpolation is
fast but requires a large amount of memory.  A solution to the memory
consumption of the latter is to use an equidistant frequency grid, so that the
interpolation weights and indices only depend on the ray direction and spatial
grid point, but not on the frequency%
\footnote{This idea was suggested by J.~de la Cruz Rodr\'{\i}guez.}.

The requirement to properly sample to absorption profile sets the grid spacing
to ${\sim}1$\,km\,s$^{-1}$.  Because PRD lines typically have wide wings, such
sampling leads to thousands of knots. This is undesirable because of the
corresponding increase in computing time.

It turns out that one can keep the advantages of an equidistant grid but not
have the disadvantage of the low speed by performing the formal solution only
for selected knots (which  we call \textit{real}).
The other knots (which we call \textit{virtual}) are ignored.
We keep all knots real close to nominal line center in order to resolve the
Doppler core, and have mainly virtual knots in the line wing where a coarse
frequency resolution is acceptable.
We can then still use frequency-independent interpolations coefficients, at the
price of a small computational and memory overhead to track and store real and
virtual knot locations and certain relations between them.
Usually, $\sim$90\,\% of the real knots are in the line core.

The number of operations required to perform the forward transform depends on
whether the real frequency knot is in the line core or in the line wings.  The
number of operations to perform the forward transform on all $N_\nu$ knots is
around $2 C_\blacktriangleright N_\nu$, where $1 < C_\blacktriangleright \la 10$
is a factor that depends on $N_\nu$ and on the fine grid resolution.  The
$t_\blacktriangleright$ scales as $O(1)$ but the number of operations is higher
than for precomputed interpolation.

In the backward transform, the algorithm is essentially the same.
The total number of operations for all frequency knots is
$2 C_\blacktriangleleft N_\nu$, where
$1 < C_\blacktriangleleft < C_\blacktriangleright \la 10$ so that
$t_\blacktriangleleft$\ scales as $O(1)$.

The interpolation indices and weights are the same for both the forward and
backward transforms.
To store them we need $ 2 N_\mu N_\mathrm{X} N_\mathrm{Y} N_\mathrm{Z} $
four-byte numbers resulting in 6\,MiB plus some auxiliary variables amounting to
another $0.3$\,MiB.

\subsubsection{Timing experiments}
\label{subsubsec:timing-transforms}
%
\begin{figure}
  \includegraphics[width=\columnwidth]{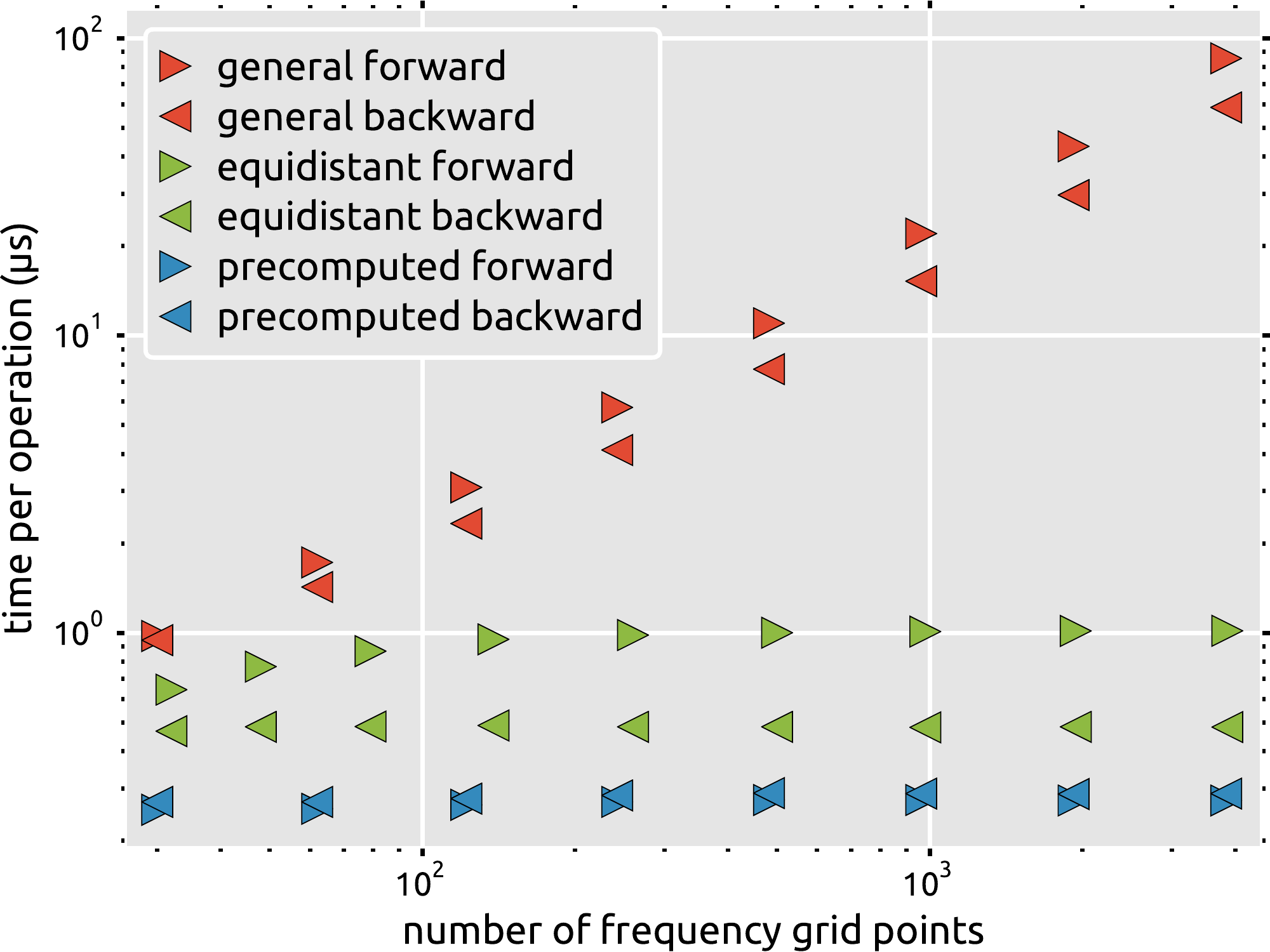}
  \caption{Dependence of the mean interpolation time spent per frequency per
    direction at spatial coordinate point on the total number of frequency
    points in one PRD line profile.  Times are given for the forward
    ($\blacktriangleright$) and the backward ($\blacktriangleleft$)
    transforms for three different algorithms: general (red), precomputed
    (blue), and precomputed equidistant (green) interpolation.  The measurements
    were done for the \ion{Mg}{II} h\&k lines in the 1D~Bifrost model
    atmosphere.}
  \label{fig:timing}
\end{figure}
We measured the performance of the three algorithms in the 1D~Bifrost model
atmosphere using the 5-level \ion{Mg}{II} model atom with the two overlapping
h\&k lines treated in PRD.  The number of frequency points in each line profile
was varied from from 15 to 1\,900.  We used a ten-point Gauss-Legendre
quadrature for the angle integrals.  The calculations were run on a computer
with
Intel\textsuperscript{\textregistered}
Xeon\textsuperscript{\textregistered}
CPU E5-2697 v3 2.60~GHz cores.

Figure~\ref{fig:timing} shows our measurements for both transforms using the
three interpolation algorithms.  In the general algorithm, the total computing
time scales as $O(N_\nu^2)$ , which makes the algorithm unsuitable for wide
chromospheric lines with many frequency points.

Contrary to that, the total time for the precomputed and the precomputed
equidistant algorithms scales linearly as $O(N_\nu)$.  The equidistant algorithm
is slower than the precomputed one, but of the same order.  The extra
book-keeping required for the equidistant interpolation causes a small
performance penalty compared to the precomputed case.  This is however a small
price for the large saving in memory consumption compared to the precomputed
algorithm.

We preferably use the precomputed algorithm in 1D and small 3D model
atmospheres, while for big 3D model atmospheres we use the equidistant one.

\subsection{Truncated frequency grid}
\label{subsec:truncate}
%
\begin{figure*}[t]
  \begin{minipage}{\textwidth}
    \includegraphics[width=0.6812\textwidth,trim=0    0 0 0,clip=true]{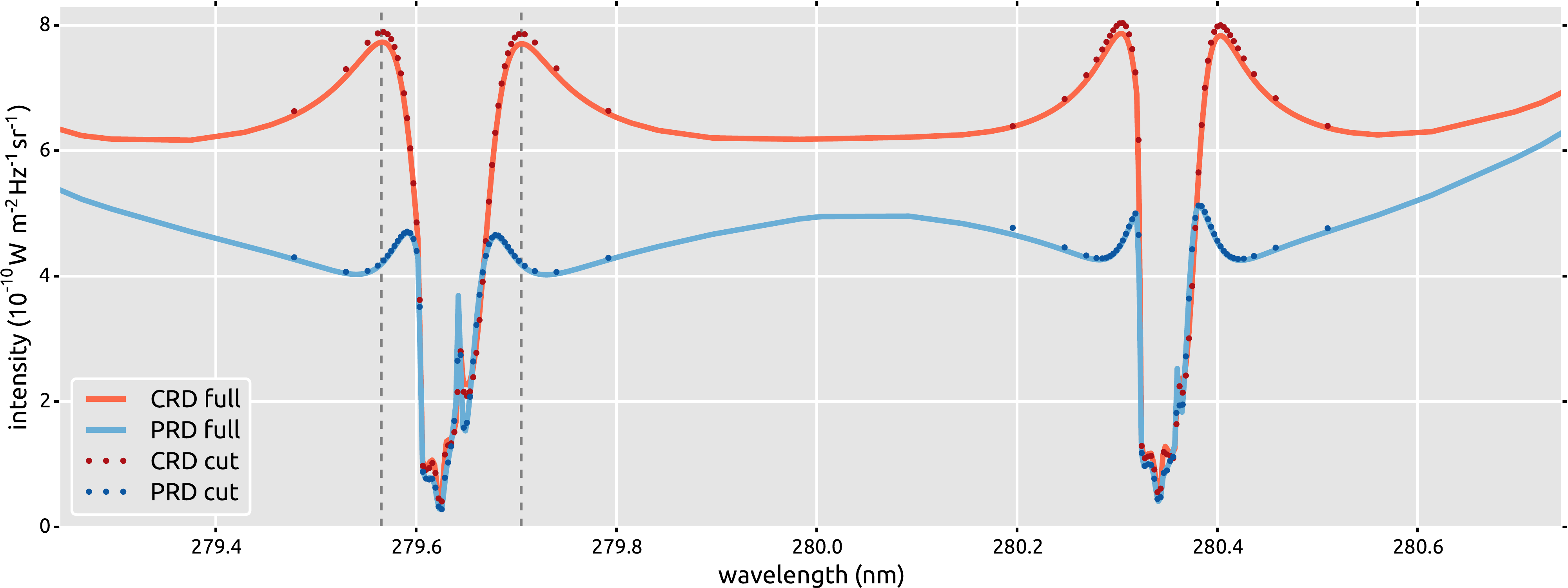}\hfil
    \includegraphics[width=0.3188\textwidth,trim=13mm 0 0 0,clip=true]{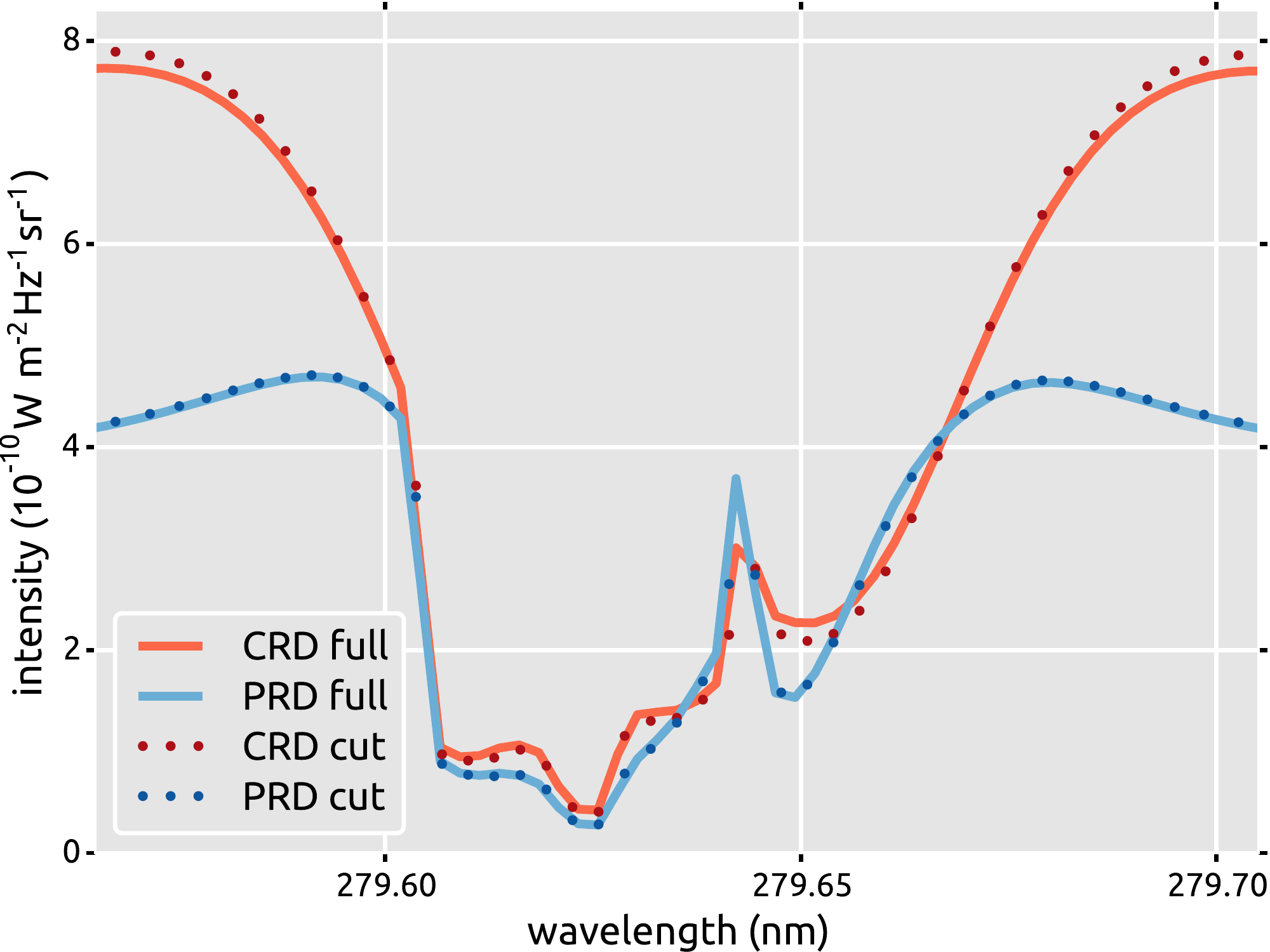}
  \end{minipage}
  \caption{Intensity profiles of the \ion{Mg}{II} h\&k lines at
    $ \mu_\mathrm{Z} = 0.953 $ computed on fully-winged (solid) or truncated
    (dots) frequency grids in CRD (red) or PRD (blue).
    Left: wide wavelength range showing both the k~279.6\,nm and the
    h~280.3\,nm lines.
    Dashed vertical lines show the wavelength range for the right panel.
    Right: close-up of the k~line core.
    The model atmosphere is 1D~Bifrost.}
  \label{fig:truncated-grid}
\end{figure*}
Only the core and inner wings of strong chromospheric lines are formed in the
chromosphere, which is the region where observers of such lines are mainly
interested in.  It is therefore interesting to see whether ignoring the extended
wings has an influence on the intensity in the central part of the line.  If
this is not the case we can lower the number of frequency points in the line
profile and so speed up the computations.

Therefore we tested whether a truncation of the profile wings makes a
difference.  For the \ion{Mg}{II} h\&k lines, we performed a computation using
equidistant interpolation where we resolve the core on a fine frequency grid up
to $\pm 70$\,km\,s$^{-1}$ around the line center, and use the coarse grid out to
$\pm 160$\,km\,s$^{-1}$, so that we only cover the core of the line profiles
from just outside the h$_1$ and k$_1$ minima.

In Fig.\,\ref{fig:truncated-grid} we show the differences between the truncated
and the fully-winged profiles of the \ion{Mg}{II} h\&k lines in CRD and PRD.
In CRD, there are small differences around the k$_2$ emission peaks while in PRD
differences are negligible.  We thus conclude that we can use a truncated grid.

The main reason why truncated grids work for chromospheric lines is that the
density in the chromosphere is so low that the damping wings are weak, so that
the extinction profile is well-approximated by a Gaussian.  The averages over
the extinction profile of the form
\begin{equation}
  \oint\mspace{-11mu}\int
  \dotsi
  \varphi( \vec{n}, \nu )\,
  \dd\nu
  \frac{\dd\Omega}{4\pi}
\end{equation}
that enter into the non-LTE problem are thus dominated by a range of few Doppler
widths around the nominal line center in the local comoving frame.  As long as
the frequency grid covers this range, the radiative transfer in the chromosphere
is accurate.

In CRD there is a larger inaccuracy because photons absorbed in the line wings
tend to be re-emitted in the line core due to Doppler diffusion
\citep{2014tsa..book.....H}.
In PRD, this effect is weak, because the complete redistribution component of
the redistribution function from Eq.\,\eqref{eq:Rkji} is strongly reduced
because the coherency fraction $ \gamma $ is nearly equal to one, so that the
wings and the core are effectively decoupled and do not significantly influence
each other.  This is a rare instance where the inclusion of PRD effects actually
makes computations easer instead of harder.

\section{Results}
\label{sec:results}
\subsection{Comparison to RH}
\label{subsec:intensity-comparison}
%
\begin{figure*}
  \begin{minipage}{\textwidth}
    \includegraphics[width=0.3518\textwidth,trim=0      14mm 0 0,clip=true]{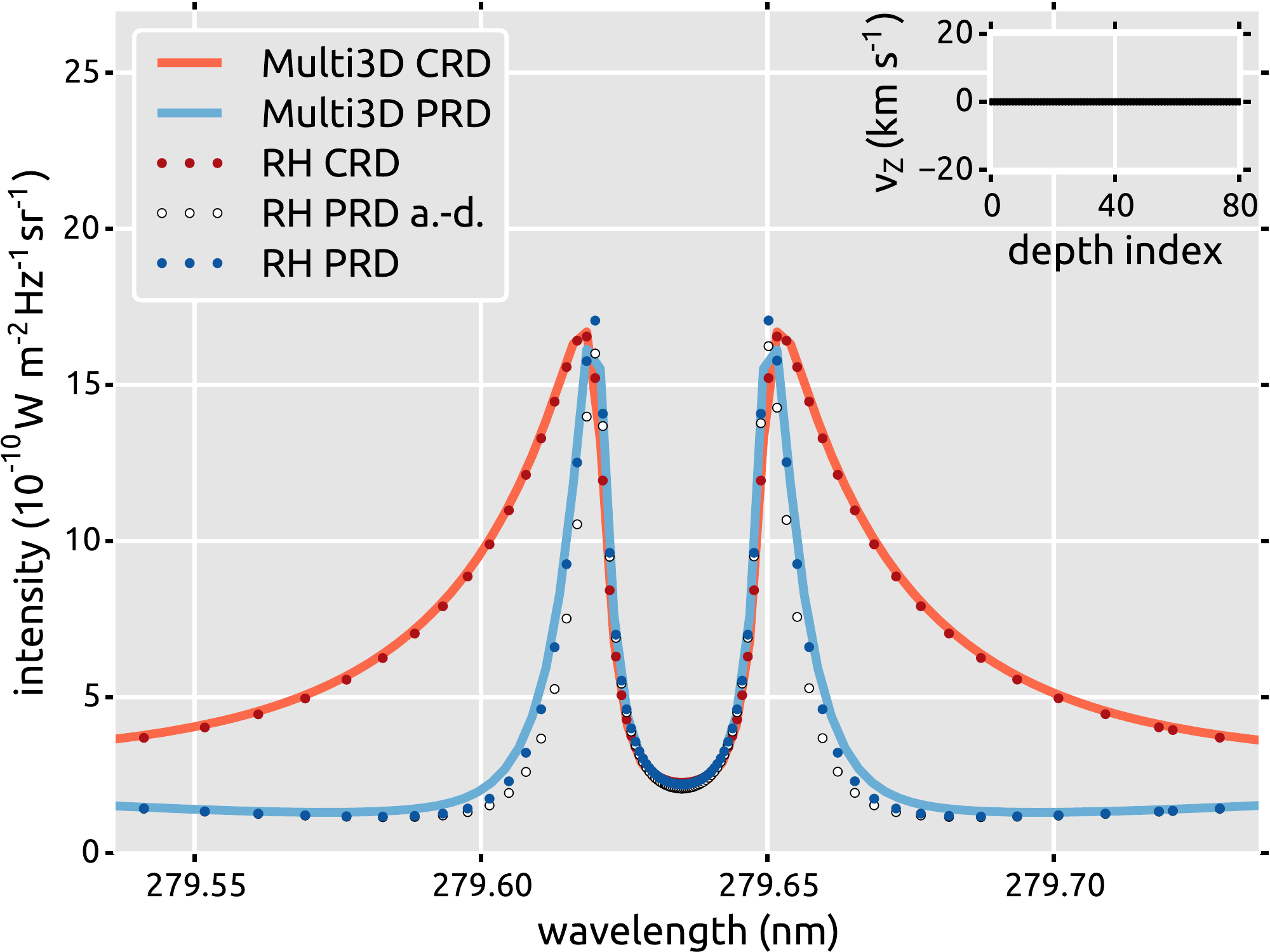}\hfil
    \includegraphics[width=0.3232\textwidth,trim=16.5mm 14mm 0 0,clip=true]{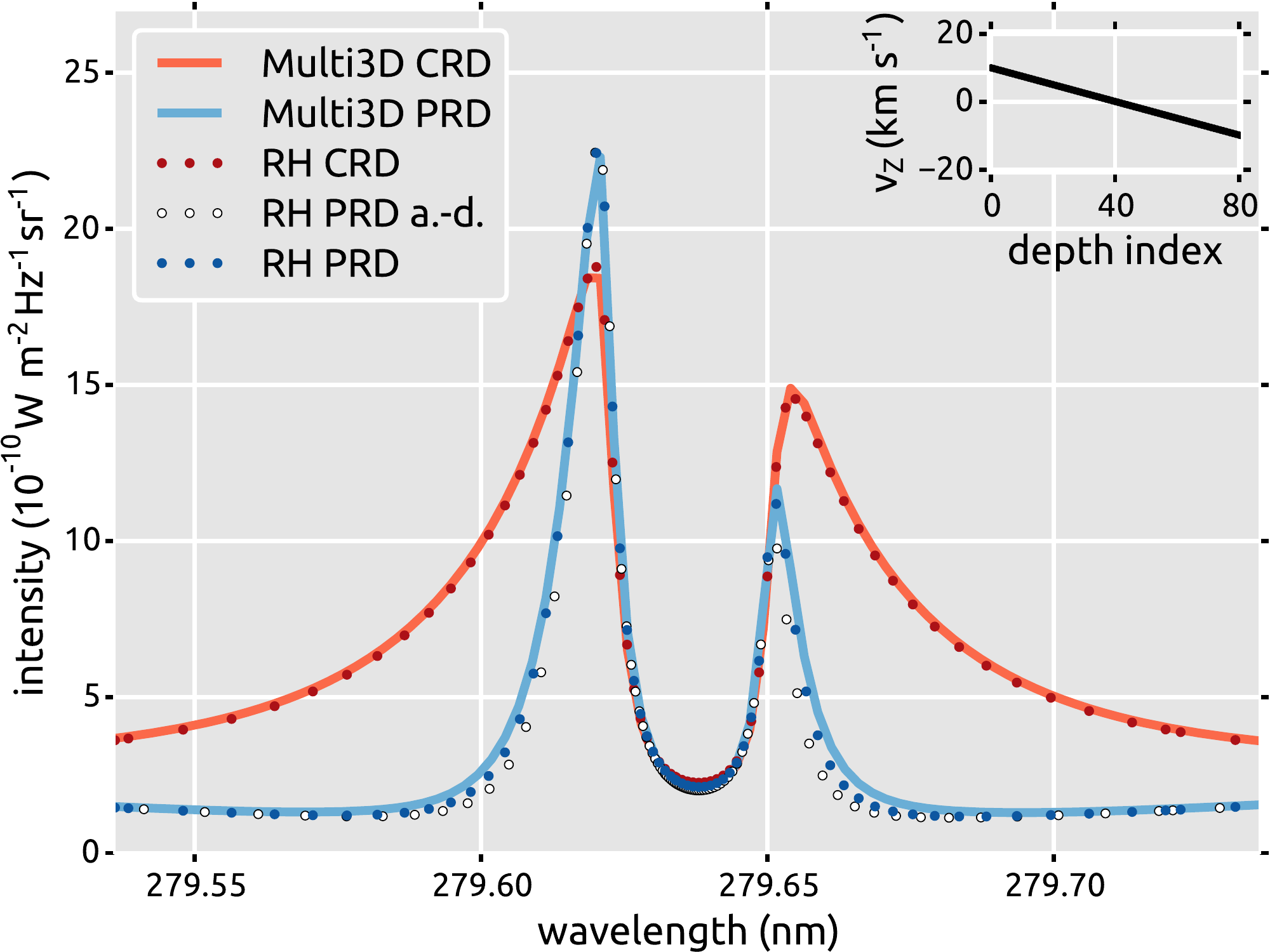}\hfil
    \includegraphics[width=0.3232\textwidth,trim=16.5mm 14mm 0 0,clip=true]{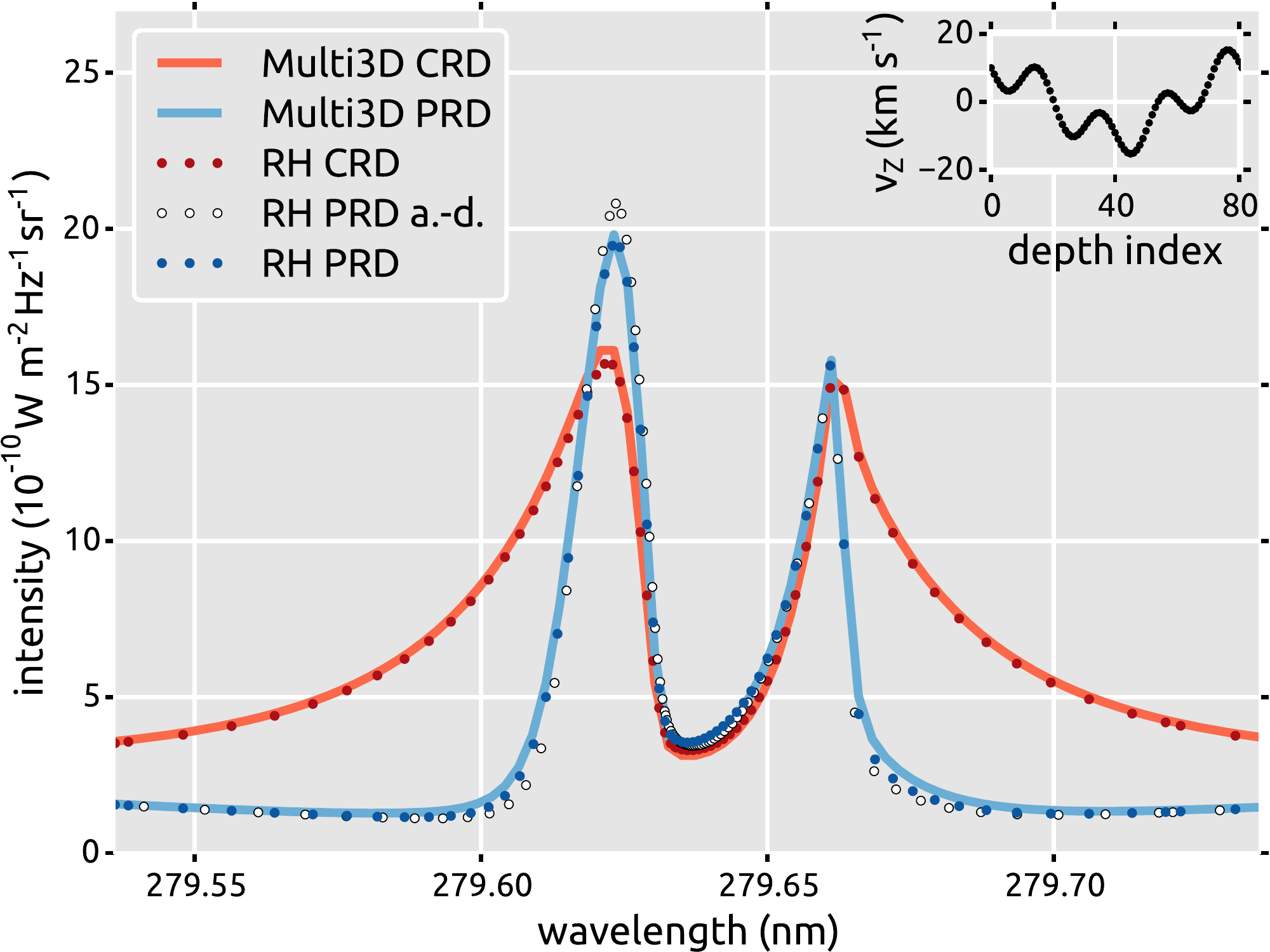}
    \\
    \includegraphics[width=0.3518\textwidth,trim=0      0    0 0,clip=true]{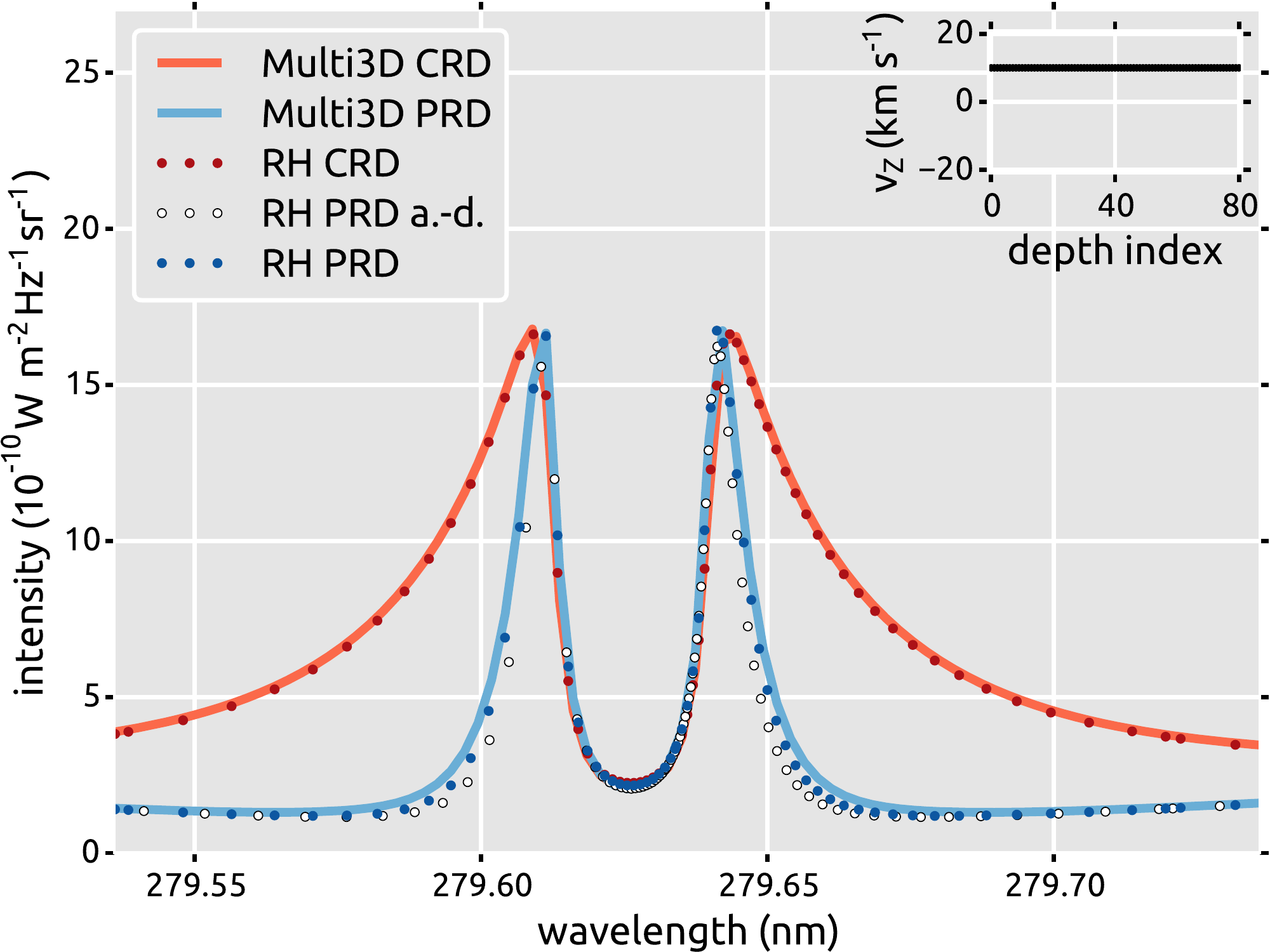}\hfil
    \includegraphics[width=0.3232\textwidth,trim=16.5mm 0    0 0,clip=true]{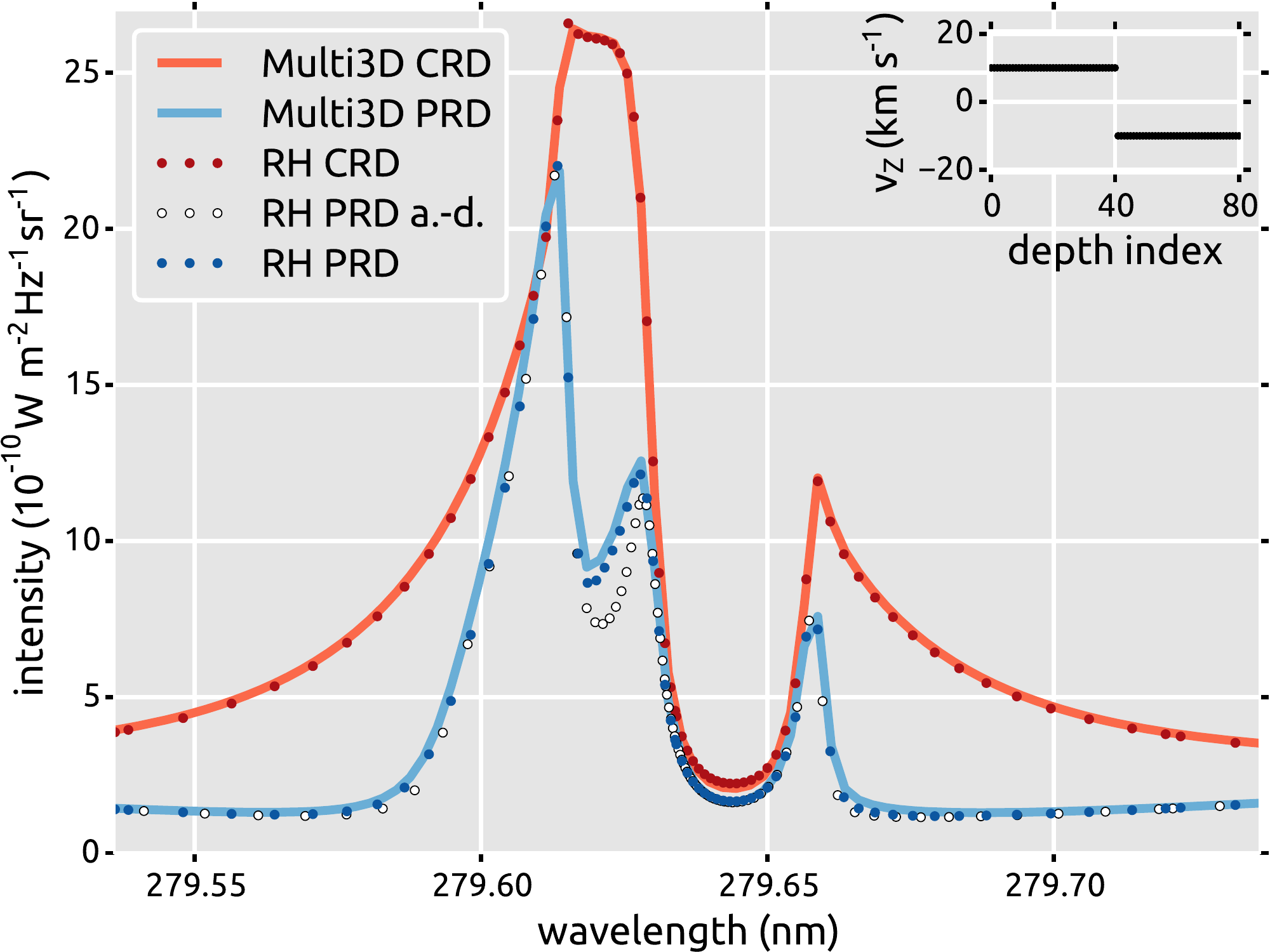}\hfil
    \includegraphics[width=0.3232\textwidth,trim=16.5mm 0    0 0,clip=true]{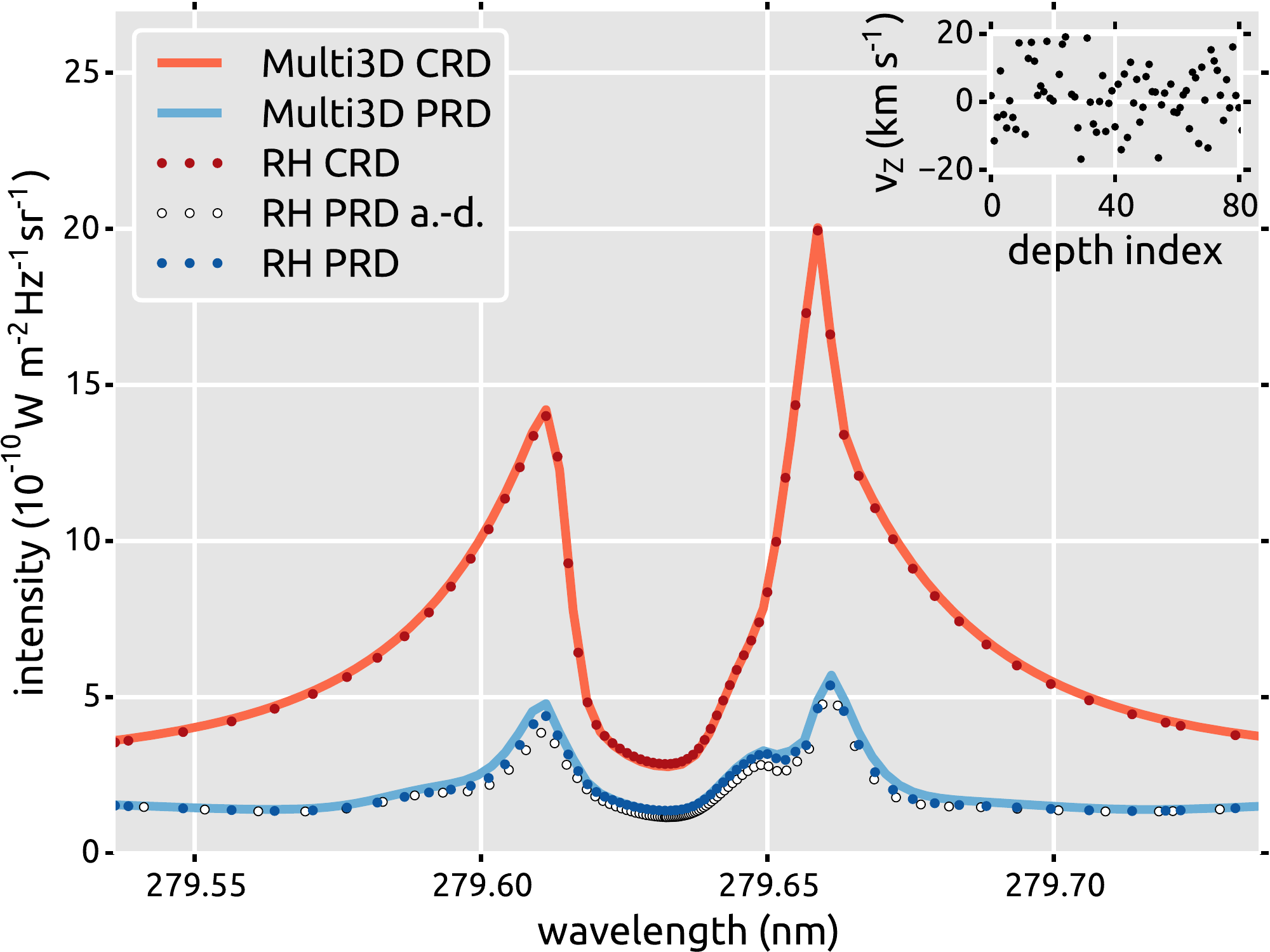}
  \end{minipage}
  \caption{Intensity profiles of the \ion{Mg}{II} h~279.64\,nm line at
    $ \mu_\mathrm{Z} = 0.953 $ computed using the Multi3D (solid) or the RH
    (dots) codes in CRD (red) or PRD (blue, white).  PRD profiles are computed
    using either the hybrid approximation (blue) or full angle-dependent
    redistribution (white).  The model atmosphere is FAL-C with modified
    vertical velocities.  The profiles are shown for six different velocity
    configurations illustrated by the inset plots.}
  \label{fig:m3d-vs-rh_falc}
\end{figure*}
\begin{figure}
  \includegraphics[width=\columnwidth]{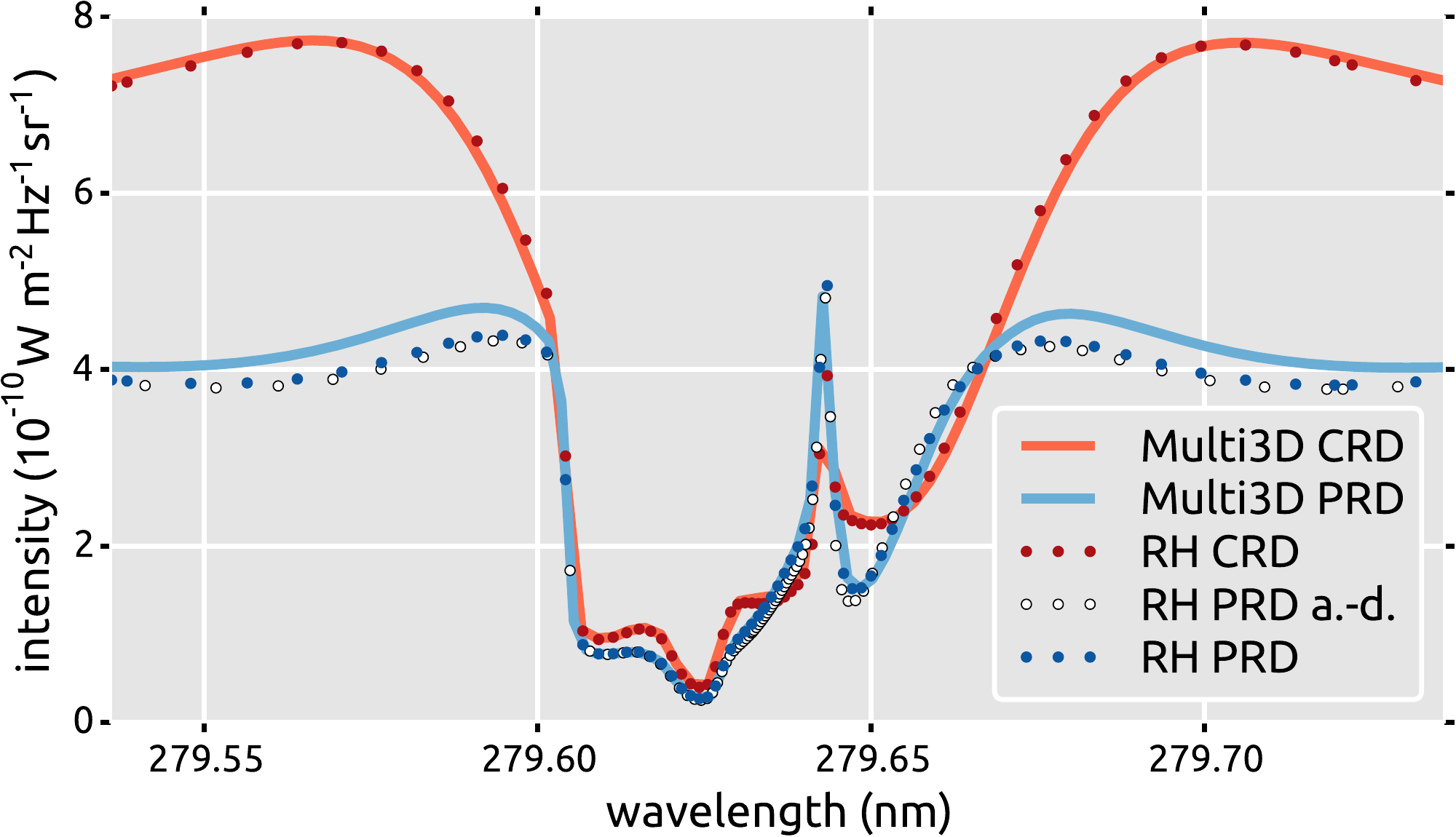}
  \caption{Same as in Fig.\,\ref{fig:m3d-vs-rh_falc} but for the 1D~Bifrost
    model atmosphere shown in
    Fig~\ref{fig:1d-atmospheres}.}
  \label{fig:m3d-vs-rh_1dbifrost}
\end{figure}
To validate our method and implementation, we simulated the \ion{Mg}{II} h\&k
lines in CRD and PRD using both the Multi3D and the RH code, and compared the
resulting intensities.  The PRD calculations with RH were done using both the
angle-dependent and the hybrid PRD algorithms, while Multi3D computations were
done only using the hybrid algorithm.

Intensity profiles of both \ion{Mg}{II} h\&k lines were calculated in the FAL-C
model atmosphere with modified vertical velocity distributions given in the
insets in Fig.\,\ref{fig:m3d-vs-rh_falc}.  The simplest cases are those with
constant velocities, the most extreme is that with the random normal
distribution.  Note that the case of a random velocity distribution only
illustrates that the method is stable in extreme situations.

Both codes produce CRD intensity profiles in a good agreement.  Comparison of
the RH computations using the hybrid and angle-dependent algorithms shows that
the hybrid algorithm is accurate for all cases except the strong toy-shock wave.
In such strong gradients the assumption of isotropy of the radiation in the
comoving frame is not accurate.

Both codes yield the same hybrid PRD intensities except for some small
differences around the inner wings close to the line core (see PRD intensity
values at 279.58--279.61\,nm and 279.65--279.68\,nm in
Fig.\,\ref{fig:m3d-vs-rh_falc}).  This effect is caused by slight differences in
frequency grids used by the codes and differences in line broadening recipes
that lead to differences in coherency fraction $\gamma$.

We performed the same computations for the 1D~Bifrost column.  Corresponding
profiles are given in Fig.\,\ref{fig:m3d-vs-rh_1dbifrost}.  Also here both codes
and algorithms yield excellent agreement.

\subsection{Convergence properties}
\label{subsec:convergence}

Numerical PRD operations are computationally expensive.  In order to minimize
the computing time we investigated how many PRD subiterations are needed to lead
to a converged solution, how the number of subiterations influences the
convergence speed of the ALI iterations, and whether convergence acceleration
using the Ng method
\citep{1974JChPh..61.2680N} 
%
can be applied.

\subsubsection{Number of PRD subiterations}
\label{subsubsec:nprd}
%
\begin{figure*}
  \begin{minipage}{\textwidth}
    \includegraphics[width=0.2758\textwidth,trim=0    14mm 0 0,clip=true]{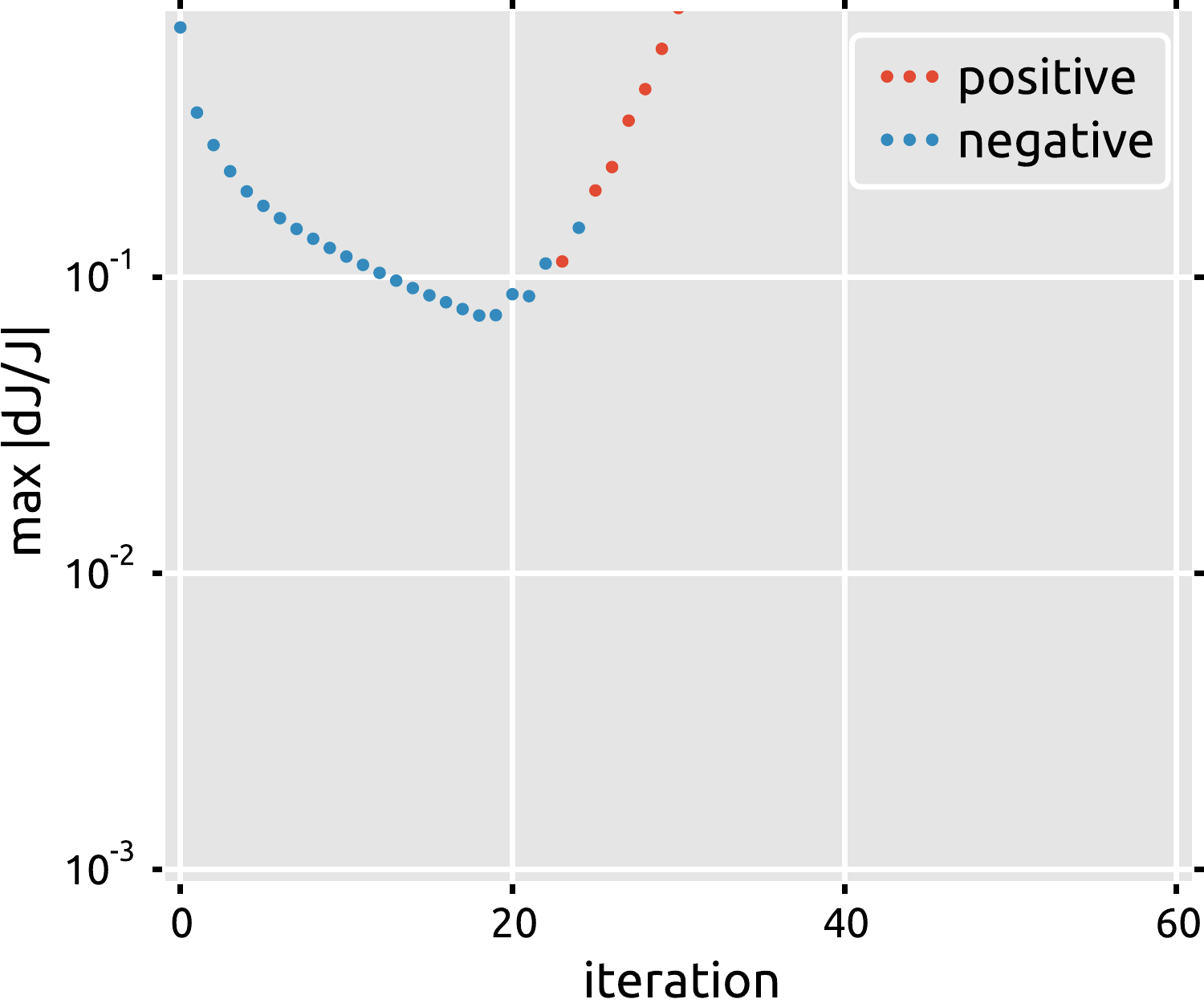}\hfil
    \includegraphics[width=0.2414\textwidth,trim=19mm 14mm 0 0,clip=true]{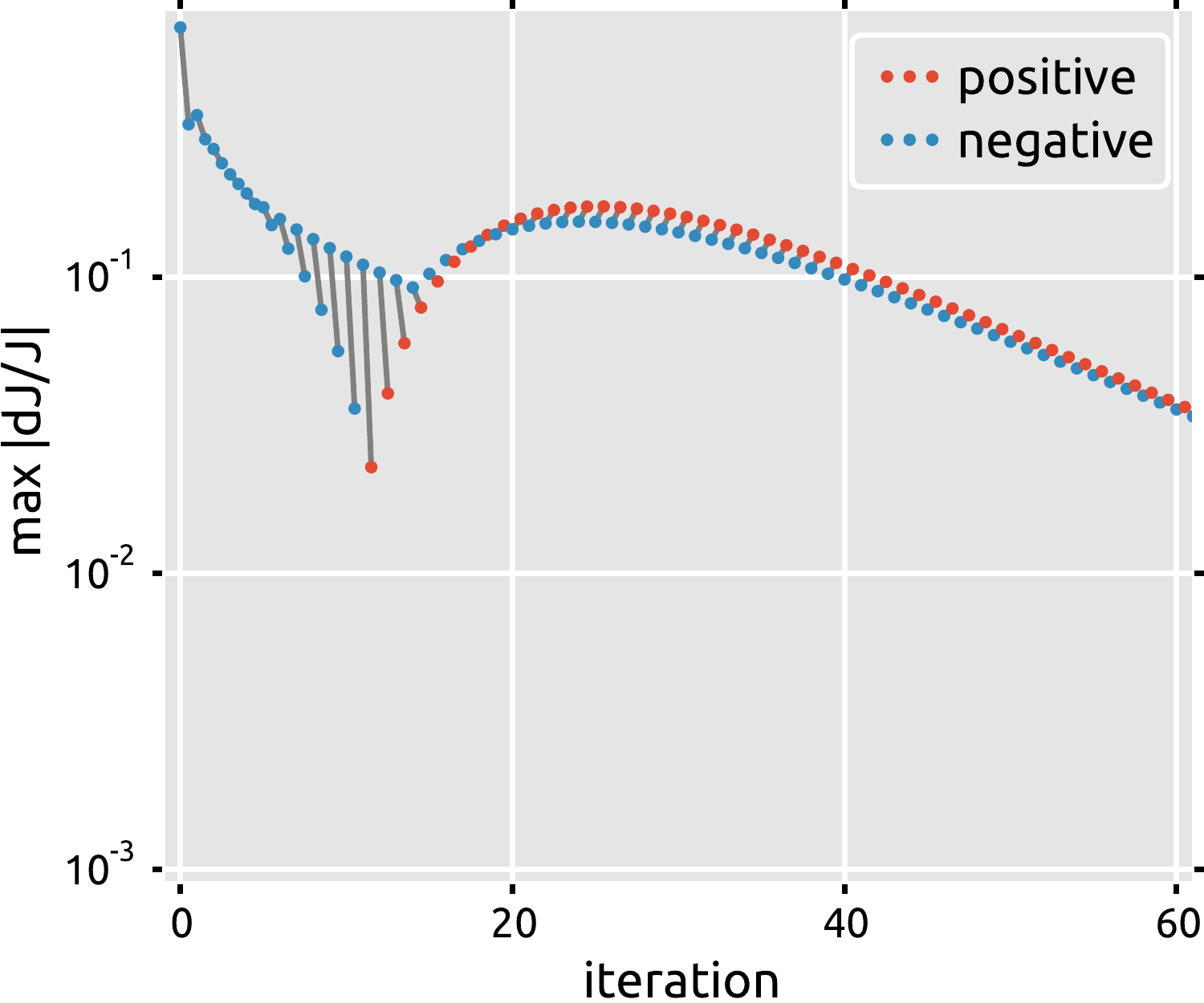}\hfil
    \includegraphics[width=0.2414\textwidth,trim=19mm 14mm 0 0,clip=true]{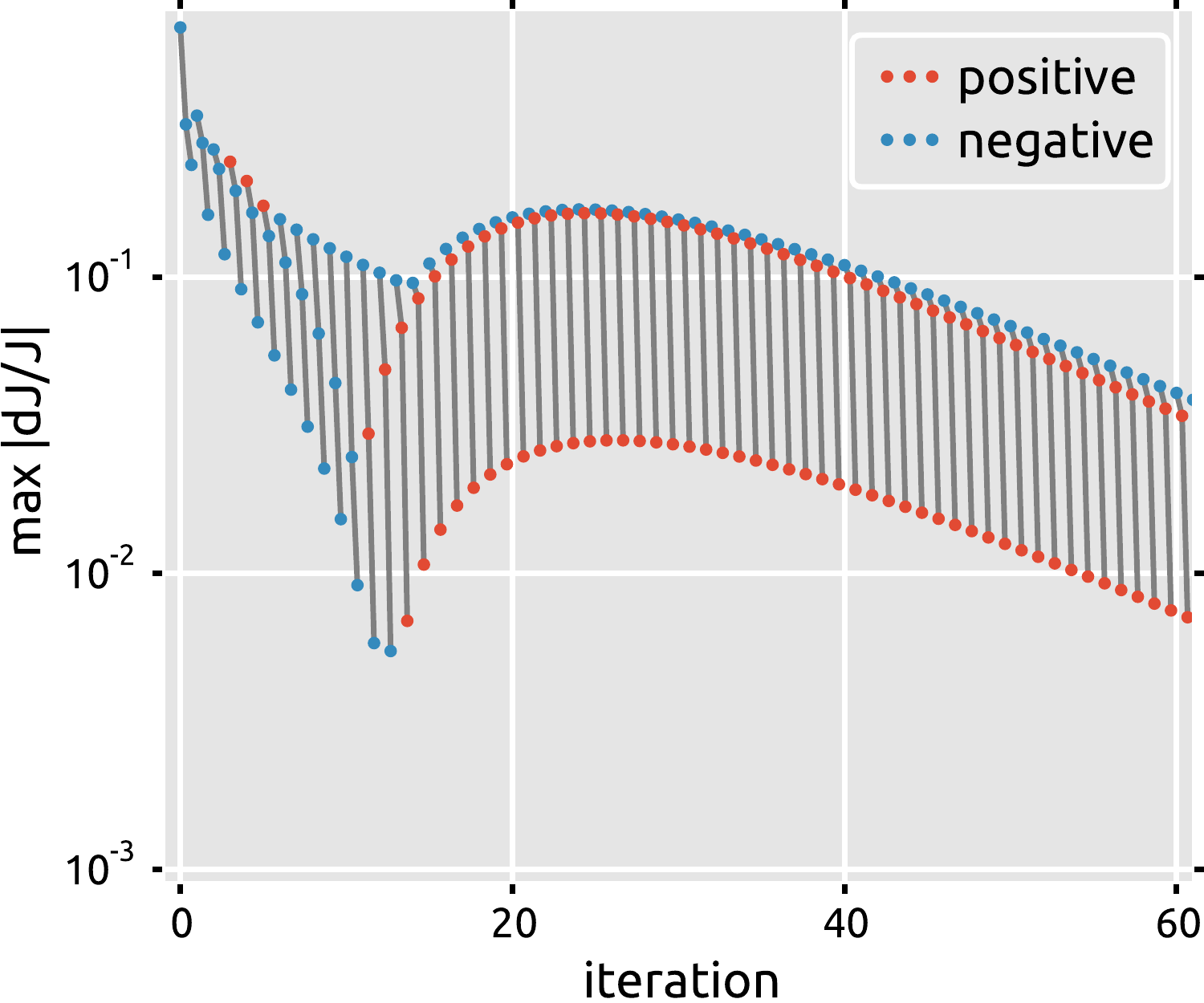}\hfil
    \includegraphics[width=0.2414\textwidth,trim=19mm 14mm 0 0,clip=true]{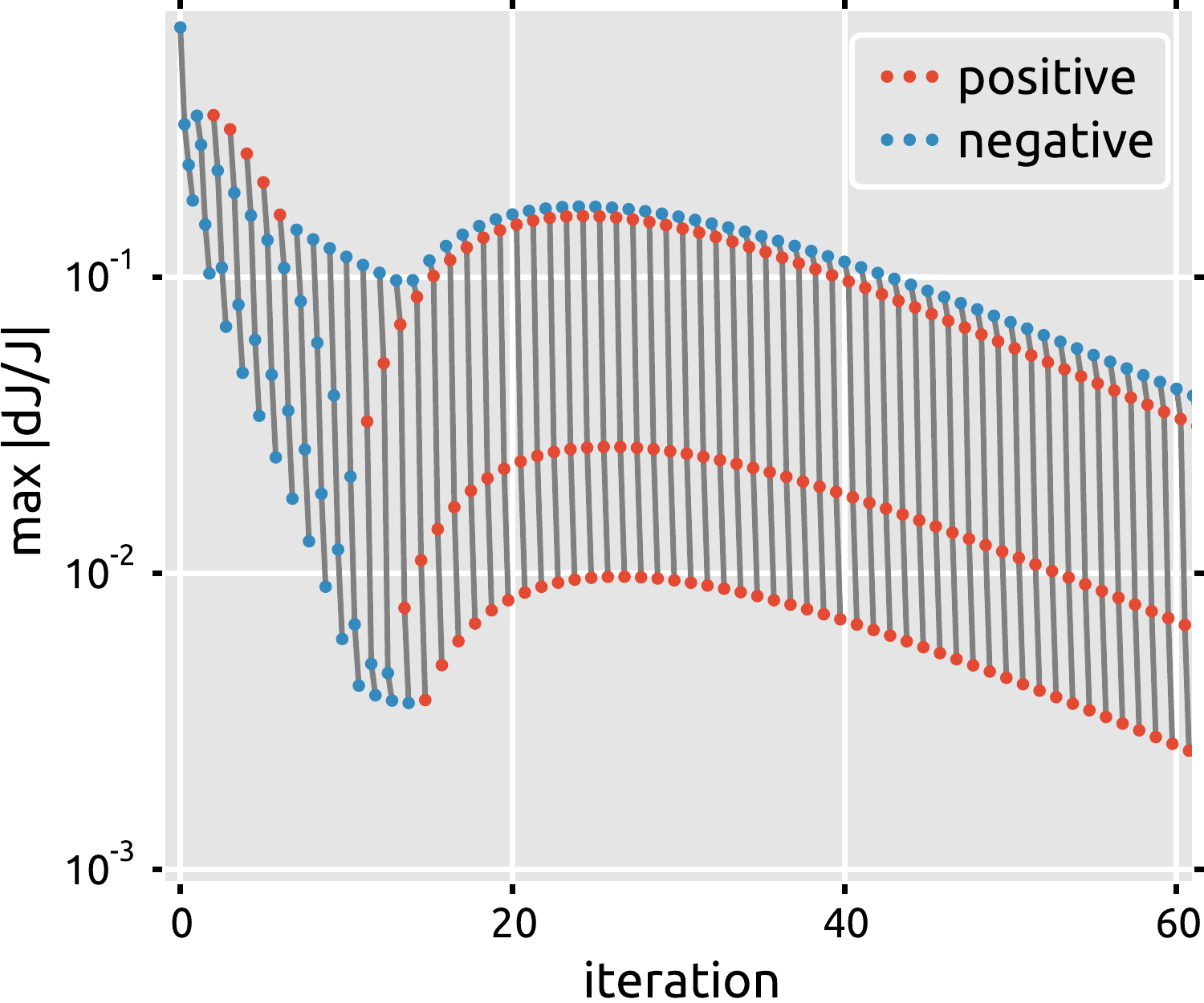}
    \\
    \includegraphics[width=0.2758\textwidth                             ]{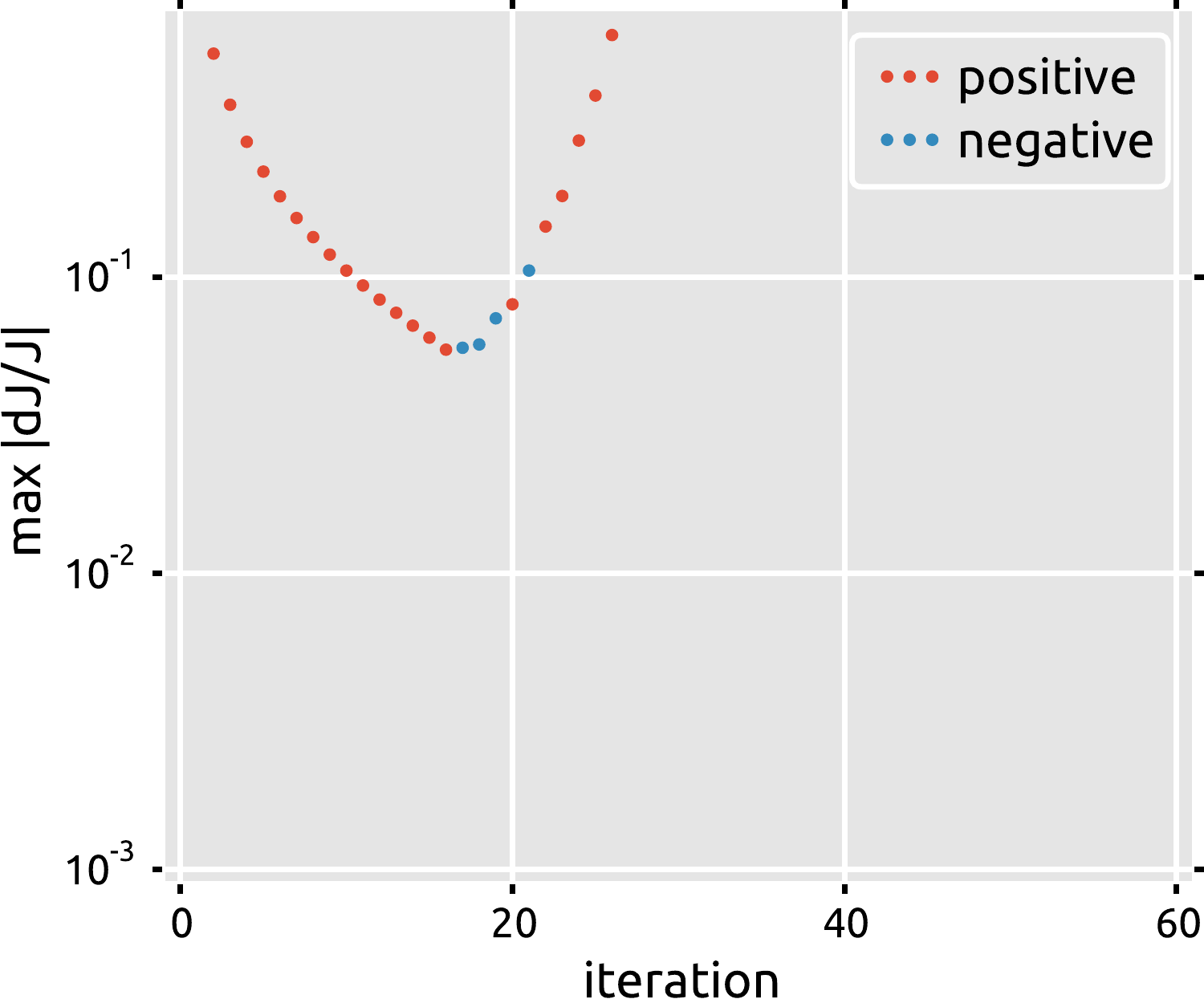}\hfil
    \includegraphics[width=0.2414\textwidth,trim=19mm 0    0 0,clip=true]{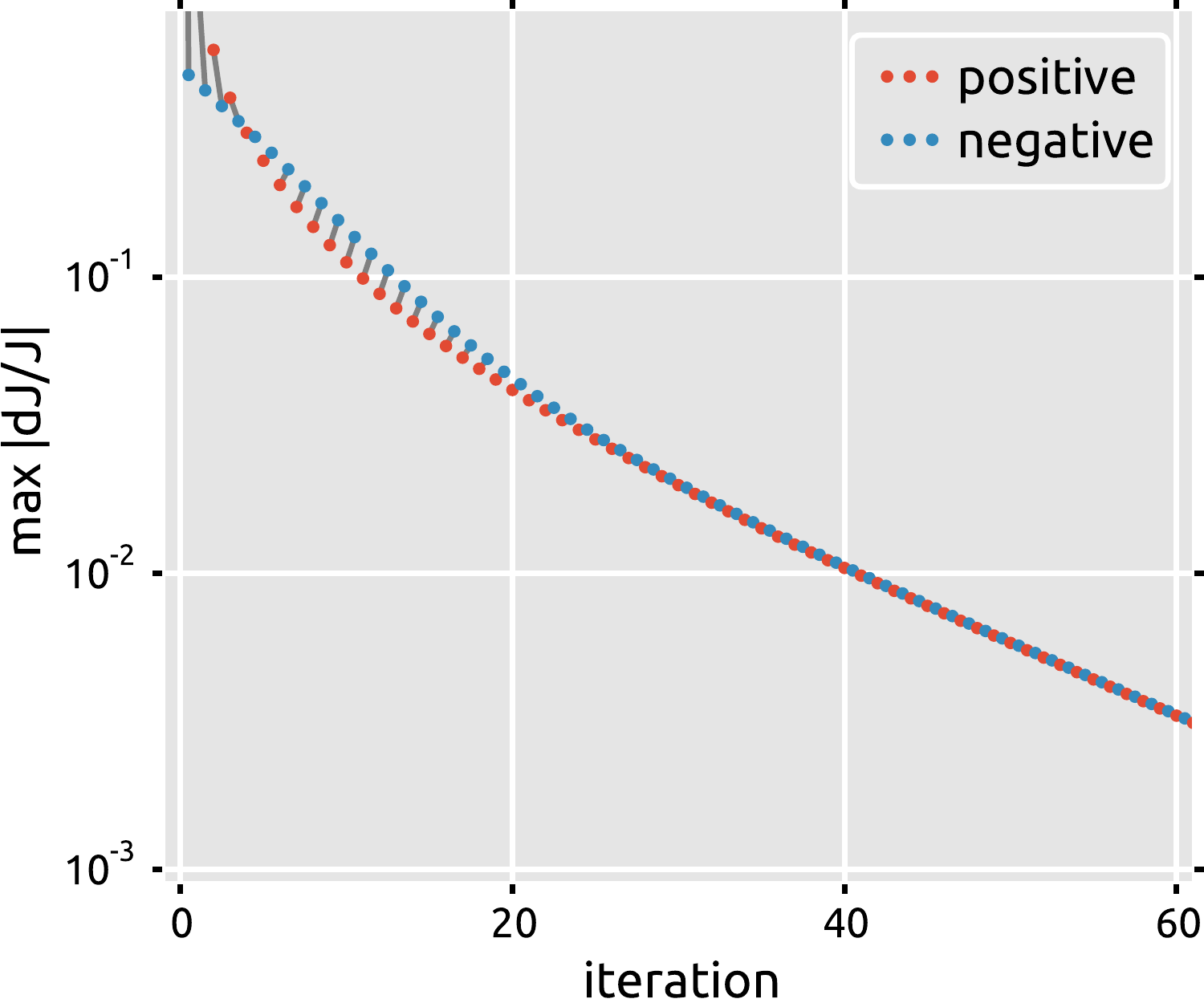}\hfil
    \includegraphics[width=0.2414\textwidth,trim=19mm 0    0 0,clip=true]{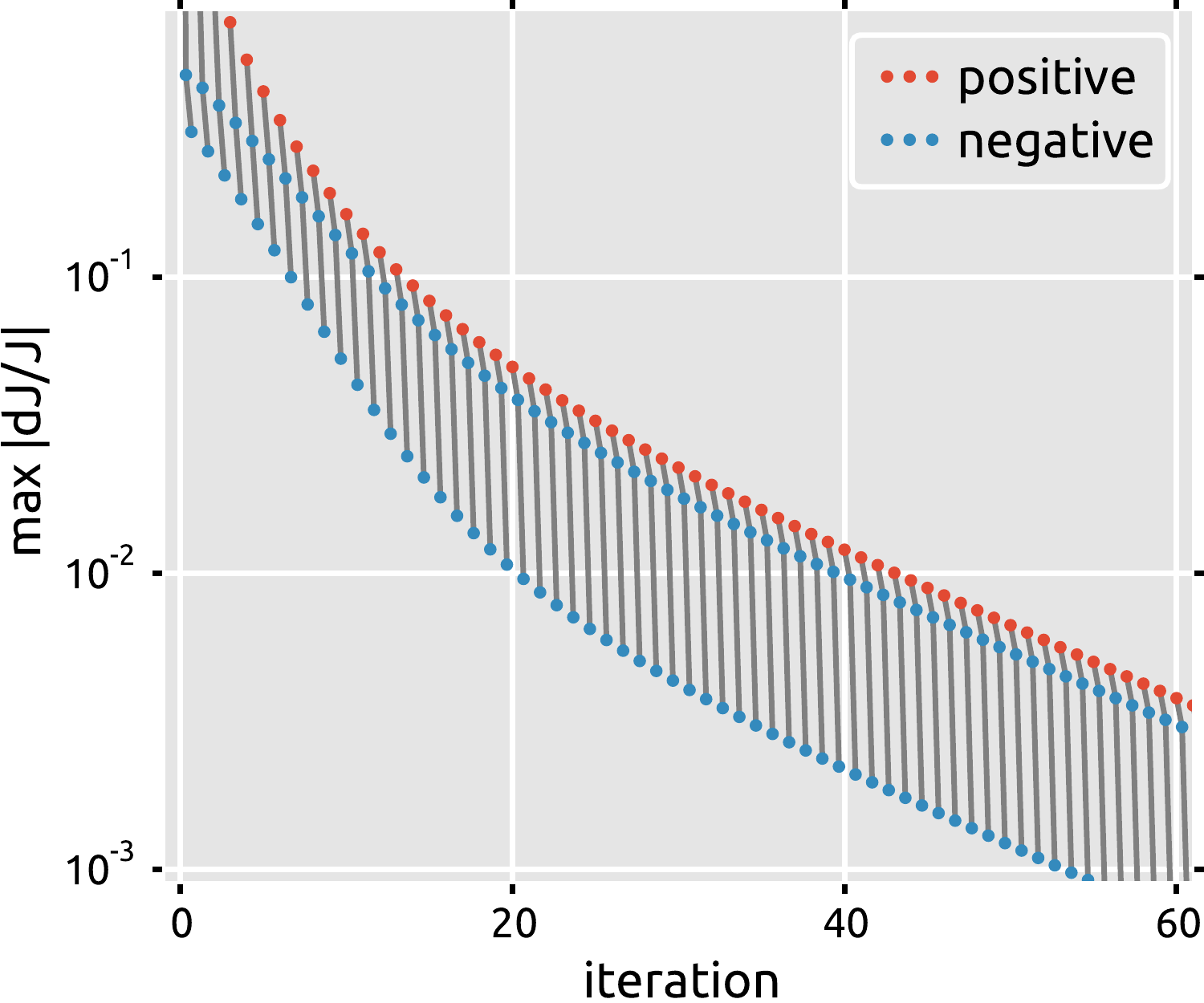}\hfil
    \includegraphics[width=0.2414\textwidth,trim=19mm 0    0 0,clip=true]{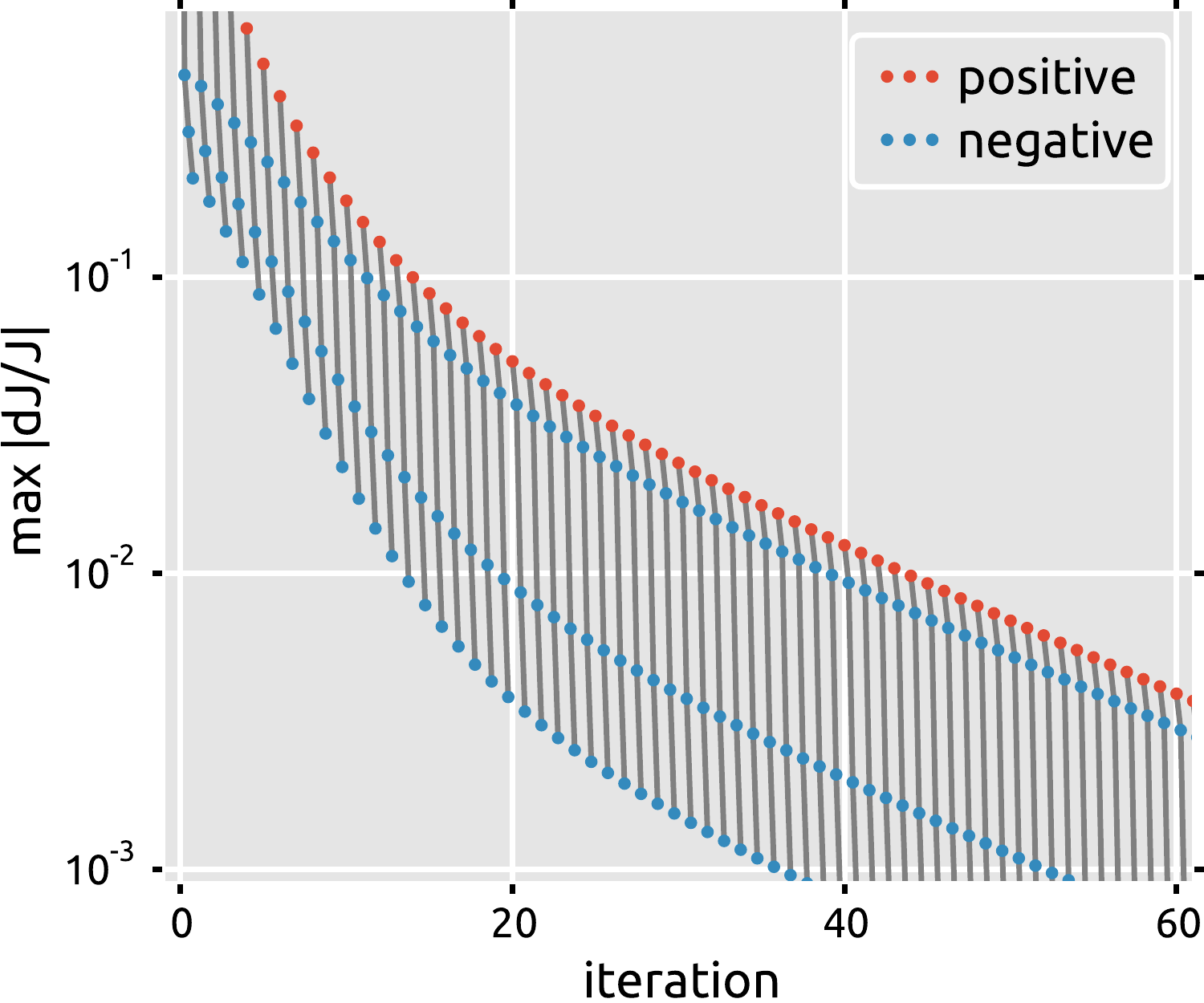}
  \end{minipage}
  \caption{Iteration behavior of the maximum absolute relative change in mean
    intensity ($\max|\delta J/J|$) for the 5-level \ion{Mg}{II} model atom with
    the k and the h lines treated in PRD in the FAL-C model atmosphere.  Initial
    level populations are initialized to LTE populations (top row) or
    initialized using the zero radiation approximation (bottom row).  Red
    markers indicate a positive $\max \delta J/J$, and blue markers a negative
    value.  Grey lines connect values of successive PRD subiterations within one
    ALI-iteration.  The number of PRD subiterations ranges from one to four
    (panels from left to right).}
  \label{fig:dj2j_mgii_falc}
\end{figure*}
\begin{figure*}
  \begin{minipage}{\textwidth}
    \includegraphics[width=0.5\textwidth,trim=0 14mm 0 0,clip=true]{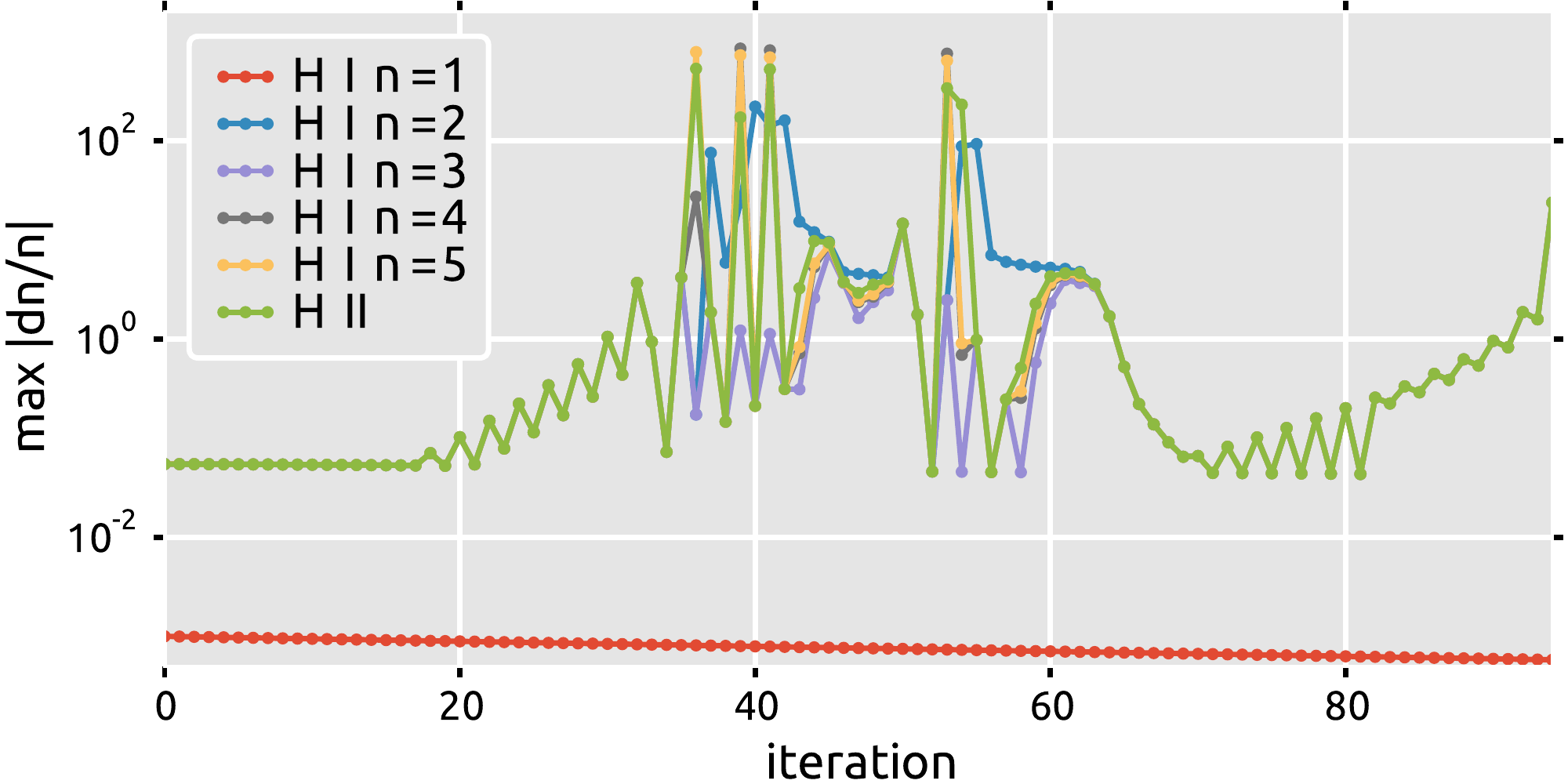}\hfil
    \includegraphics[width=0.5\textwidth,trim=0 14mm 0 0,clip=true]{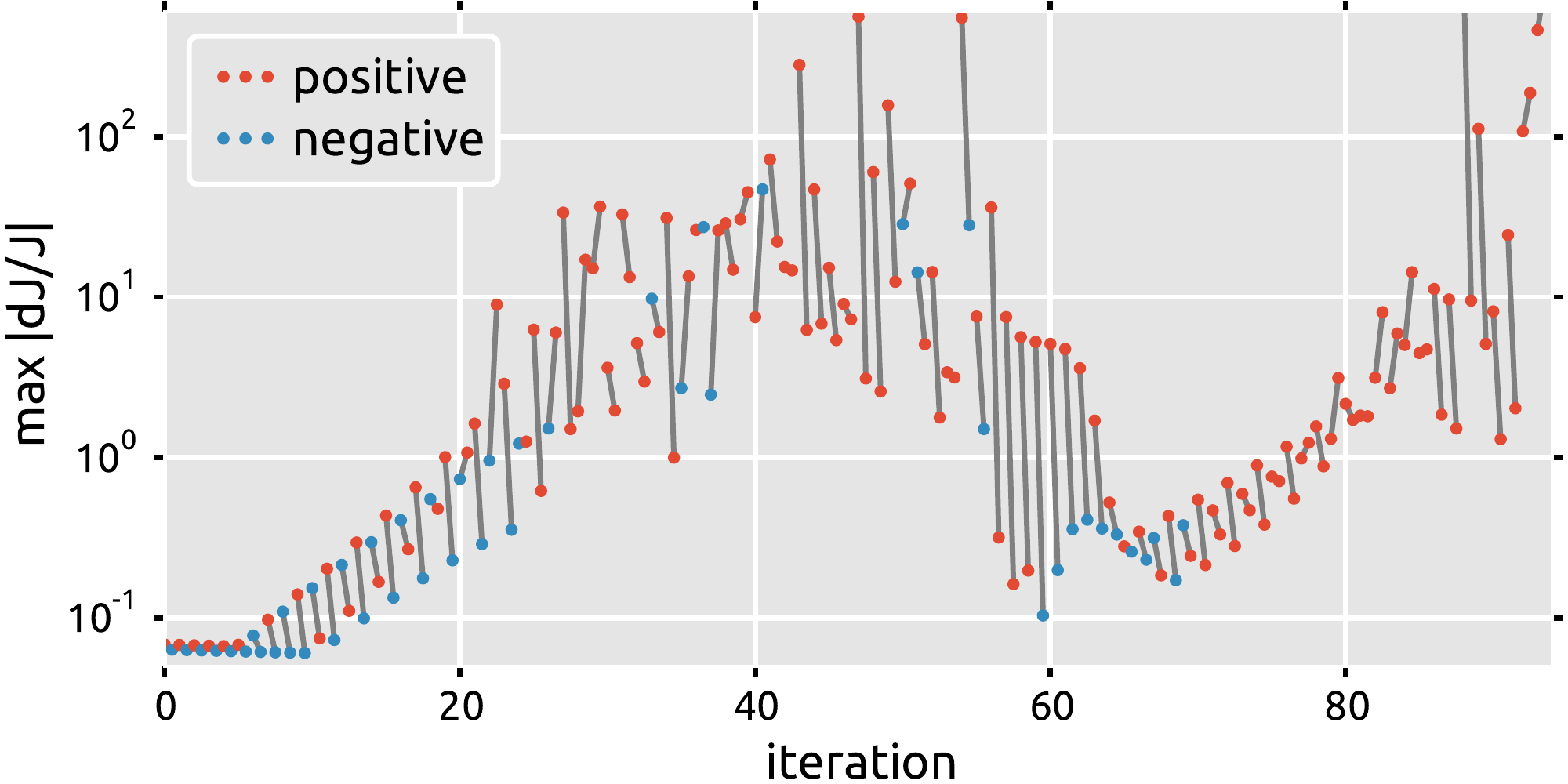}
    \\
    \includegraphics[width=0.5\textwidth]{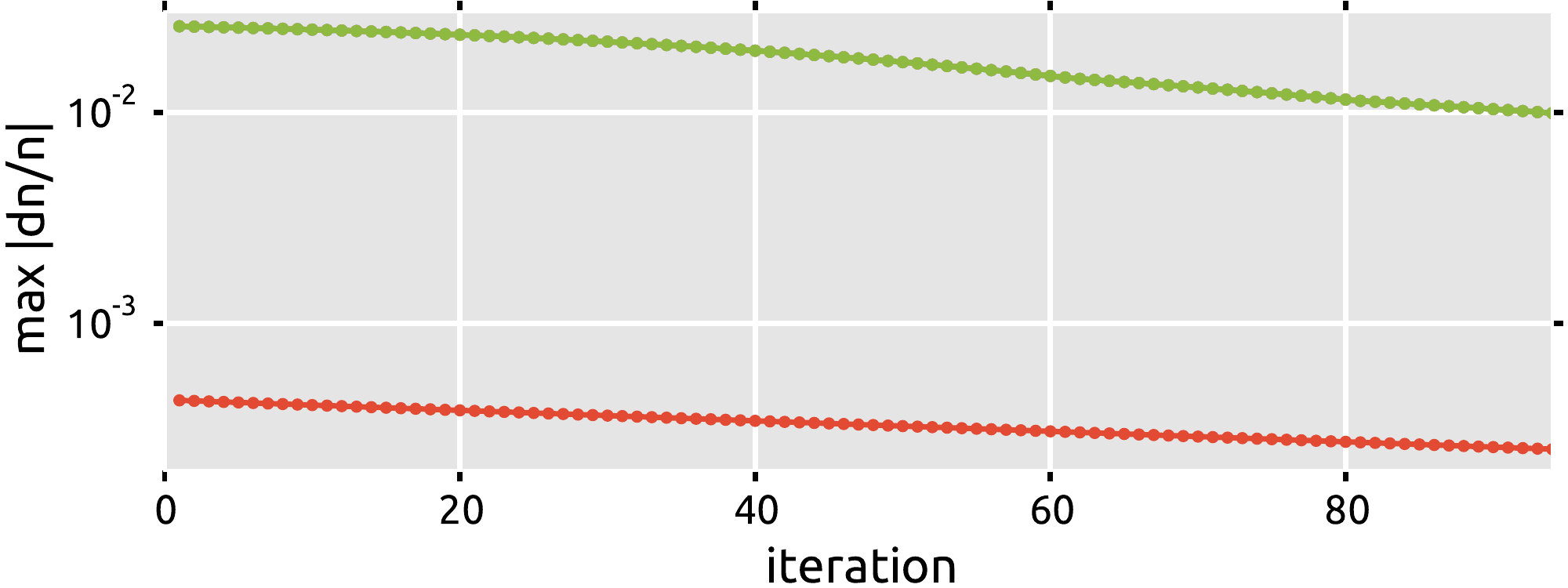}\hfil
    \includegraphics[width=0.5\textwidth]{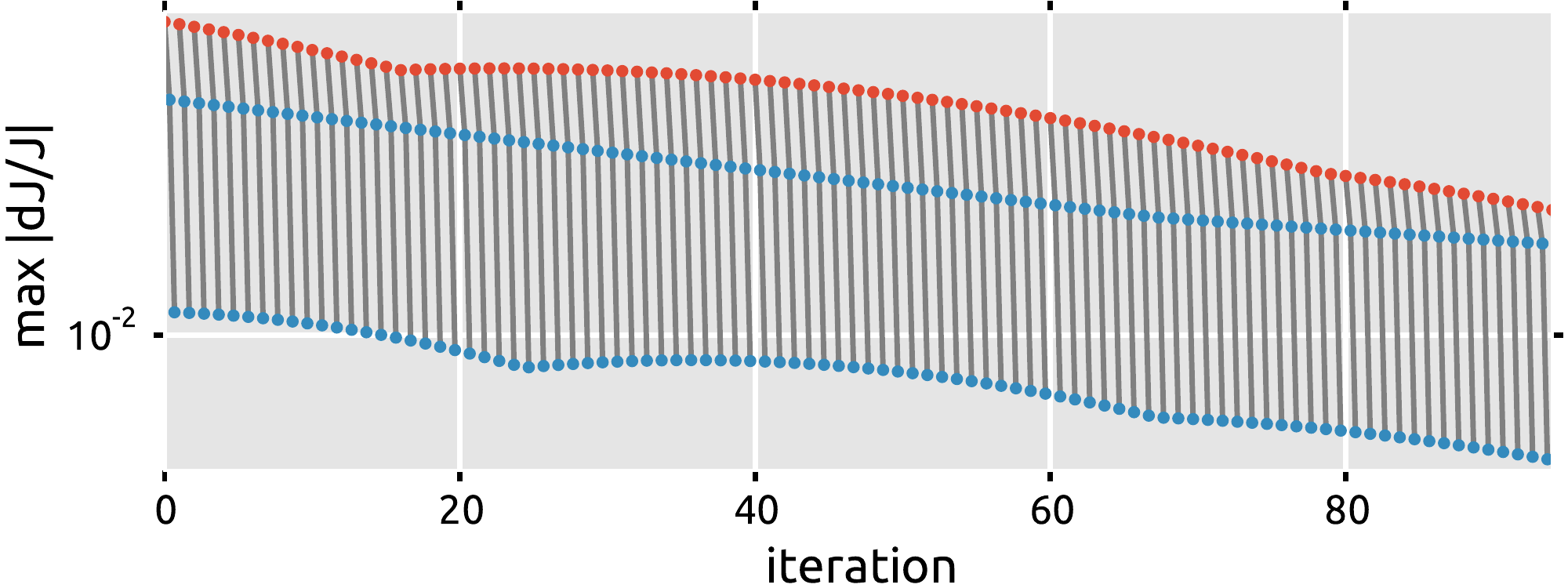}
  \end{minipage}
  \caption{Iteration behavior of the maximum absolute relative changes in atomic
    level population $\max|\delta n/n|$ (left) and in angle-averaged intensity
    $\max|\delta J/J|$ (right) in the 3D~Bifrost model atmosphere for the
    6-level hydrogen model atom with the Ly\,$\alpha$ line treated in PRD.  The
    top row shows the behavior with two subiterations, the bottom row with three
    subiterations.  Note that the $\max|\delta n/n|$ values for the excited
    states of \ion{H}{I} and the \ion{H}{II} continuum overlap.}
  \label{fig:dn2n-dj2j_h6_3d}
\end{figure*}
\begin{table}
  \caption{Number of accelerated $\Lambda$-iterations needed to converge to a
    $\max|\delta n/n| \leq 10^{-4}$ solution against the number of PRD
    subiterations for different model atoms, model atmospheres, and
    approximations for initial atomic level populations.}
  \label{tab:nali-vs-nprd}
  \centering
  \newcommand{ \zerorad }{ \begin{tabular}{ @{}c@{} }
                            zero\\radiation
                           \end{tabular} }
  \newcommand{ \prdsub }{ \begin{tabular}{ @{}c@{} }
                            PRD\\subiterations
                          \end{tabular} }
  \begin{tabular}{r c | c c c c}
    \hline\hline
    \multicolumn{2}{r}{model atmosphere:}    & \multicolumn{2}{c}{FAL-C}
                                             & \multicolumn{2}{c}{1D~Bifrost} \\
    \multicolumn{2}{r}{initial populations:} & LTE & \zerorad
                                             & LTE & \zerorad                 \\
    \hline
    \multicolumn{2}{r}{}       & \multicolumn{4}{c}{5-level \ion{Mg}{II}}     \\
    \hline
    \multirow{3}{*}{ \prdsub } & \phantom{{>}}1 &  -- &  -- &  -- &  --       \\
                               & \phantom{{>}}2 & 168 & 124 & 695 & 438       \\
                               &           {>}3 & 167 & 123 & 694 & 437       \\
    \hline
    \multicolumn{2}{r}{}       & \multicolumn{4}{c}{6-level \ion{H}{I}}       \\
    \hline
    \multirow{3}{*}{ \prdsub } & \phantom{{>}}1 &  -- &  -- &  -- &  --       \\
                               & \phantom{{>}}2 & 247 & 141 & 922 & 532       \\
                               &           {>}3 & 247 & 140 & 917 & 526       \\
    \hline
  \end{tabular}
  \tablefoot{A lack of convergence is indicated by ``--''.}
\end{table}
We tested the convergence of accelerated $\Lambda$-iterations (ALI) for the
\ion{Mg}{II} and the \ion{H}{I} model atoms with PRD lines in the FAL-C and the
1D~Bifrost model atmospheres for different numbers of PRD subiterations.
Similar but simpler tests were done when solving the non-LTE PRD problem in the
3D~Bifrost model.  We considered the solution to be converged if
$\max|\delta n/n| \leq 10^{-4}$.  The atomic level populations are initialized
either using the LTE or the zero radiation approximation%
\footnote{Populations are obtained by solving the system of statistical
          equilibrium equations with the radiation field set to zero throughout
          the atmosphere \citep{1986UppOR..33.....C}.  B.\,Lites introduced this
          idea in 1983.}.
We show examples of such tests in
Figs.\,\ref{fig:dj2j_mgii_falc}--\ref{fig:dn2n-dj2j_h6_3d} and
Table~\ref{tab:nali-vs-nprd}.

Figure~\ref{fig:dj2j_mgii_falc} shows the behavior of the maximum absolute
relative change in angle-averaged intensity, $\max|\delta J/J|$, for the
\ion{Mg}{II} model atom in the FAL-C model atmosphere, as a function of the
number of ALI iterations.  With one PRD subiteration the solution diverges.
With two or more subiterations the solution converges.  For iterations starting
with LTE populations there is a maximum in $\max|\delta J/J|$ after about 25
iterations, while $\max|\delta J/J|$ decreases monotonically when the
populations are initialized using the zero radiation approximation.

When the correction to the atomic level populations in non-LTE PRD become
monotonically decreasing, the major correction of the redistributed intensity is
performed during the first two PRD subiterations.  If the first PRD subiteration
corrects the redistributed intensity by some amount, then, on average, the
second PRD subiteration produces a slightly smaller correction but with an
opposite sign.  Each next PRD subiteration decreases this correction by almost
an order of magnitude while keeping the same sign.  Therefore, the major
adjustment to make the intensities consistent with the level populations occurs
during the first two or three PRD subiterations.

In Table~\ref{tab:nali-vs-nprd} we summarize our experiments that investigated
the dependence of the number of iterations required to reach a converged
solution on the number of PRD sub-iterations and the initial populations.
A converged solution is reached faster for the zero-radiation inital condition
than for LTE.  The solution converges in practically the same number of ALI
iterations independently of the number of PRD subiterations.  As long as the
number of subiterations is sufficient to lead to a converged solution, then the
number of PRD subiterations has no influence on the convergence rate.

We also investigated the convergence properties as a function of the number of
PRD subiterations in the 3D~Bifrost model atmosphere, for both the \ion{Mg}{II}
and \ion{H}{I} model atoms.  This atmosphere has large velocity, density, and
temperature gradients in the chromosphere, and represents a much tougher test
for the algorithm than the 1D~atmospheres.  We found that for \ion{Mg}{II} we
obtain a converged solution using two PRD subiterations, but for \ion{H}{I} we
needed three subiterations.  In Fig.\,\ref{fig:dn2n-dj2j_h6_3d} we illustrate
the difference between two and three subiterations for \ion{H}{I}.  It shows the
corrections in atomic level populations and the radiation field, starting with
initial conditions obtained from the previous incomplete run and using three PRD
subiterations.  The top row shows the behavior after we switch to two
subiterations.  Two subiterations are not enough to make the radiation field
consistent with the level populations, and from iteration~5 the corrections to
the radiation field become larger and larger.  At iteration~16 the corrections
to the level populations begin to increase as well, and the solution starts to
diverge.  In the bottom row, we show the computation with three PRD
subiterations, where the corrections to both the level populations and the
radiation field decrease steadily.

In summary, we found the following empirical rules:
\begin{itemize}
  \item One PRD subiteration is never enough to converge for any model atom in
        any model atmosphere.
  \item Two PRD subiterations are enough for convergence in the \ion{Mg}{II}
        lines.  Three PRD subiterations must be done for convergence in the
        \ion{H}{I} lines.
  \item The number of accelerated $\Lambda$-iterations needed to achieve a
        converged solution does not depend on the number of PRD subiterations.
\end{itemize}

\subsubsection{Convergence speed}
\label{subsubsec:convergence-prd}
%
\begin{table}
  \caption{Computational expenses needed to run either the \ion{Mg}{II} or the
    \ion{H}{I} model atoms in the 3D Bifrost snapshot using the 3D~formal
    solution of the transfer equation.  Time is measured for a computational
    subdomain of $N_\mathrm{X} N_\mathrm{Y} N_\mathrm{Z} = 32^3$ grid points and
    is rounded to s(econds), m(inutes), h(ours), or d(ays).  The calculations
    were performed on a Cray XC40 supercomputer.}
  \label{tab:ndex-tdex}
  \centering
   \begin{tabular}{l c c c c}
    \hline\hline
    model atom: & \multicolumn{2}{c}{5-level \ion{Mg}{II}}
                & \multicolumn{2}{c}{6-level \ion{H}{I}} \\
          case: & CRD & PRD
                & CRD & PRD \\
    \hline
    $N_\nu^\mathrm{total}$ & \multicolumn{2}{c}{457} & \multicolumn{2}{c}{628}\\
    $N_\nu^\mathrm{PRD}$   & \multicolumn{2}{c}{306} & \multicolumn{2}{c}{204}\\
    \hline
    $N_\dex$               & 20         & 70         & 90         & 200       \\
    $t_\iter(32^3)$        & 4\,m       & 9\,m       & 5\,m       & 12\,m     \\
    $t_\dex(32^3)$         & 80\,m      & 11\,h      & 7\,h       & 2\,d      \\
    \hline
  \end{tabular}
\end{table}
We measured how many iterations and how much time are needed to obtain a
converged CRD or PRD non-LTE solution for the \ion{Mg}{II} and the \ion{H}{I}
model atoms in the 1D and the 3D~model atmospheres.  We estimated how many
iterations $N_\mathrm{dex}$ are needed to decrease $\max |\delta n/n|$ by an
order of magnitude, how much time $t_\mathrm{iter}$ is spent per one full
iteration including PRD subiterations, and how much time $t_\mathrm{dex}$ is
spent per $N_\mathrm{dex}$ iterations.  Table~\ref{tab:ndex-tdex} summarizes our
results for the 3D~Bifrost snapshot.

From our measurements we found that
\begin{itemize}
  \item In the 3D model atmosphere, $N_\mathrm{dex}$ is a few times larger in
        PRD than in CRD.
  \item The computational time per full iteration increases in PRD mostly due to
        the extra PRD subiterations.  It is about 2~times bigger for magnesium
        (2~PRD subiterations) and 2.5--5 times bigger for hydrogen (3~PRD
        subiterations).
  \item For a typical $32^3$ subdomain, computational time per decade change in
        $\max|\delta n/n|$ is of the order of half a day to two days if PRD
        effects are computed using the hybrid approximation with equidistant
        grid interpolation.
\end{itemize}

For a typical PRD run in a 3D~model atmosphere, iteration until
$\max|\delta n/n| \leq 10^{-4}$ is often required to get intensities accurate
better than 1\%.  For our 3D~model atmosphere with $252 \times 252 \times 496$
grid points, this means that one reaches a converged solution using 1024\,CPUs
in ${\sim}2$~days for \ion{Mg}{II} and ${\sim}8$~days for \ion{H}{I},
corresponding to roughly 50\,000 and 200\,000~CPU hours.

These numbers are sufficiently small to make it possible to model PRD lines on
modern supercomputers either in a single model or a time-series of snapshots
from large 3D radiation-MHD simulations.  This is not possible if full
angle-dependent PRD is used. Table~1 in
\citet{2012A&A...543A.109L}, 
shows that the full angle-dependent algorithm is almost two orders of magnitude
slower and much more memory consuming than the hybrid algorithm, requiring
millions of CPU hours for a single snapshot.

\subsubsection{Convergence acceleration}

We tested whether convergence acceleration using the algorithm by
\citet{1974JChPh..61.2680N} 
can be applied to 3D~PRD calculations.  We used 2\textsuperscript{nd} order Ng
acceleration with 5 iterations between acceleration steps.

We found that Ng acceleration does not work if it is applied too soon after
starting a new computation.  It only works if it is started when both
$ \max|\delta n/n| $ and $ \max|\delta J/J| $ are steadily decreasing, such as
in the bottom panels of Fig.\,\ref{fig:dn2n-dj2j_h6_3d}.  The steeper the
decrease rate, the better the acceleration will work.  If the acceleration is
started too soon, then the convergence rate may be slower than without applying
acceleration, or the computation will diverge.

We observed that acceleration yields better results if the atomic level
populations are initialized with the zero radiation approximation than if they
are initialized with LTE populations.  Initializing the populations with a
previously converged CRD solution also yields good acceleration results,
especially in 3D~model atmospheres.

Finally, we noticed that acceleration generally works reliably in 1D~model
atmospheres and for our \ion{Mg}{II} atom in the 3D~model atmosphere.  For the
\ion{H}{I} atom in the 3D atmosphere, we find that using Ng acceleration does
not significantly improve the convergence rate.

\subsection{Application to the \ion{Mg}{II}~h\&k lines}
\label{subsec:mgii-kh-3d}

To illustrate the joint action of 3D and PRD effects on emerging line profiles we
calculated intensity profiles of the \ion{Mg}{II} h\&k lines in the 3D~Bifrost
snapshot in CRD and PRD using both 1D and 3D formal solvers.  We compared
results from the 3D~PRD case (which is the most realistic) to results from the
two approximations of 3D~CRD and 1D~PRD.  In 3D~CRD, line intensities are
computed neglecting effects of partial redistribution but including effects of
the horizontal radiative transfer.  In 1D~PRD, only effects of partial
redistribution are present, while the radiative transfer is performed treating
each vertical column as a plane-parallel atmosphere.

For all the pixels in each computation we determined the wavelength position and
intensities of the k$_\mathrm{2v}$, k$_3$, and k$_\mathrm{2r}$ spectral features
using the algorithm described in
\citet{2013ApJ...778..143P}. 
The intensities are mostly used to diagnose temperatures in the chromosphere,
while the wavelength positions and separations between the features are used to
measure either velocities or velocity gradients
\citep[see][]{2013ApJ...772...90L,
              2013ApJ...778..143P}. 

Here we verify two statements made by
\citet{2013ApJ...772...89L}. 
These authors stated that the central depression features (h$_3$ and k$_3$) of
the \ion{Mg}{II} h\&k lines are only weakly affected by PRD and are mostly
influenced by effects of the horizontal radiative transfer.  They also stated
that the emission peaks (h$_\mathrm{2v}$, k$_\mathrm{2v}$, h$_\mathrm{2r}$, and
k$_\mathrm{2r}$) are mostly controlled by PRD effects while effects of the
3D~radiative transfer play a minor role.  They thus argued that modeling of the
emission peaks could be done assuming 1D~PRD while modeling of the central
depressions could be done assuming 3D~CRD.  They gave arguments based on 1D~PRD
calculations and 3D~CRD to support these claims because they could not perform
3D~PRD calculations.

\subsubsection{Imaging} 
\label{subsubsec:mgii-k-imaging}
%
\begin{figure*}[!t]
  \begin{minipage}{\textwidth}
    \includegraphics[width=0.32666\textwidth,trim=0    13.5mm 40mm 0   ,clip=true]{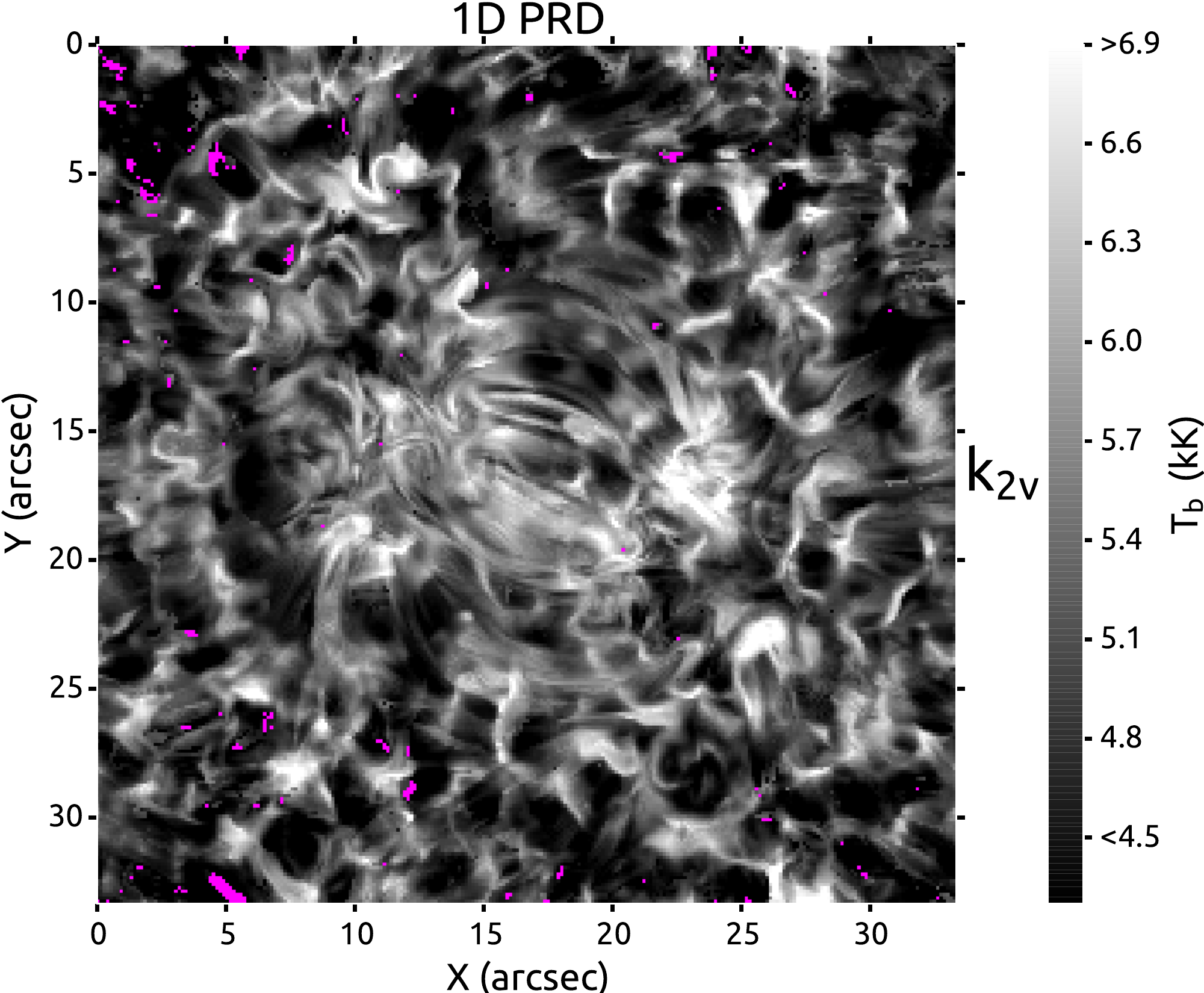}\hfil
    \includegraphics[width=0.29664\textwidth,trim=15mm 13.5mm 40mm 0   ,clip=true]{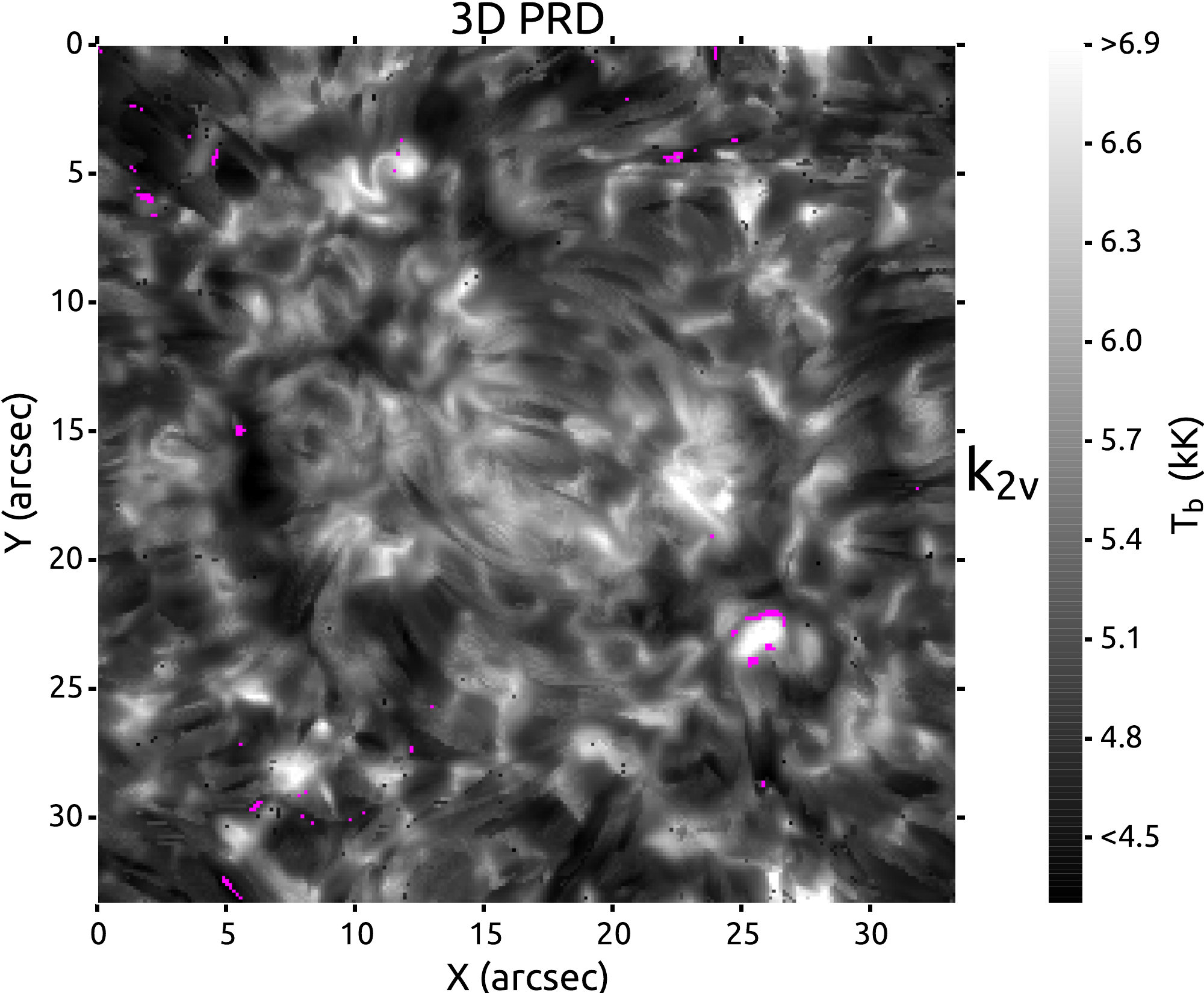}\hfil
    \includegraphics[width=0.37670\textwidth,trim=15mm 13.5mm 0    0   ,clip=true]{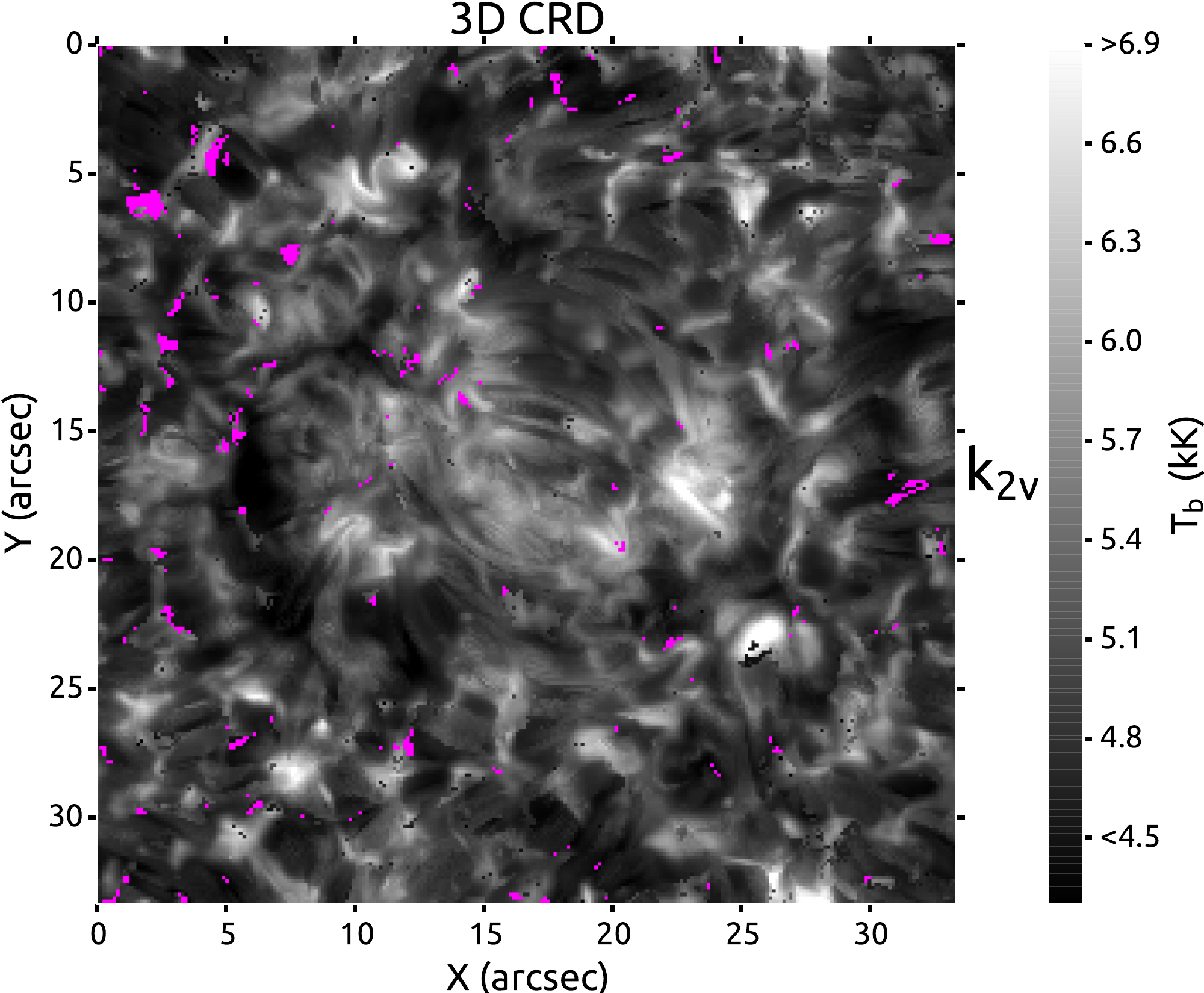}
    \\
    \includegraphics[width=0.32666\textwidth,trim=0    13.5mm 40mm 5.6mm,clip=true]{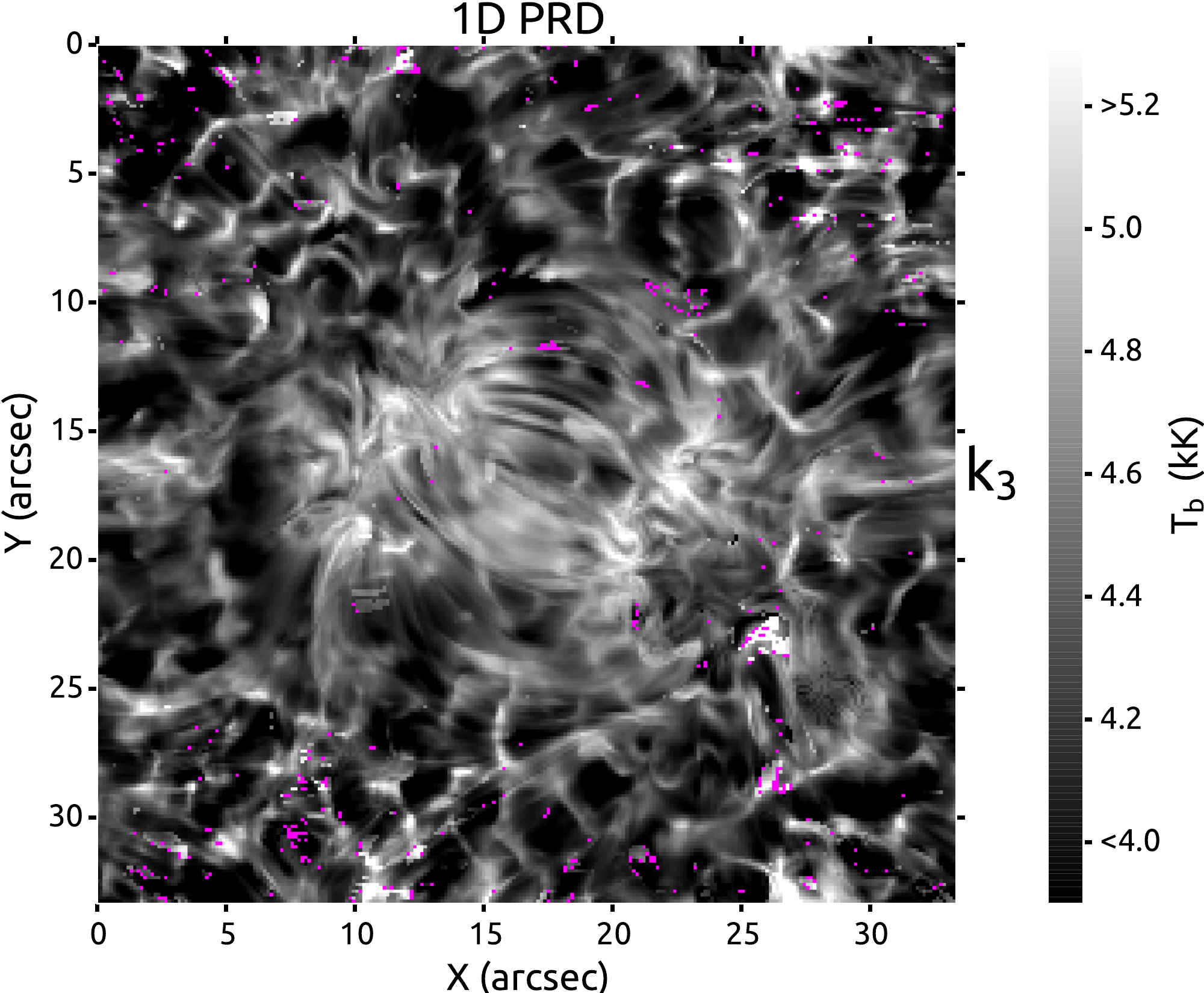}\hfil
    \includegraphics[width=0.29664\textwidth,trim=15mm 13.5mm 40mm 5.6mm,clip=true]{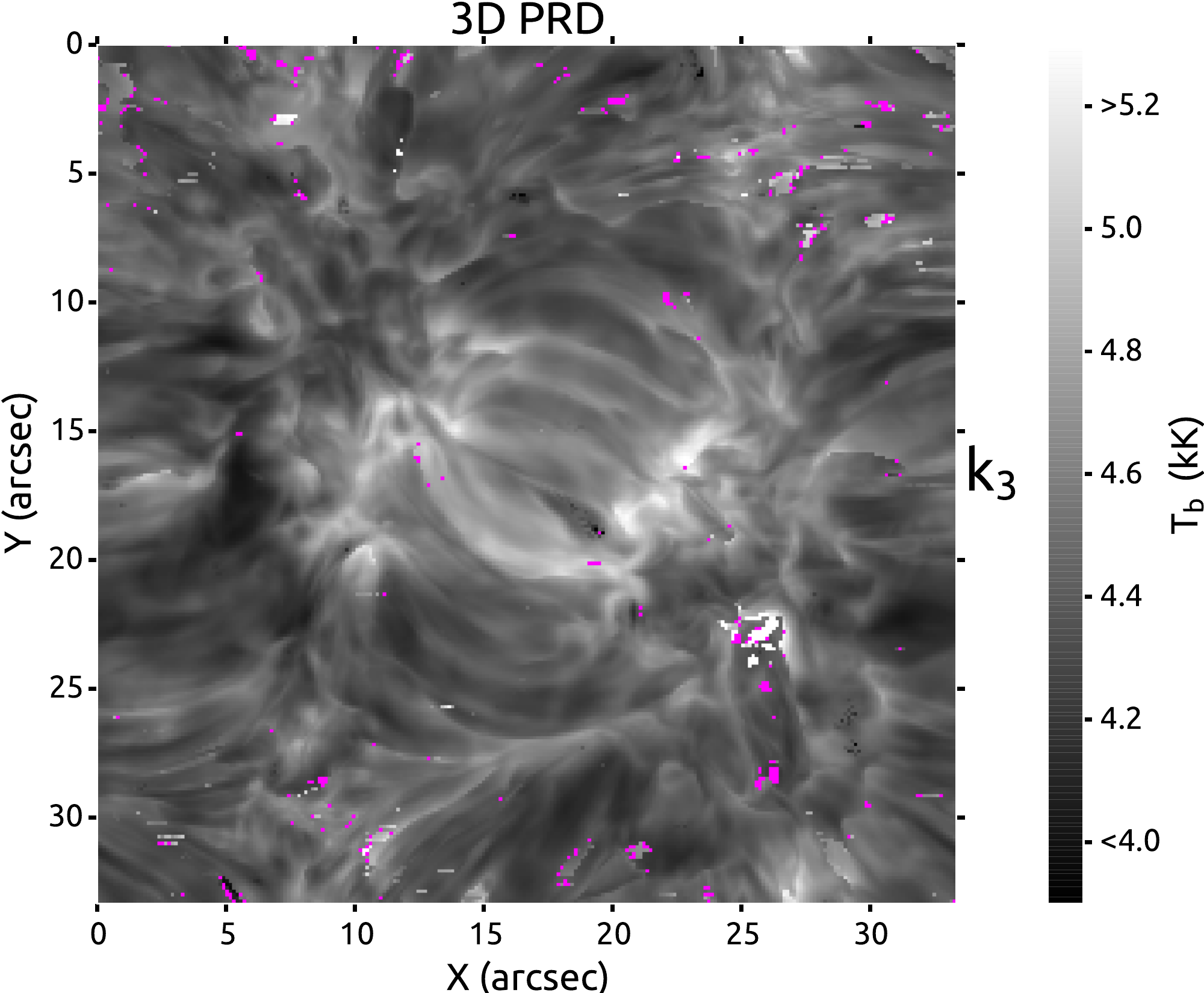}\hfil
    \includegraphics[width=0.37670\textwidth,trim=15mm 13.5mm 0    5.6mm,clip=true]{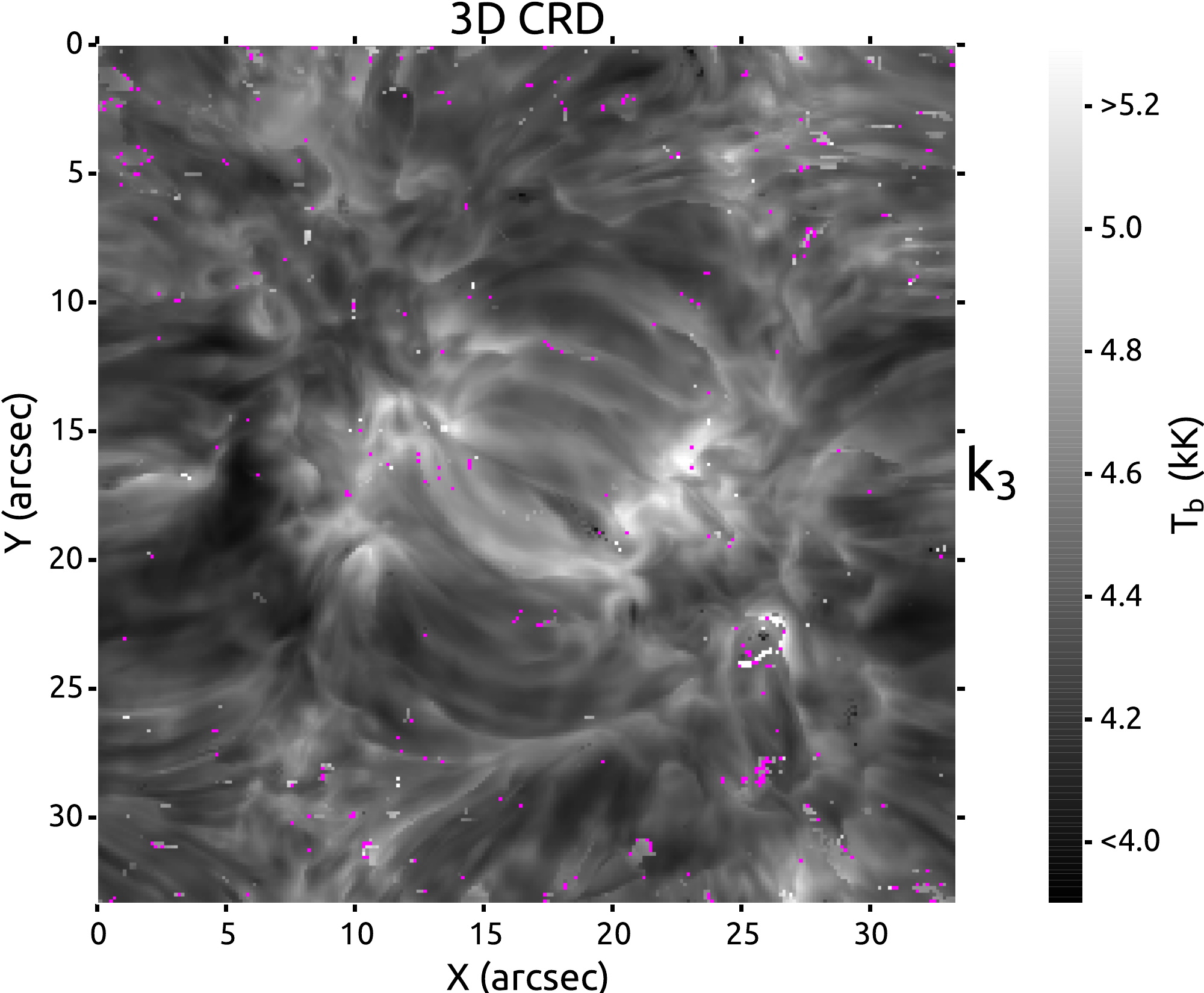}
    \\
    \includegraphics[width=0.32666\textwidth,trim=0    0    40mm 5.6mm,clip=true]{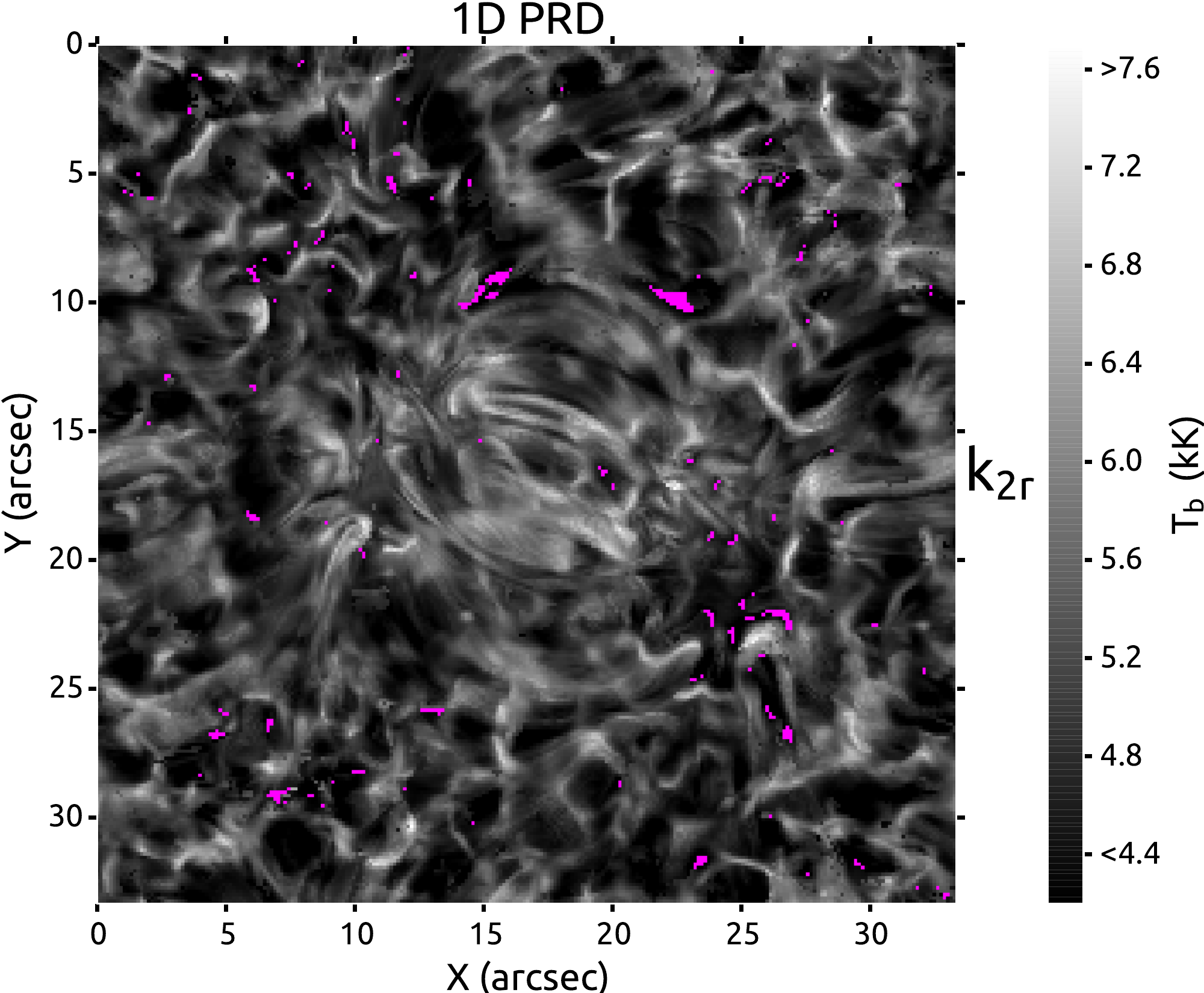}\hfil
    \includegraphics[width=0.29664\textwidth,trim=15mm 0    40mm 5.6mm,clip=true]{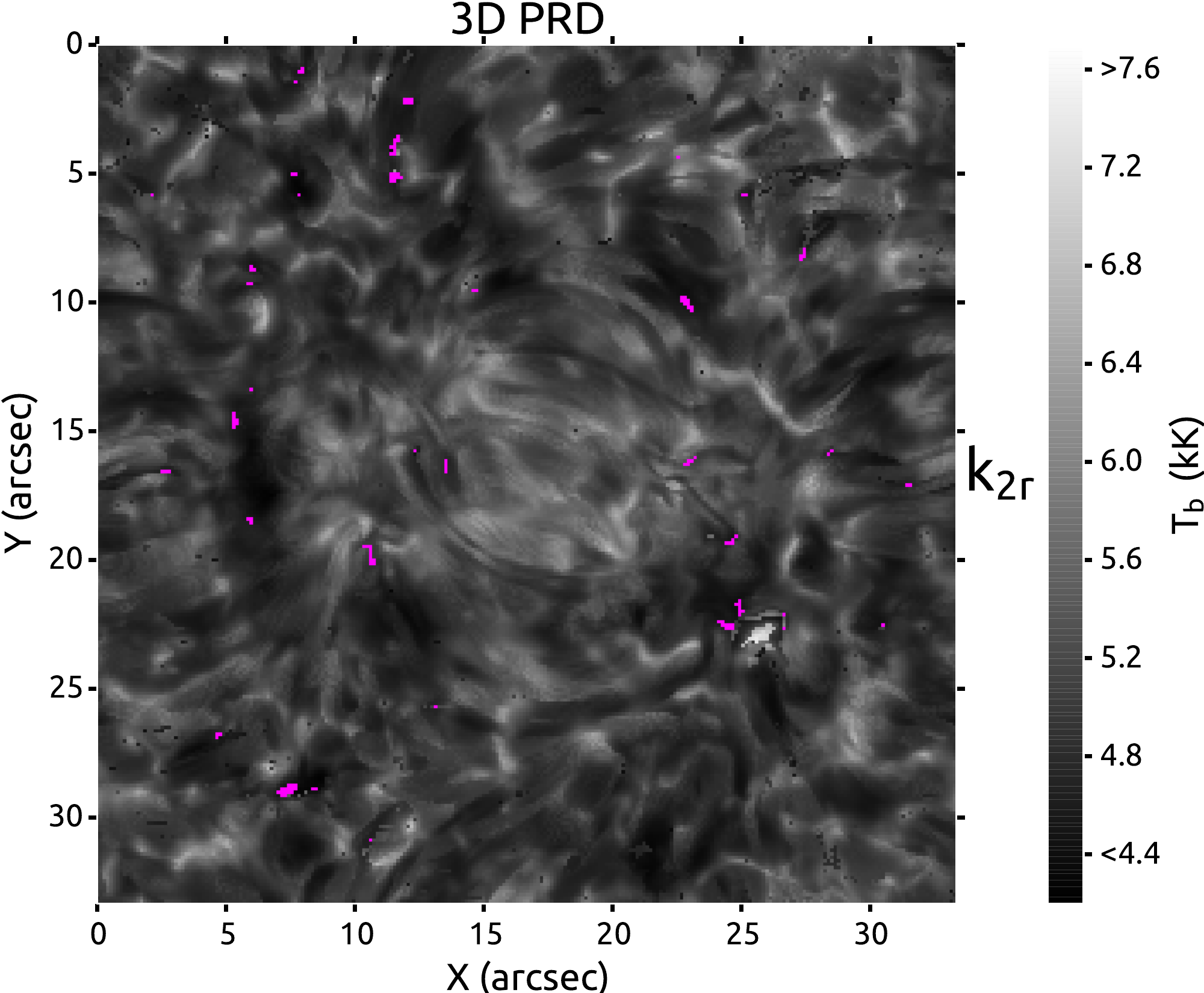}\hfil
    \includegraphics[width=0.37670\textwidth,trim=15mm 0    0    5.6mm,clip=true]{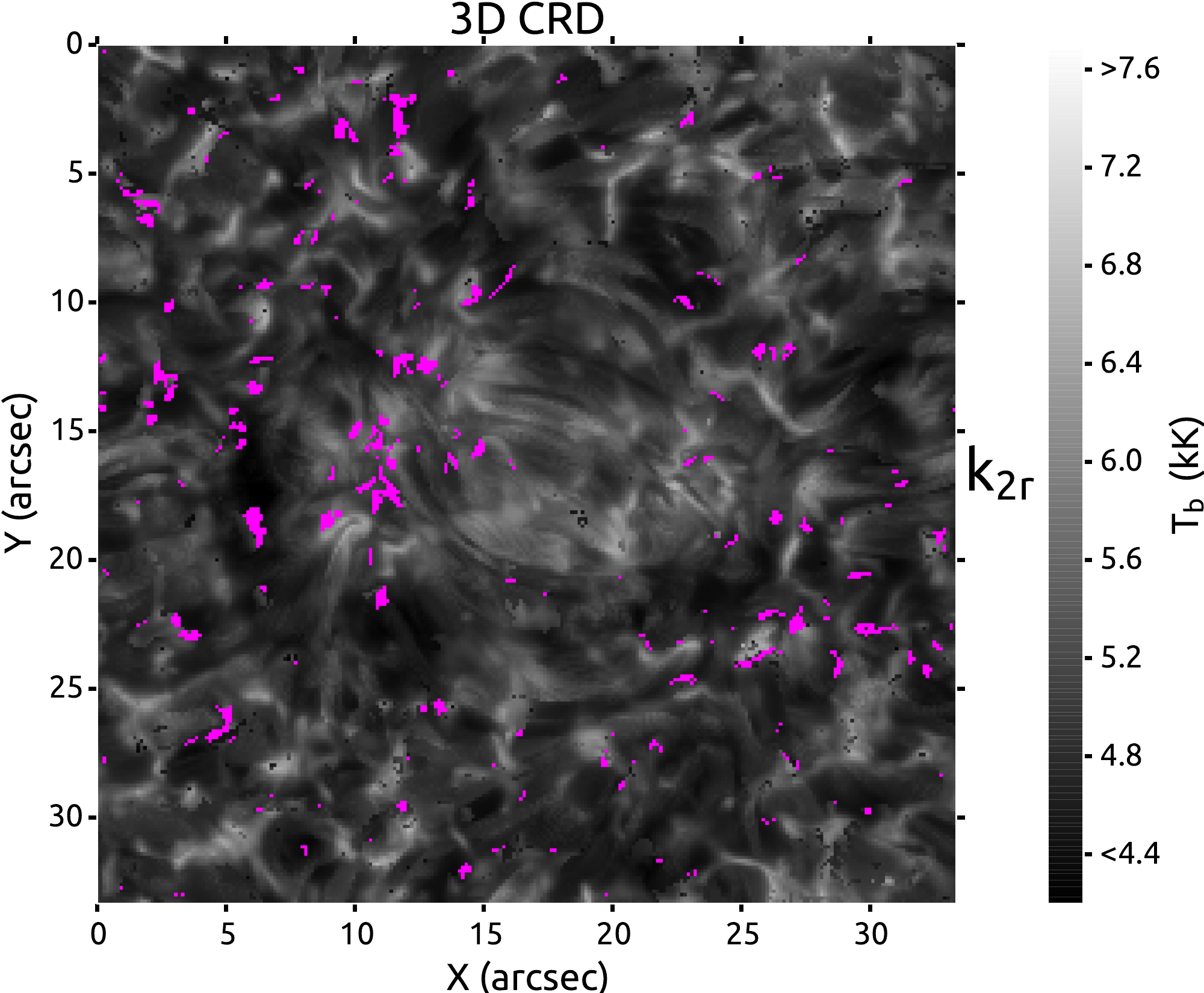}
  \end{minipage}
  \caption{Images in the \ion{Mg}{II} k~line computed from the 3D~Bifrost
    snapshot at $\mu_\mathrm{Z} = 1$.  Rows: blue emission peak k$_\mathrm{2v}$
    (top), central depression k$_\mathrm{3}$ (middle), and the red emission peak
    k$_\mathrm{2r}$ (bottom).  Columns: computations in 1D~PRD (left), 3D~PRD
    (middle), and 3D~CRD (right).  The intensity is shown as a brightness
    temperature: $B_\nu(T_\mathrm{b}) = I_\nu$.  Fuchsia-colored spots indicate
    coordinates where a feature is not present or misidentified.  The brightness
    scale for all images in a row is identical and indicated in the colorbar on
    the right side.}
  \label{fig:mgii-2d-t}
\end{figure*}
\begin{figure*}[!t]
  \begin{minipage}{\textwidth}
    \includegraphics[width=0.3333\textwidth]{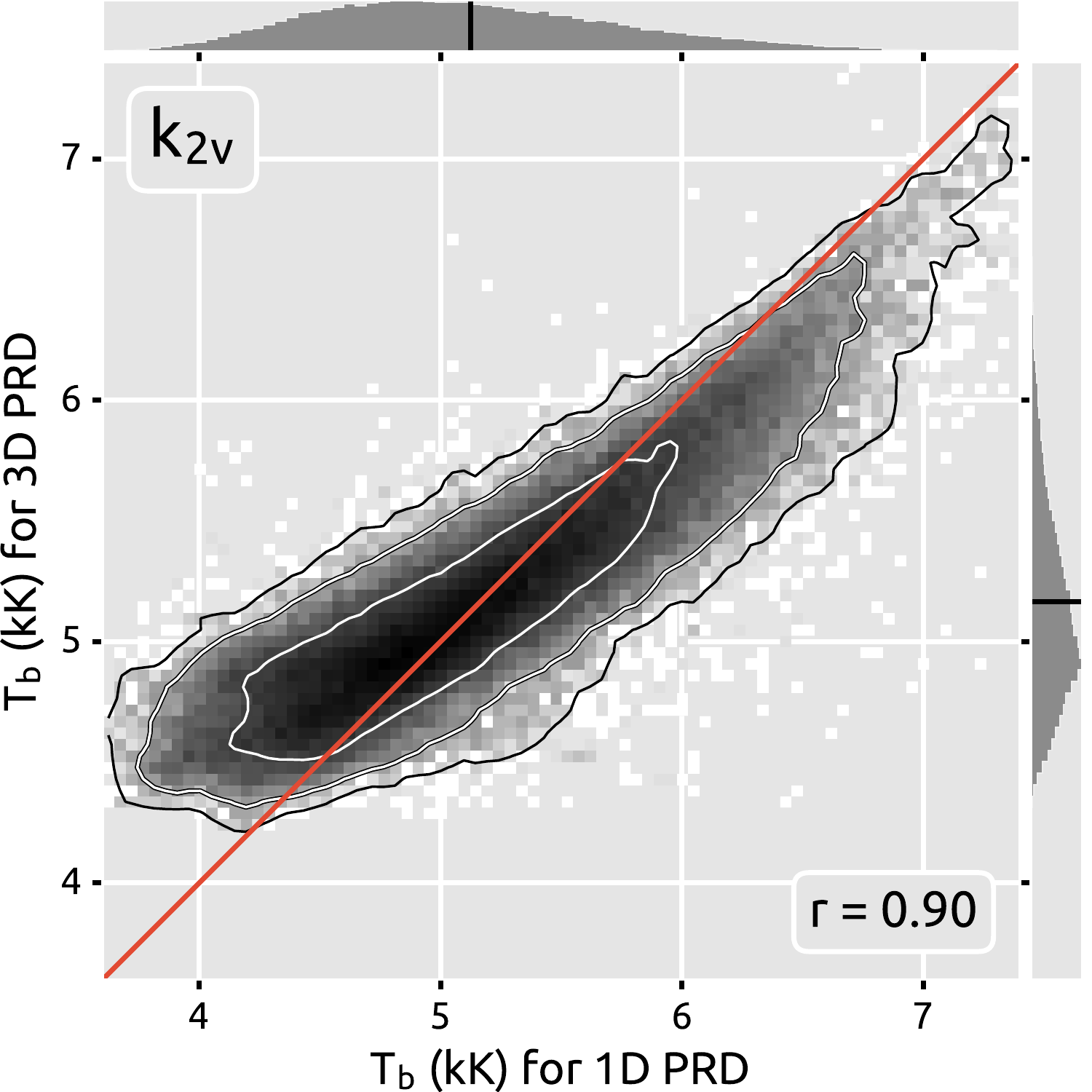}\hfil
    \includegraphics[width=0.3333\textwidth]{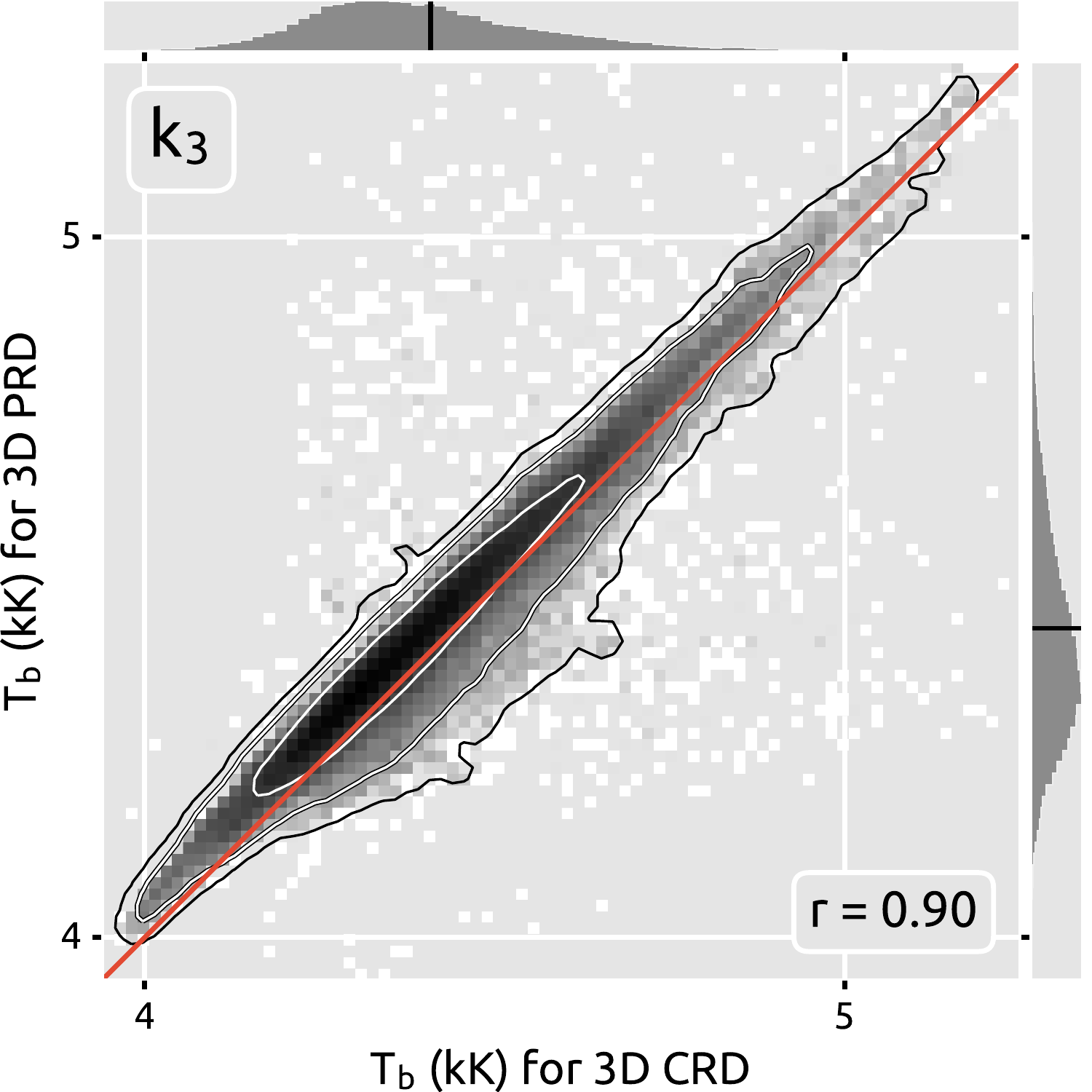}\hfil
    \includegraphics[width=0.3333\textwidth]{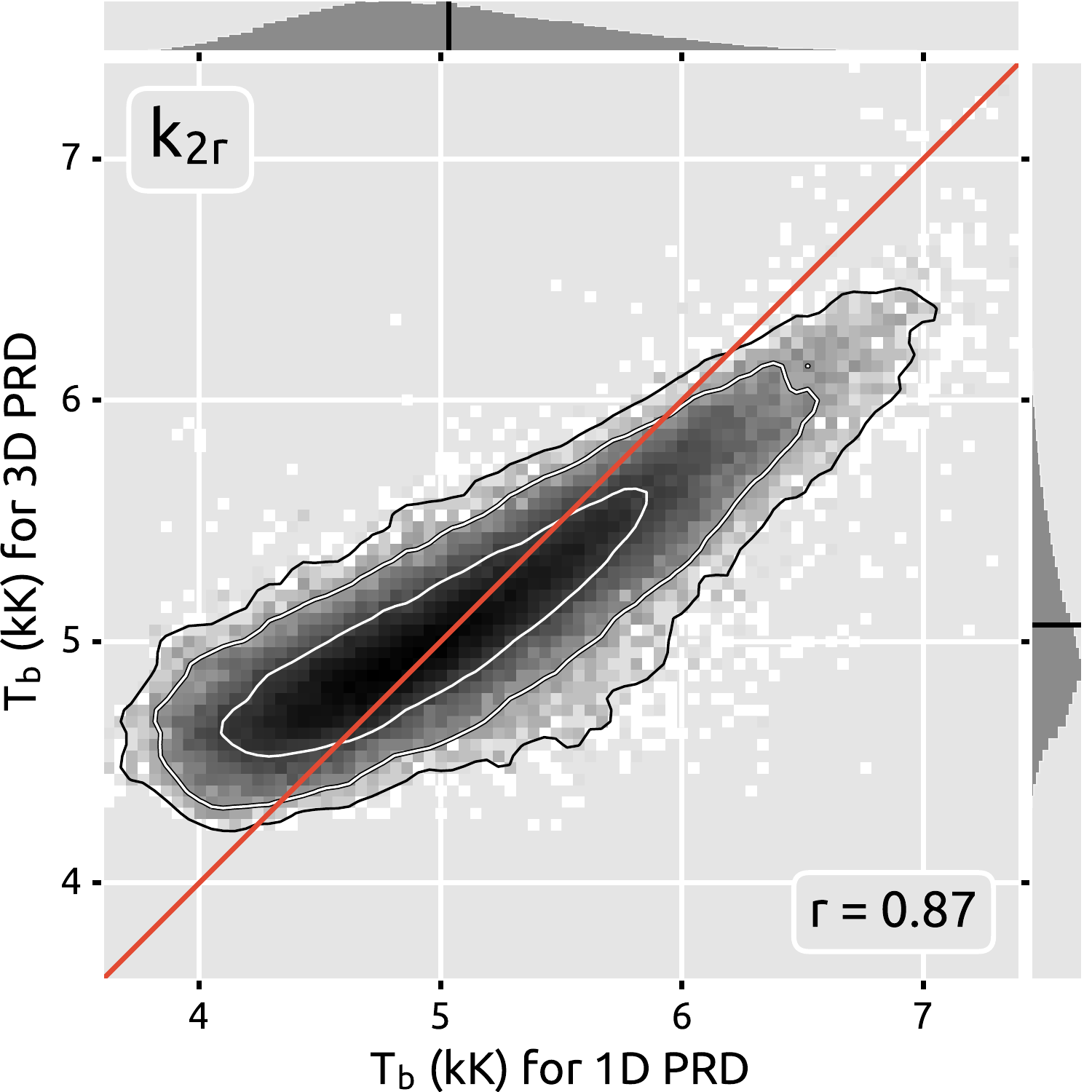}
  \end{minipage}
  \caption{Joint probability density functions of the brightness temperature
    in \ion{Mg}{II} k$_\mathrm{2v}$ (left), k$_3$ (center), and k$_\mathrm{2r}$
    (right) computed in different cases.  Horizontal axes are 1D~PRD (left and
    right panels) and 3D~CRD (center).  Vertical axes are all 3D~PRD.  The color
    of the JPDFs displays the logarithm of the number of counts ranging from
    white (smallest) to black (biggest).  The three nested iso-contours
    encompass regions with 68\%, 95\%, and 99\% of all pixels.  The red lines
    denote $x = y$.  The small side plots show 1D~histograms for the horizontal
    (top) or vertical (right) axes with mean values indicated by short black
    lines.  The Pearson correlation coefficient $r$ is given in an inset in each
    panel.}
  \label{fig:mgiik-correlations}
\end{figure*}
Figure~\ref{fig:mgii-2d-t} shows images of the brightness temperature in
k$_\mathrm{2r}$, k$_3$, and k$_\mathrm{3r}$ for the 1D~PRD, 3D~PRD, and 3D~CRD
computations.  These features are not always present and sometimes the automated
fitting routine misidentifies them.  These locations are indicated with
fuchsia-colored pixels.

We see that the k$_3$ images computed in 3D~CRD and 3D~PRD (middle row of
Fig.\,\ref{fig:mgii-2d-t}) indeed look similar. As was already shown by
\citet{2013ApJ...772...89L}, 
the 1D~PRD image appears very different than the images computed in 3D.

For the emission peak images (top and bottom rows) we see substantial
differences between the 1D~PRD and 3D~PRD computations.  The 1D~PRD images have
substantially higher contrast, and bright structures have much sharper edges
than in the 3D~PRD computations.  Interestingly there is also a significant
structural similarity between 3D~PRD and 3D~CRD.

Figure~\ref{fig:mgiik-correlations} illustrates the same points by showing
joint probability density functions (JPDFs) between brightness temperatures
computed in 3D~PRD and those computed in 1D~PRD for the emission peaks and
3D~CRD for the central depression.  The k$_3$ brightness temperature in 3D~PRD
is typically slightly higher than predicted by the 3D~CRD computation.

For k$_\mathrm{2v}$ and k$_\mathrm{2r}$ we see that the contrast for the 3D~PRD
computation is lower than for the 1D~PRD case.  The 1D~PRD computations
underestimate the brightness where $T_\mathrm{b} < 5$\,kK, and overestimate it
where $T_\mathrm{b} > 5$\,kK.

\subsubsection{Average line profiles and center-to-limb variation}
\label{subsubsec:mgii-k-T-and-v}
%
\begin{figure*}
  \sidecaption
  \begin{minipage}{120mm}
    \includegraphics[width=0.5165\textwidth,trim=0    14.5mm 0 0,   clip=true]{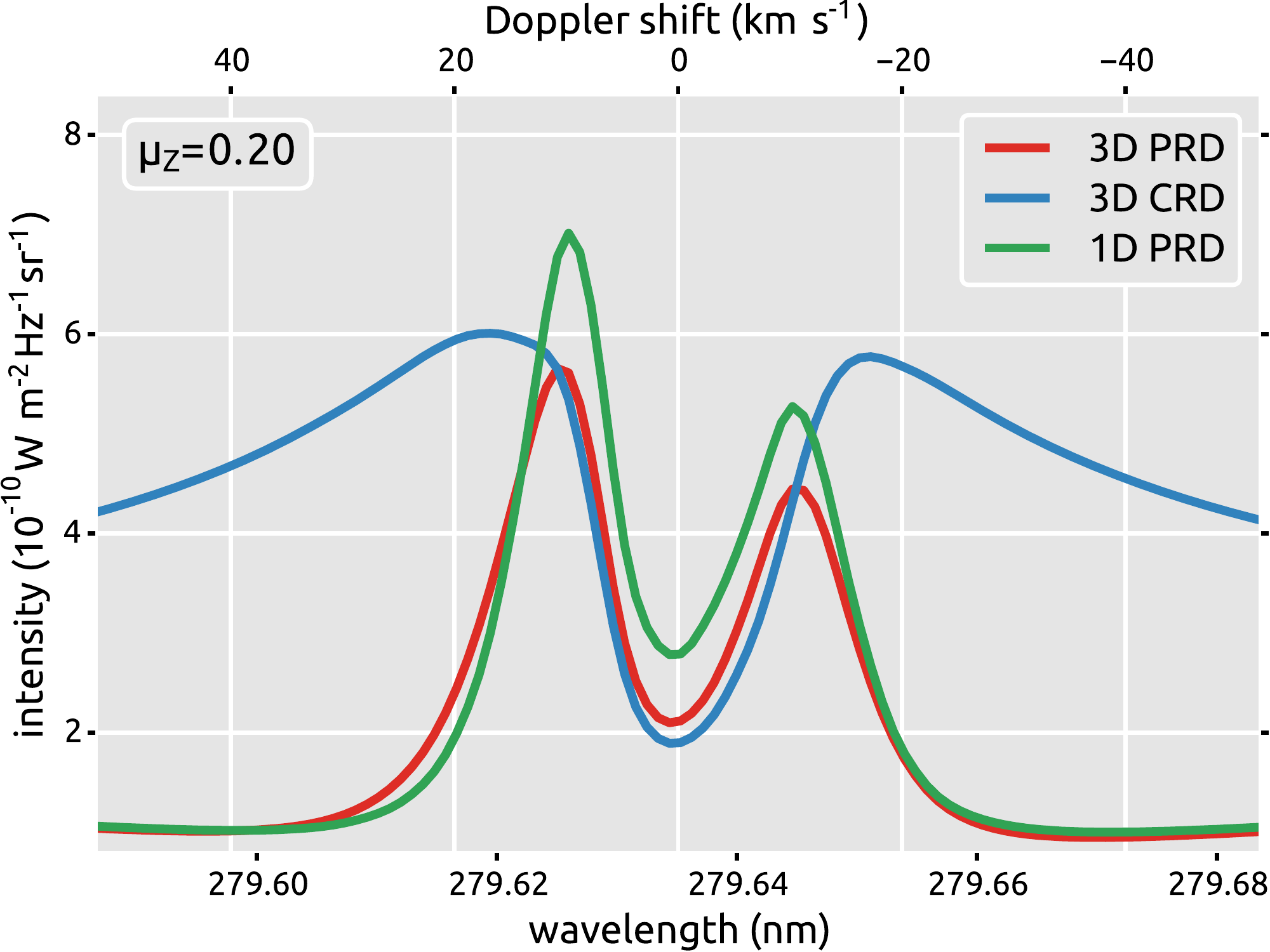}\hfil
    \includegraphics[width=0.4809\textwidth,trim=14mm 14.5mm 0 0,   clip=true]{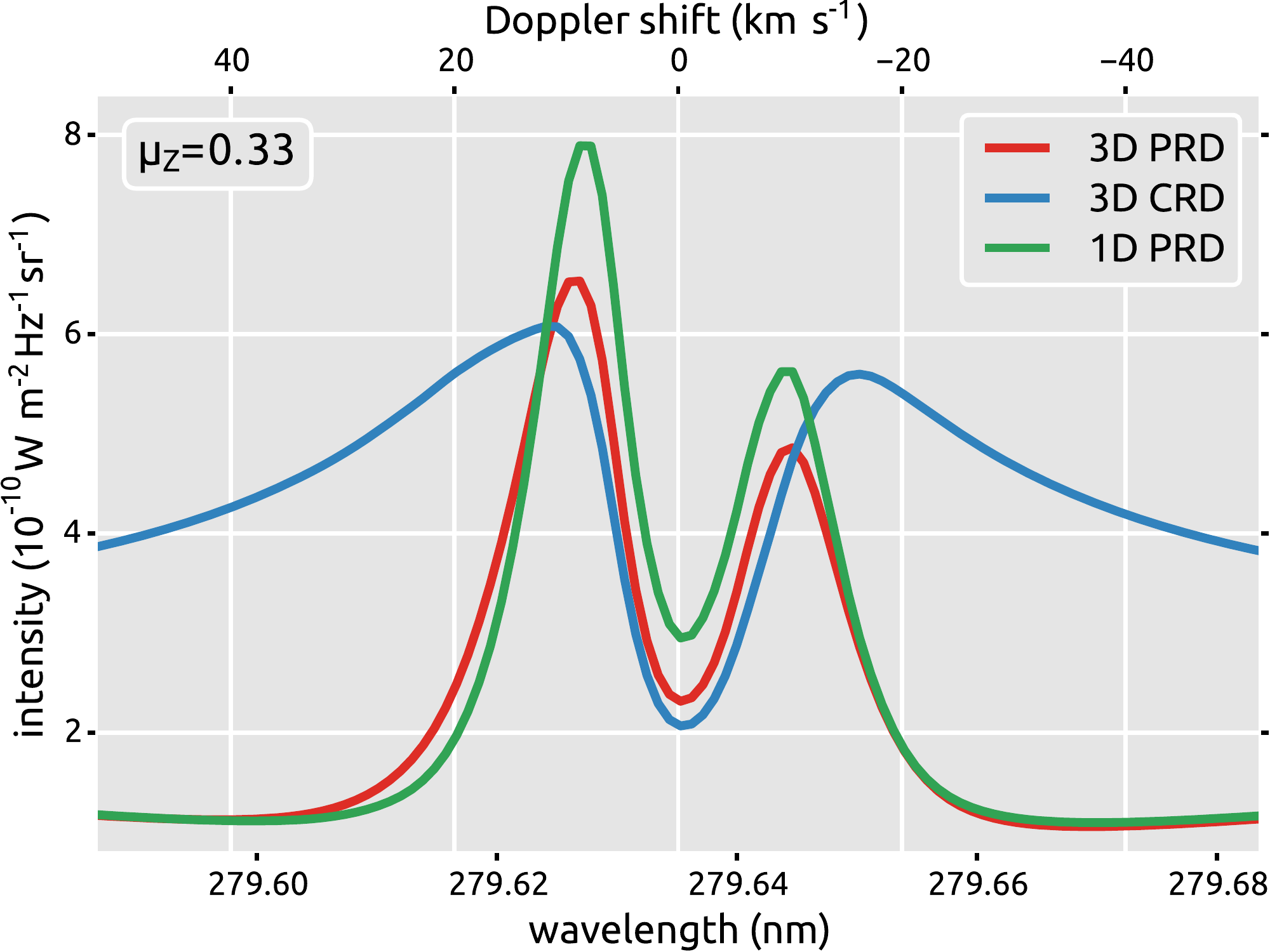}
    \\
    \includegraphics[width=0.5165\textwidth,trim=0    0    0 13.5mm,clip=true]{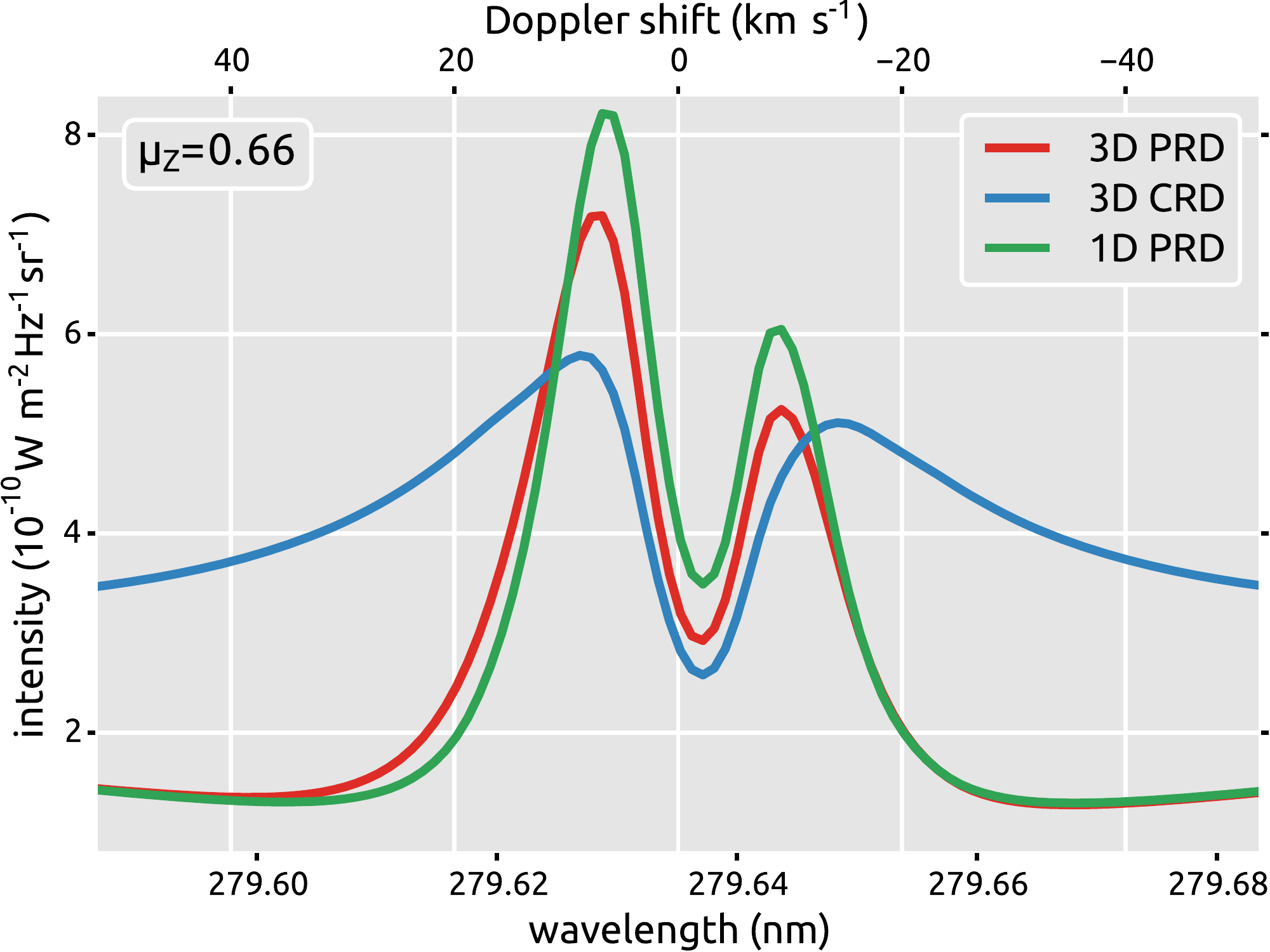}\hfil
    \includegraphics[width=0.4809\textwidth,trim=14mm 0    0 13.5mm,clip=true]{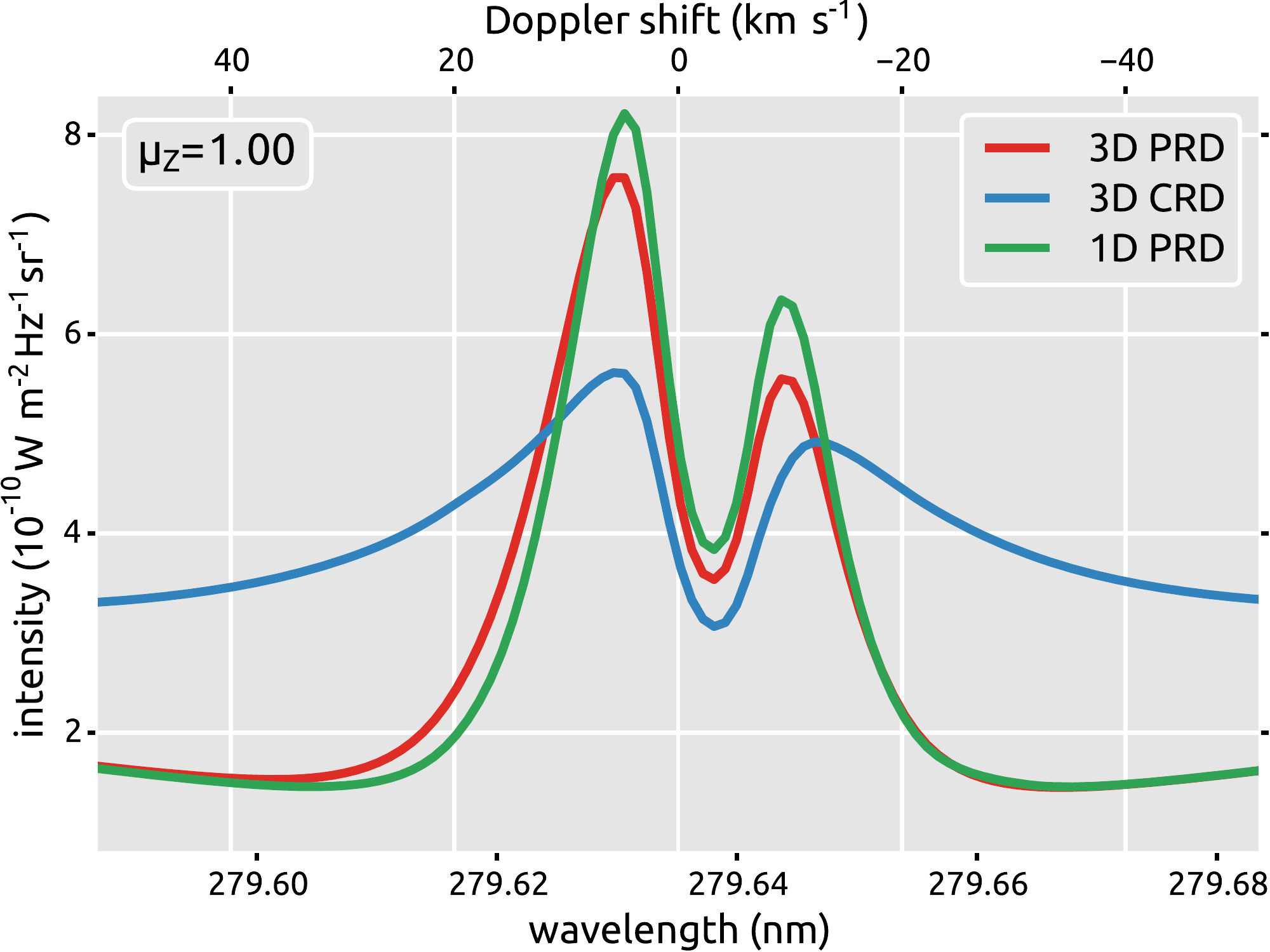}
  \end{minipage}
  \caption{\ion{Mg}{II} k~line profiles for spatially-averaged intensity in
    3D~PRD (red), 3D~CRD (blue), and 1D~PRD (green) computed at different
    $ \mu_\mathrm{Z} $ angles in the 3D~Bifrost model atmosphere.}
  \label{fig:mgiik-meani-clv-mu}
\end{figure*}
\begin{figure*}
  \begin{minipage}{\textwidth}
    \includegraphics[width=0.34878\textwidth,trim=0    0 0 0, clip=true]{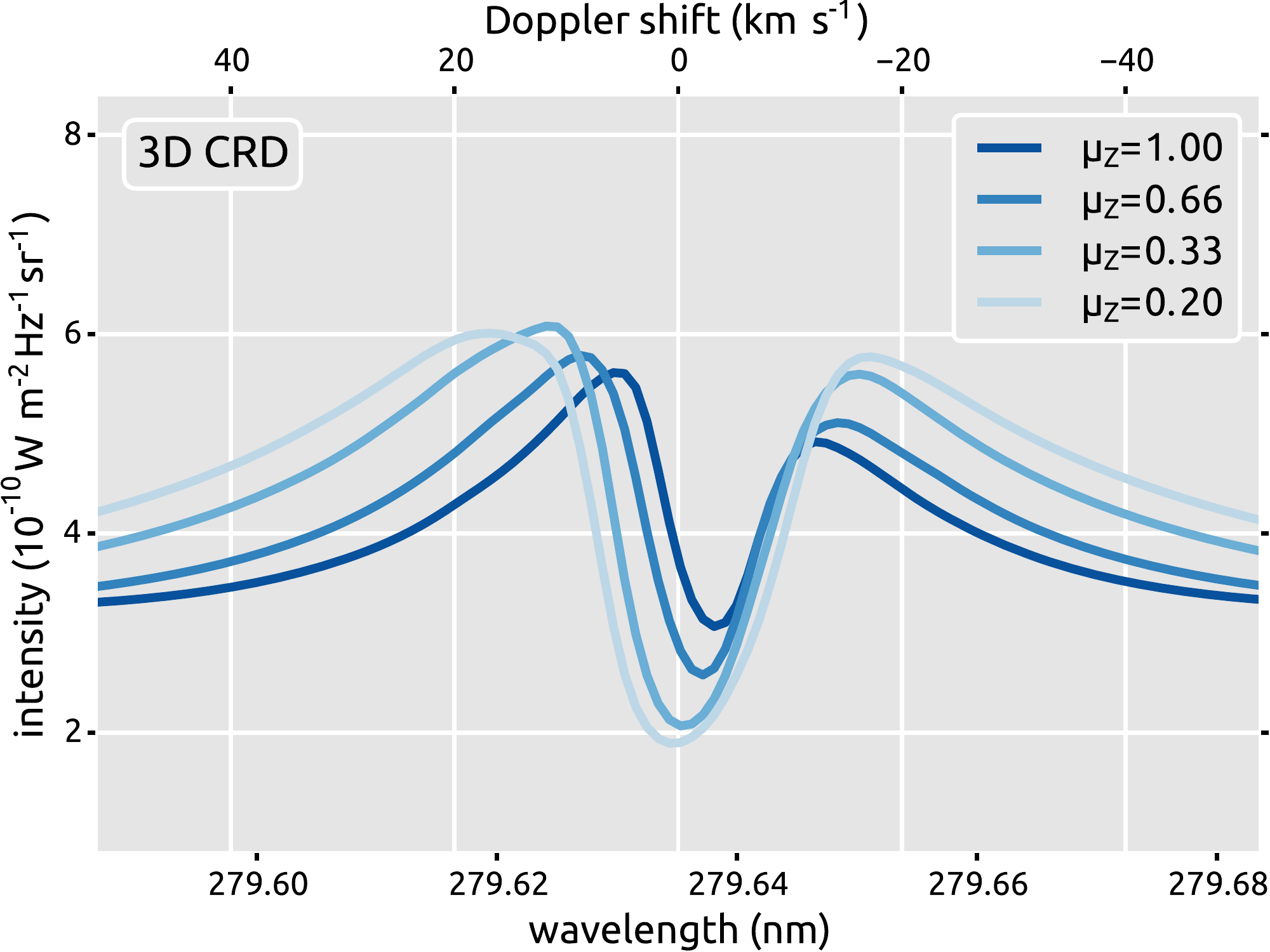}\hfil
    \includegraphics[width=0.32475\textwidth,trim=14mm 0 0 0, clip=true]{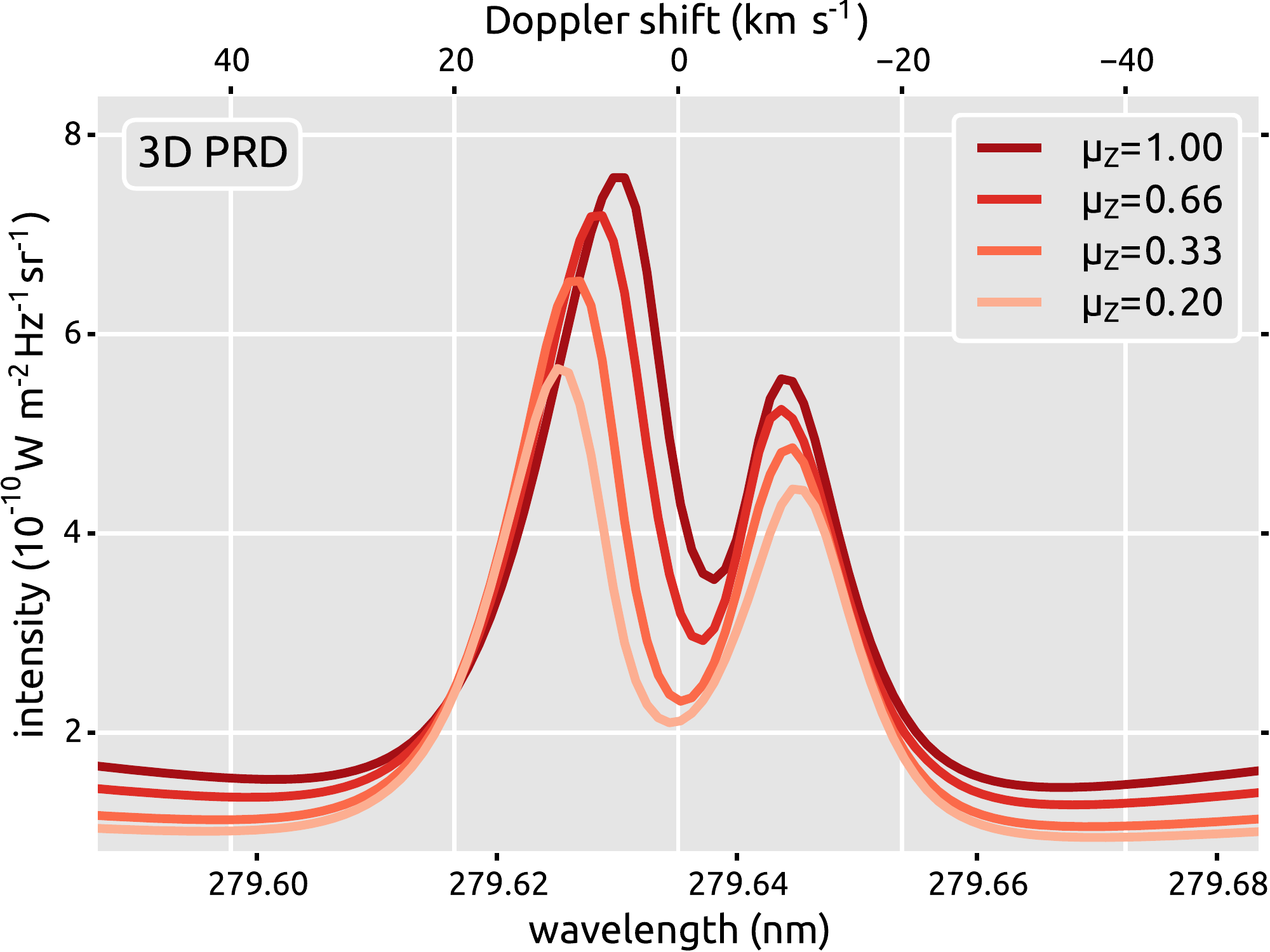}\hfil
    \includegraphics[width=0.32475\textwidth,trim=14mm 0 0 0, clip=true]{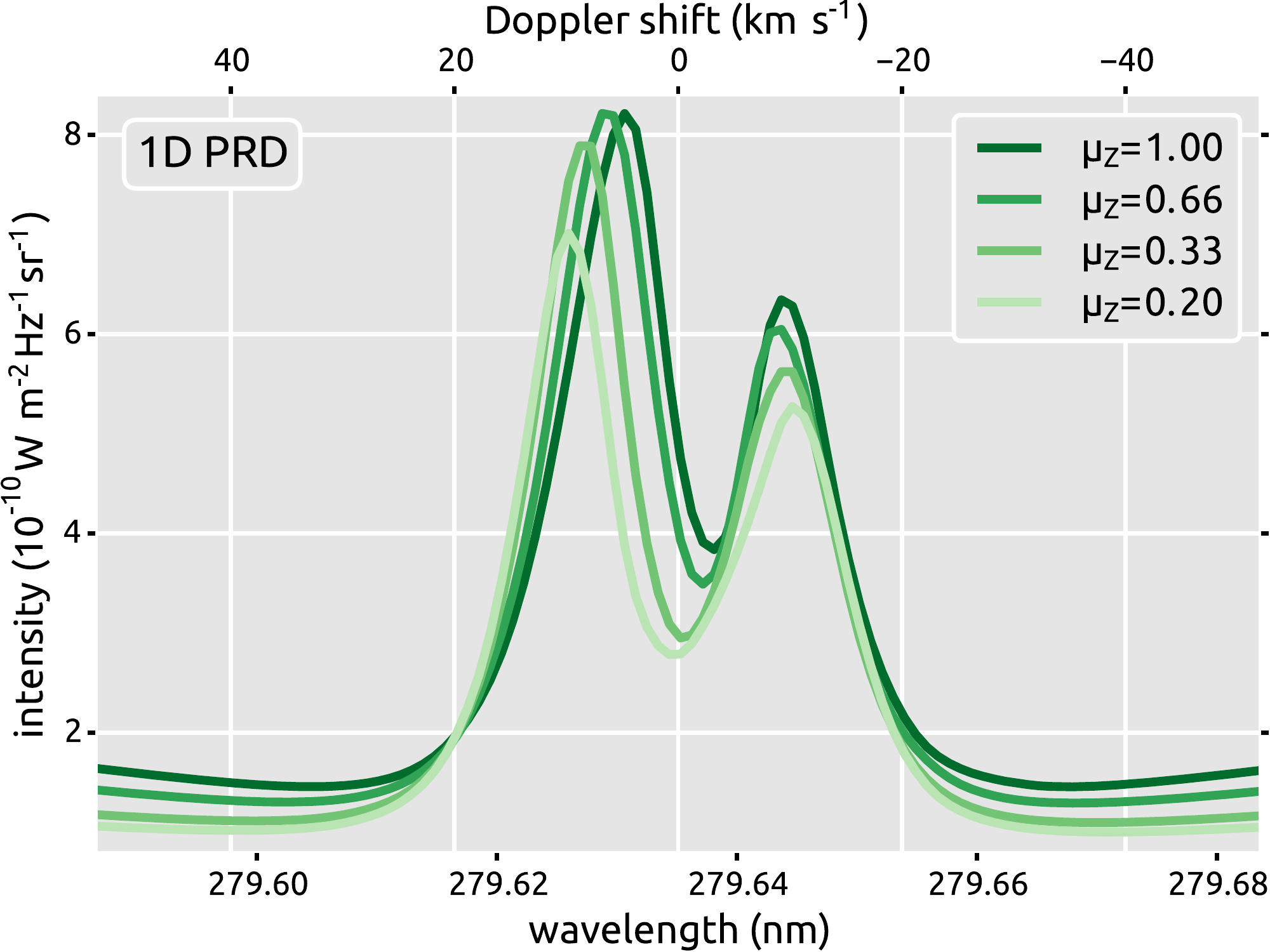}
  \end{minipage}
  \caption{Same as in Fig.\,\ref{fig:mgiik-meani-clv-mu}, but the line profiles,
    computed at different angles, are grouped separately for 3D~CRD, 3D~PRD, and
    1D~PRD.}
  \label{fig:mgiik-meani-clv}
\end{figure*}
Figures~\ref{fig:mgiik-meani-clv-mu} and \ref{fig:mgiik-meani-clv} display
the center-to-limb variation of the spatially-averaged intensity profiles of the
\ion{Mg}{II} k~line.

First, we discuss how PRD and 3D effects change the wavelength positions of the
k$_\mathrm{2v}$, k$_3$, and k$_\mathrm{2r}$ features.
The 1D~PRD approximation correctly predicts the wavelength positions of all the
features, while the 3D~CRD approximation does this only for the k$_3$ feature.
Going towards the limb, we see similar blueshifts of the k$_3$ feature in
3D~CRD, 1D~PRD, and 3D~PRD.
We see the same effect in the k$_\mathrm{2v}$ peak in both PRD cases but a much
stronger shift in 3D~CRD.
Going towards the limb, we see a slight redshift of the k$_\mathrm{2r}$ peak in
both PRD cases and a strong redshift in 3D~CRD.
The peak separation $ \lambda(\mathrm{k_{2r}}) - \lambda(\mathrm{k_{2v}}) $
slightly increases towards the limb in 3D~PRD and it behaves similar in 1D~PRD.
However, in 3D~CRD it strongly increases.

In short, one can use 1D~PRD as a valid approximation for the wavelength
positions of the k$_\mathrm{2v}$ and the k$_\mathrm{2r}$ emission peaks as well
as for the k$_3$ absorption feature.
The 3D~CRD approximation is accurate only for the wavelength of the k$_3$
feature.

Second, we show how PRD and 3D effects change the intensities of the line
profiles including the features.

As expected, the 3D~CRD approximation is not valid for the intensities in the
inner wings and in the emission peaks.
It also underestimates the intensities of the line core at disk center, but this
difference gets smaller towards the limb.

The 1D~PRD approximation is very good for the inner wings, but it overestimates
the intensities of all the features and this effect increases towards the limb.

We see asymmetries in the line profiles in all three cases.

In all three cases we see that the intensities of the k$_\mathrm{2v}$ peak is
larger than the intensities of k$_\mathrm{2r}$.
This is caused by asymmetries in vertical velocity field in the simulation, for
example by upward-propagating shocks
\citep[cf.][for a discussion of this effect in the \ion{Ca}{II} H\&K lines]
      {1997ApJ...481..500C}. 

There is an interesting asymmetry in the flanks of the emission peaks in the
3D~PRD case that is not present in the 1D~PRD case (compare the red and green
curves Fig.~\ref{fig:mgiik-meani-clv-mu} between 279.60 nm and 279.63 nm).
A similar asymmetry is present in  3D~CRD, except that it happens further away
from the line center because of the wider emission peaks.
This asymmetry in the 3D computations must be caused by interaction between the
velocity field in the simulation and the horizontal transport of radiation, and
should be investigated in more detail.

In short, the intensities of the inner wings in the average line profile are
well reproduced by the 1D~PRD approximation.
The line core is however not modeled accurately by 1D PRD.
The 3D~CRD approximation reproduces only the line core intensity towards the
limb.

In summary, we have shown for the \ion{Mg}{II}~k~line that its profile, as well
as the profile features, are influenced by both PRD and 3D effects.
The 1D PRD approximation reproduces the wavelength position of the features, and
can be used to accurately compute intensities in the inner wings.
The intensity and the wavelength position of k$_3$ can be computed using 3D~CRD
at disk center, but not towards the limb.
Accurate quantitative modeling of the whole line profile and its center-to-limb
variation requires 3D~PRD.

\section{Discussion and Conclusions}
\label{sec:discussion-conclusions}

In this paper we presented an algorithm for 3D non-LTE radiative transfer
including PRD effects in atmospheres with velocity fields.
It improves on the algorithm described by
\citet{2012A&A...543A.109L} 
by discretizing PRD line profiles on a frequency grid with intervals being
integer multiples of a chosen fine resolution.  Such a grid allows using fast
linear interpolation with pre-computed indices and weights stored with a small
memory footprint.  This permits a relatively fast solution of the 3D~non-LTE
problem, costing 50\,000--200\,000~CPU hours to reach a converged solution in a
model atmosphere with $252 \times 252 \times 496$ grid points.

We investigated the numerical properties of the algorithm in realistic use case
scenario's and found that it is stable, applicable to strongly scattering lines
like Ly\,$\alpha$, and sufficiently fast to be practically useful for the
computation of time-series of snapshots from radiation-MHD simulations.

We presented an application to the formation of the \ion{Mg}{II} h\&k lines in
the solar chromosphere.  We found that the conclusion of
\citet{2013ApJ...772...89L} 
that the k$_\mathrm{2v}$ and k$_\mathrm{2r}$ emission peaks for disk-center
intensities can be accurately modeled using 1D~PRD computations to be incorrect.
The emission peaks are affected by both PRD and 3D~effects, and should be
modeled using both 3D and PRD~effects simultaneously.

We also showed that the center-to-limb variation in 1D~PRD, 3D~CRD, and 3D~PRD
is markedly different.  The method presented in this paper will allow comparison
of numerical models and observations of PRD-sensitive lines not only for
disk-center observations but also for observations towards the limb.

We are planning to perform a detailed comparison of the center-to-limb variation
of the \ion{Mg}{II} h\&k lines as predicted by numerical simulations to IRIS
observations, such as those by
\citet{2015ApJ...811..127S}. 
%

\begin{acknowledgements}
  Computations were performed on resources provided by the Swedish National
  Infrastructure for Computing (SNIC) at the National Supercomputer Centre at
  Link\"oping University, at the High Performance Computing Center North at
  Ume\aa\ University, and at PDC Centre for High Performance Computing (PDC-HPC)
  at the Royal Institute of Technology in Stockholm.  This study has been
  supported by a grant from the Knut and Alice Wallenberg Foundation
  (CHROMOBS).
\end{acknowledgements}

\bibliographystyle{aa}

\begin{appendix}

\section{Algorithms for the forward and backward transforms}
\label{sec:appendix}
\subsection{Definitions and notations}

The notations adopted in this Appendix are as follows.
Parentheses {(} and {)} denote functional dependence, e.g.,
$ \rho( \vec{n}, q) $, while brackets {[} and {]} denote array indexing, e.g.,
$ q[i] $.
Both notations should not be mistaken with intervals of real numbers, e.g.,
$ x \in {[ x_\mathrm{A}, x_\mathrm{B} )} $ means
$ x_\mathrm{A} \leq x < x_\mathrm{B} $.
A different notation is used for ordered ranges of integers, e.g.,
$ l = m, \dotsc, n $.
Index tuples, which subscript arrays, are always named explicitly not to confuse
them with open intervals, e.g., index tuple $ (x, y, z, \mu, i) $.
Sections of 1D-arrays are denoted by their initial and terminal subscripts
separated by a colon, e.g., $ q[i_\red{:}i_\blue] $.
Names of auxiliary arrays are given in typewriter font, e.g.,
$ \mathtt{map}[i] $, while other quantities, stored in arrays, are given in
roman font, e.g., $ \vec{\varv}[x, y, z] $.
The symbol $ \mapsto $ means interpolation, the symbol $ \gets $ is an
assignment, and the symbol $ \overset{+}{\gets} $ indicates an increment.

As mentioned in Sect.\,\ref{subsec:hybrid-approximation}, we deal with the
inertial frame and the comoving frame.
Frequency-dependent quantities from both frames are Doppler-shifted in
frequency with respect to each other due to the velocity field
$ \vec{\varv}[x, y, z] $ given in the model atmosphere.
The forward transform converts the specific intensity $ I $ from the inertial
frame to the comoving frame by taking into account Doppler shifts of
frequencies, while the backward transform does the same in the opposite
direction for the profile ratio $ \rho $.
All quantities in the comoving frame are marked with a superscript
$ {\!}^{\star\!} $: specific intensity $ I^{\star\!} $, angle-averaged intensity
$ J^{\star\!} $, profile ratio $ \rho^{\star\!} $, and frequency $ q^{\star\!}$.
The same quantities in the inertial frame have no superscript.
The only exception is for line frequencies, which always appear in indexed array
form as either $ q[i] $ or $ q[j] $, where different subscripts are used to tell
to which of the frames $ q $ belongs.
For brevity, we say `inertial/comoving $ X $' instead of `quantity $ X $ in the
inertial/comoving frame'.

Computations are performed in a Cartesian subdomain of
$ N_\mathrm{X} N_\mathrm{Y} N_\mathrm{Z} $ grid points with indices
$ x = x_\mathrm{min}, \dotsc, x_\mathrm{max} $,
$ y = y_\mathrm{min}, \dotsc, y_\mathrm{max} $, and
$ z = z_\mathrm{min}, \dotsc, z_\mathrm{max} $.
Angle-dependent quantities are represented on an angle quadrature with $ N_\mu $
rays having directions $ \vec{n}[\mu]\colon \mu = 1, \dotsc, N_\mu $.
Frequency-dependent quantities are discretized on a grid of $ N_\nu $ points
having frequencies $ q[i]\colon i = i_\red, \dotsc, i_\blue $
(see Sect.\,\ref{subsubsect:doppler-transform} for the definition of $ q $).
Subscript $ i $ denotes frequency points in the inertial grid, while subscript
$ j $ denotes them in the comoving grid, such that their ranges and frequency
grid values are identical: $ i_\red = j_\red $, $ i_\blue = j_\blue $, and
$ q[i] = q[j] $ if $ i = j $.
For the equidistant interpolation
(see Sect.\,\ref{subsec:equidistant-interpolation}), we use two extra grids with
two different subscripts: $ m $ for the fine inertial grid, and $ n $ for the
fine comoving grid.

We use three short-hands `knot', `lie', and `match' with the following meanings.
Following textbooks on interpolation methods, we refer to frequency grid points
as \textit{knots}, e.g., saying `comoving knot $i$' we mean `frequency $ q[i] $
in the comoving frame'.
Saying `inertial knot $ i $ lies on the comoving interval $ {[ j, j+1 )} $', we
mean that the frequency $ q[i] $ in the inertial frame is between two
frequencies $ q^{\star\!}[j] $ and $ q^{\star\!}[j + 1] $ in the comoving frame,
i.e., $ q^{\star\!}[j] \leq q[i] < q^{\star\!}[j + 1] $.
Saying `inertial knot $ i $ matches the comoving knot $ j $', we mean that
$ q[i] = q^{\star\!}[j] $.

In the listings of algorithms, we implicitly use an advantage of the iteration
loop control in Fortran do-statements to avoid if-statements: if the increment
is positive/negative and the starting index is larger/smaller than the ending
index, then no iterations are performed.
The same applies to array sections: if the increment is positive/negative and
the lower bound is larger/smaller than the upper bound, then no array elements
are indexed.
We also use the Fortran feature that array subscripts can start at any integer
number.

\subsubsection{Doppler transform}
\label{subsubsect:doppler-transform}

Usually, velocities in the upper atmosphere of the Sun do not exceed
150\,km\,s$^{-1}$ so that $ \beta \equiv \varv / c < 5\cdot 10^{-4} $ and we can
adopt a non-relativistic approximation for the Doppler transform in both
directions neglecting $ O(\beta^2) $ terms:
\begin{align}
  \nu^{\star\!} &
    = \nu\phantom{^{\star\!}}
    \Bigl( 1 - \dfrac{ \vec{n}\cdot\vec{\varv} }{ c } \Bigr),
    \label{eq:doppler-nu-forward}\\
  \nu\phantom{^{\star\!}} &
    = \nu^{\star\!}
    \Bigl( 1 + \dfrac{ \vec{n}\cdot\vec{\varv} }{ c } \Bigr).
    \label{eq:doppler-nu-backward}
\end{align}
We also neglect aberration of light so that $ \vec{n}^{\star\!} = \vec{n} $.

For convenience, we express frequency $ \nu $ as dimensionless frequency
displacement $ q $ given in Doppler units
$ \Delta\nu_\mathrm{D} = \nu_0\varv_\mathrm{B} / c $:
\begin{equation} \label{eq:q}
  q \equiv
    \dfrac{ \nu - \nu_0 }{ \Delta\nu_\mathrm{D} }
    =
    \Bigl( \dfrac{ \nu }{ \nu_0 } - 1 \Bigr)
    \dfrac{ c }{ \varv_\mathrm{B} },
\end{equation}
with $ \nu_0 $ the line center frequency, and
$ \varv_\mathrm{B} $ the characteristic broadening velocity, which we typically
set to a value of a few km\,s$^{-1}\!$.

Using frequencies $ q $, we can linearize the Doppler transform in
Eqs.\,\eqref{eq:doppler-nu-forward}--\eqref{eq:doppler-nu-backward}:
\begin{align}
  q^{\star\!} & = q\phantom{^{\star\!}} - q_\mu,
  \label{eq:doppler-q-forward}
  \\
  q\phantom{^{\star\!}} & = q^{\star\!} + q_\mu,
  \label{eq:doppler-q-backward}
\end{align}
with the Doppler shift along direction $ \vec{n}[\mu]$ given by a scalar product
\begin{equation}
  q_\mu =
    \dfrac{ \vec{n}[\mu]\cdot\vec{\varv}[x, y, z] }
          { \varv_\mathrm{B}                      }.
  \label{eq:q_mu}
\end{equation}
The linear approximation in
Eqs.\,\eqref{eq:doppler-q-forward}--\eqref{eq:doppler-q-backward}
is valid if $ |q|\,\varv_\mathrm{B} \ll c $, which is true even for the most
extreme situation of the \ion{H}{I} Ly\,$\alpha$ line.

\subsubsection{Piecewise linear interpolation}

Since we are restricted by available computing time, we employ the fastest and
the easiest method: piecewise linear interpolation and constant extrapolation%
\footnote{If more accurate method is needed, the algorithms given in
          Sect.\,\ref{subsec:equidistant-interpolation} for equidistant linear
          interpolation can be easily adapted to shifted linear interpolation
          \citep{2004ITIP...13..710B}. 
}.
We especially benefit from it when we use equidistant frequency grids.

Given data samples $ y_\mathrm{L} $ and $ y_\mathrm{R} $ on two knots L and R
with frequencies $ q_\mathrm{L} $ and $ q_\mathrm{R} $, the interpolant value
$ y $ at frequency $ q \in {[ q_\mathrm{L}, q_\mathrm{R} )} $ is a linear
combination of both samples
\begin{equation}
  y = \varw\,y_\mathrm{L} + (1 - \varw)\,y_\mathrm{R},
\end{equation}
where knot L contributes with the left-hand-side ({l.h.s.})\ weight
\begin{equation}
  \varw = \frac{ q_\mathrm{R} - q\phantom{_\mathrm{R}} }
               { q_\mathrm{R} - q_\mathrm{L}           }
\end{equation}
and knot R contributes with the right-hand-side ({r.h.s.})\ weight
\begin{equation}
  1 - \varw = \frac{ q\phantom{_\mathrm{R}} - q_\mathrm{L} }
                   { q_\mathrm{R}           - q_\mathrm{L} }.
\end{equation}
An advantage of linear interpolation over other schemes is that only one weight
$ \varw $ has to be known, the other one always complements $ \varw $ to one.

Constant extrapolation is used if $ q $ is beyond the interpolation range
$ {[ q_\mathrm{A}, q_\mathrm{Z} )} $ set by the outermost knots A and Z.
If $ q < q_\mathrm{A} $, then $ y = y_\mathrm{A} $, and if
$ q_\mathrm{Z} \leq q $, then $ y = y_\mathrm{Z} $.

When necessary, we call a standard function
\begin{equation} \label{eq:interpolate}
  Y \gets \mathrm{interpolate}\big( X, \vec{x}, \vec{y} \big),
\end{equation}
which evaluates an interpolant $ Y $ at a given coordinate $ X $ for data values
$ \vec{y} $ sampled on knots with coordinates $ \vec{x} $.
It uses a correlated bisection search
\citep[see p.\,111]{1992nrfa.book.....P}: 
\begin{equation} \label{eq:locate}
  l \gets \mathrm{locate}\big( X, \vec{x}, l^\prime \big),
\end{equation}
which returns the index $l$ of the array $ \vec{x} $ such that
$ \vec{x}[l] \leq X < \vec{x}[l + 1] $.
The input parameter $ l^\prime $ is used to initialize the search.

\subsubsection{Interval coverage}
%
\begin{figure}
  \includegraphics[width=\columnwidth]{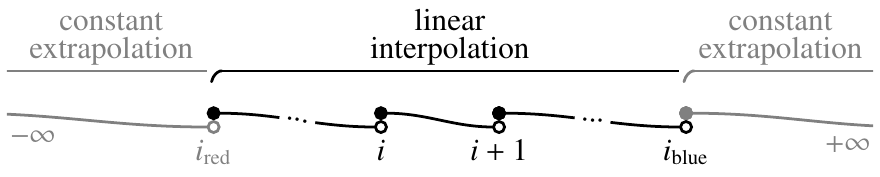}
  \caption{Adopted coverage of the interpolation intervals.  Linear
    interpolation is performed on internal right-open intervals, given in black.
    Constant extrapolation is performed on outer semi-infinite intervals, given
    in gray.  Lines are sloped to show their end points separately.}
  \label{fig:interval-coverage}
\end{figure}
Given a monotonically ordered set of knots
$ \{ i\colon i = i_\red, \dotsc, i_\blue \} $
representing a line frequency grid $ q[i] $ from the leftmost knot $ i_\red $ to
the rightmost knot $ i_\blue $, we complement it by two infinite points
$ {-}\infty $ and $ {+}\infty $ for convenience.
We perform interpolation and extrapolation on right-open intervals to avoid
using values at knots twice on adjacent intervals.
That is to say, linear interpolation is carried out on
$ {[ i, i + 1 )}\colon i = i_\red, \dotsc, i_\blue - 1 $ intervals, and constant
extrapolation is performed on the $ ( {-}\infty, i_\red ) $ and
$ {[ i_\blue, {+}\infty )} $ intervals as shown in
Fig.\,\ref{fig:interval-coverage}.

\subsubsection{Forward transform}
\label{subsubsec:forward-transform}
%
\begin{figure}
  \includegraphics[width=\columnwidth]{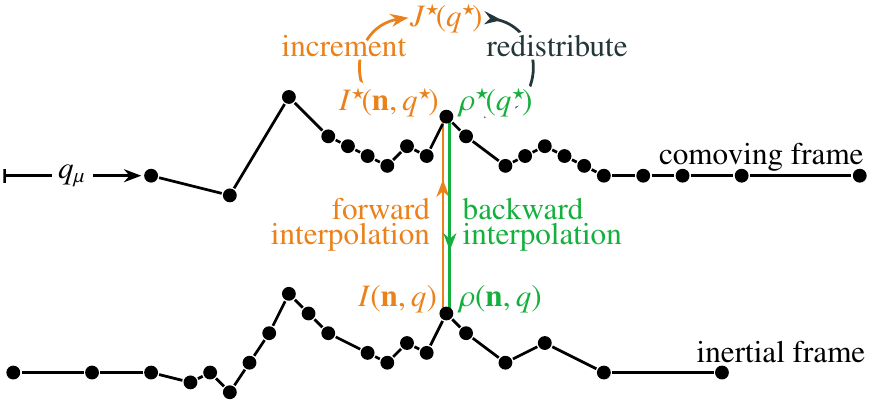}
  \caption{Doppler transform between two frequency grids: the inertial-frame
    grid $ q $ (bottom) and the comoving-frame grid $ q^{\star\!} $ (top).
    Both grids are identical, but the comoving one is displaced in increasing
    direction of frequency by Doppler shift $ q_\mu $:
    $ q^{\star\!} = q - q_\mu $.
    The forward transform (orange) consists of forward interpolation for the
    specific intensity $ I \mapsto I^{\star\!} $ and an increment of the
    angle-averaged intensity $ J^{\star\!} $.
    The backward transform (green) is backward interpolation for the profile
    ratio $ \rho^{\star\!} \mapsto \rho $.}
  \label{fig:doppler-transform}
\end{figure}
\begin{figure*}
  \includegraphics[width=\textwidth]{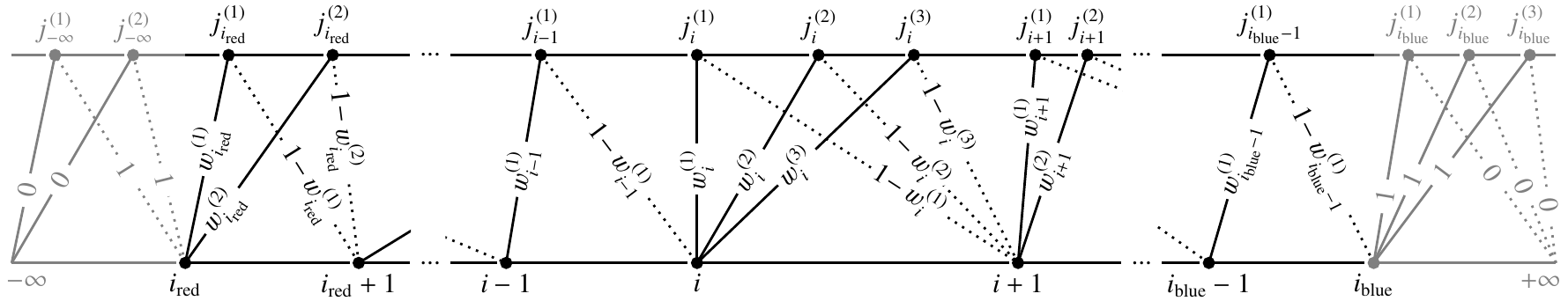}
  \caption{Illustration of the forward interpolation.
    Interpolation is performed for comoving knots $ j $ (top line) on intervals
    between the inertial knots $ i $ (bottom line).
    Each inertial interval $ {[ i, i + 1 )} $ contains zero or more comoving
    knots denoted by
    $ j_i^{\,(\alpha)}\colon \alpha = 1,\dotsc, \alpha_\mathrm{max} $.
    Corresponding {l.h.s.}\ weights $ \varw_i^{(\alpha)} $ (solid) and
    {r.h.s.}\ weights $ 1 - \varw_i^{(\alpha)} $ (dotted) are shown by sloped
    lines.
    Constant extrapolation is performed with
    $ 1 - \varw_{{-}\infty}^{(\alpha)} = 1 $ weights on the
    $ ( {-}\infty, i_\red ) $ interval, and with
    $ \varw_{i_\blue}^{(\alpha)} = 1 $ weights on the
    $ {[ i_\blue, {+}\infty )} $ interval, both drawn with gray color.
    Note that the comoving knot $ j_i^{\,(1)} $, which exactly matches the
    inertial knot $ i $, lies on the the $ {[ i, i + 1 )} $ interval, in
    agreement with the right-openness of all intervals.}
  \label{fig:forward-transform}
\end{figure*}
After the formal solution of the transfer equation is done at coordinate
$ (x, y, z) $ for direction $ \vec{n}[\mu] $ at frequency $ q[i] $, we have the
inertial specific intensity
$ I\bigl( \vec{n}[\mu], q[i] \bigr) = I[x, y, z, \mu, i] $.
Owing to the local nature of the intensity redistribution and the Doppler
transform, we drop dependencies on the spatial coordinates in the formulas given
below.

The forward transform consists of two parts (see orange notations in
Fig.\,\ref{fig:doppler-transform}).
First, the inertial specific intensity $ I\bigl( \vec{n}[\mu], q[i] \bigr) $ is
interpolated to get the comoving specific intensity
$ I^{\star\!}\bigl( \vec{n}[\mu], q^{\star\!} \bigr) $
at comoving frequency $ q^{\star\!} = q[i] - q_\mu $:
\begin{equation} \label{eq:forward-interpolation}
  I\bigl( \vec{n}[\mu], q[i] \bigr)
  \mapsto
    I^{\star\!}\bigl( \vec{n}[\mu], q^{\star\!} \bigr),
\end{equation}
This part is also called forward interpolation.
Second, the interpolated intensity
$ I^{\star\!}\bigl( \vec{n}[\mu], q^{\star\!} \bigr) $
contributes into (increments) the comoving angle-averaged intensity
$ J^{\star\!}\bigl( q^{\star\!} \bigr) $:
\begin{equation} \label{eq:forward-J*-increment}
  J^{\star\!}\bigl( q^{\star\!} \bigr)
    \overset{+}{\gets}
    \dfrac{ \omega_\mu }{ 4\pi }
    I^{\star\!}\bigl( \vec{n}[\mu], q^{\star\!} \bigr),
\end{equation}
with $ \omega_\mu / 4\pi $ the quadrature weight for direction $ \vec{n}[\mu] $.
If we combine Eq.\,\eqref{eq:forward-interpolation} with
Eq.\,\eqref{eq:forward-J*-increment} for all $ \mu $, we obtain the
angle-discretized form of Eq.\,\eqref{eq:J*}:
\begin{equation} \label{eq:J*-sum}
  J^{\star\!}\bigl( q^{\star\!} \bigr) =
    \sum_{ \mu = 1 }^{ N_\mu }
    \dfrac{ \omega_\mu }{ 4\pi }
    I\bigl( \vec{n}[\mu], q^{\star\!} + q_\mu \bigr).
\end{equation}
This equation implies first a transform of the intensity from the inertial to
the comoving frame, which numerically corresponds to an interpolation.
The problem is that the intensity $ I $ is not known for all frequencies
simultaneously, due to the way how the formal solution is performed numerically
in the Multi3D code.
Instead, we compute
Eqs.\,\eqref{eq:forward-interpolation}--\eqref{eq:forward-J*-increment}
incrementally from each $ I\bigl( \vec{n}[\mu], q[i] \bigr) $, without ever
having the intensity for other inertial frequencies and angles in memory at the
same time.

Forward interpolation
$ I\bigl[ \mu, i \bigr] \mapsto I^{\star\!}\bigl[ \mu, j \bigr] $
contributes the intensity $ I $ at the inertial knot $ i $ to intensities
$ I^{\star\!} $ at related comoving knots $ j $.
The related knots $ j $ lie on the left and right neighboring intervals on both
sides of $ i $.
We adopt the following notation for a series of the comoving knots lying on the
inertial interval $ {[ i, i + 1 )} $: $ j_i^{\,(\alpha)} $ denote the related
knots, and $ \varw_i^{(\alpha)} $ denote the corresponding
{l.h.s.}\ interpolation weights, where the subscript $ i $ is the starting index
of the inertial interval and the counter
$ \alpha = 1, \dotsc, \alpha_\mathrm{max} $ numbers all the knots in the series.
There are zero or more $ (\alpha_\mathrm{max} \geq 1) $ comoving knots, whose
indices $ j_i^{\,(\alpha)} $ and weights $ \varw_i^{(\alpha)} $ depend on the
positions and relative distances between their comoving frequencies
$ q^{\star\!} = q[j] + q_\mu $ and the inertial frequency $ q[i] $.

This is illustrated in Fig.\,\ref{fig:forward-transform}.
The knot $ i $ always has two neighbors $ i - 1 $ and $ i + 1 $ so that related
knots $ j $ are interpolated on the left $ {[ i - 1, i )} $ and right
$ {[ i, i + 1 )} $ intervals.
The inertial intensity $ I $ at the knot $ i $ is interpolated to the comoving
intensity $ I^\star $ at knots $ j_{i - 1}^{\,(\alpha)} $ using
{r.h.s.}\ weights $ 1 - \varw_{i - 1}^{(\alpha)} $ on the left interval, and at
knots $ j_{i}^{\,(\alpha)} $ using {l.h.s.}\ weights $ \varw_{i}^{(\alpha)} $ on
the right interval.
At the ends of the frequency grid, weights on the infinite intervals are always
equal to unity to force constant extrapolation:
$ 1 - \varw_{i - 1}^{(\alpha)} = 1 - \varw_{{-}\infty}^{(\alpha)} = 1 $ for
$ i = i_\red $, and
$ \varw_{i}^{(\alpha)} = \varw_{i_\blue}^{(\alpha)} = 1 $ for $ i = i_\blue $.

Plugging expressions for $ j $ and $ \varw $ into the discretized
Eqs.\,\eqref{eq:forward-interpolation}--\eqref{eq:forward-J*-increment}
we get
\begin{alignat}{2}
  J^{\star\!}\bigl[ j_{i-1}^{\,(\alpha)} \bigr]
    & \overset{+}{\gets}
    \Bigl(1 - \varw_{i-1}^{(\alpha)} \Bigr)
    \dfrac{ \omega_\mu }{ 4\pi }
    I\bigl[ \mu, i - 1 \bigr]
    && \quad\text{on } {[ i-1, i )}, \label{eq:J*-lhs}
    \\
  J^{\star\!}\bigl[ j_{i}^{\,(\alpha)} \bigr]
    & \overset{+}{\gets}
    \mspace{43mu}\varw_i^{(\alpha)}
    \dfrac{ \omega_\mu }{ 4\pi }
    I\bigl[ \mu, i \bigr]
    && \quad\text{on } {[ i, i+1 )}. \label{eq:J*-rhs}
\end{alignat}

\subsubsection{Backward transform}
\label{subsubsec:backward-transform}
%
\begin{figure}
  \includegraphics[width=\columnwidth]{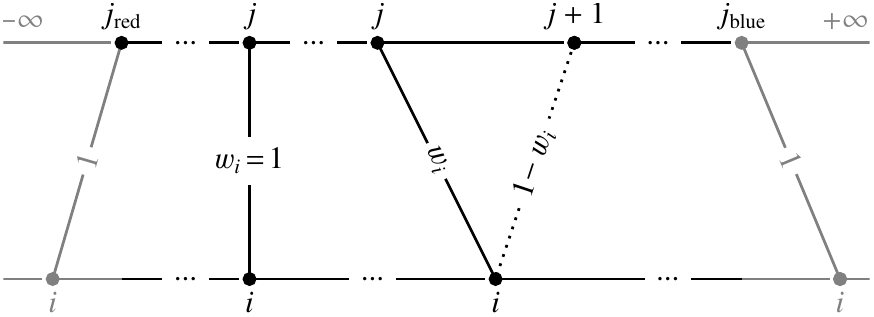}
  \caption{Illustration of the backward interpolation.
    Interpolation is performed for inertial knots $ i $ (bottom line) on
    intervals between comoving knots $ j $ (top line).
    Interpolation weights are indicated on sloped lines connecting corresponding
    knots between the frames.
    Linear interpolation is used for internal inertial knots $ i $ that lie on
    the internal comoving intervals $ {[ j, j+1 )} $ with {l.h.s.}\ $ \varw_i $
    (solid) and {r.h.s.}\ $ 1 - \varw_i $ (dotted) weights.
    Weights are named with respect to the comoving frame so that the line slopes
    are opposite to those in the forward interpolation.
    Constant extrapolation with unity weights is used for inertial knots $ i $
    exactly matching comoving knots $ j $, or for inertial knots that lie on the
    $ {( {-}\infty, j_\red )} $ or $ {[ j_\blue, {+}\infty )} $ intervals.
    Both end cases are given in gray color.}
  \label{fig:backward-transform}
\end{figure}

Once $J^{\star}$ has been fully accumulated, we compute the comoving profile
ratio $ \rho^{\star\!}\bigl( q[j] \bigr) = \rho^{\star\!}[ x, y, z, j ] $
and keep it in memory for all frequency knots $ j = j_\red, \dotsc, j_\blue $
simultaneously.
This makes the backward transform much easier than the forward one, because we
can employ a standard interpolation.

The backward transform can be directly evaluated because it is just an
interpolation from the comoving frame to the inertial frame (see green notations
in Fig.\,\ref{fig:doppler-transform}):
\begin{equation} \label{eq:backward-transform}
  \rho^{\star\!}\bigl( q[i] - q_\mu \bigr)
  \mapsto
    \rho\bigl( \vec{n}[\mu], q[i] \bigr).
\end{equation}
As follows from Fig.\,\ref{fig:backward-transform}, each
$ \rho \bigl( \vec{n}[\mu], q[i] \bigr) $ depends on either one or two comoving
knots $ j $.

It depends on two knots, if the inertial knot $ i $ lies on the comoving
interval $ {[j, j + 1)} $, where both $ j $ and $ j + 1 $ are internal knots
($ j_\red \leq j < j + 1 \leq j_\blue $), i.e.,
$ q[j] + q_\mu \leq q[i] < q[j + 1] + q_\mu $.
In this case we do linear interpolation:
\begin{equation} \label{eq:backward-two-knots}
  \rho[\mu, i]
  =
    \varw_i\, \rho^{\star\!}[j]
    +
    (1 - \varw_i)\, \rho^{\star\!}[j + 1].
\end{equation}
For the backward transform, knots and weights are always denoted by
$ j $ and $ \varw_i $ with no superscript $ (\alpha) $ to tell them from knots
$ j_i^{\,(\alpha)} $ and weights $ \varw_i^{(\alpha)} $ for the forward
transform.

It depends on one knot, if the inertial knot $ i $ matches the comoving knot
$ j $, i.e., $ q[i] = q[j] + q_\mu $.
It also depends on one knot, if the inertial knot $ i $ lies on intervals before
$ j_\red $ or after $ j_\blue $, which means $ q[i] < q[j_\red] + q_\mu $ or
$ q[j_\blue] + q_\mu \le q[i] $, because we use constant extrapolation.
Therefore, in those three cases:
\begin{alignat}{2}
  \rho[\mu, i] & = \rho^{\star\!}[j],       && \text{ or}
  \label{eq:backward-exact-match}
  \\
  \rho[\mu, i] & = \rho^{\star\!}[j_\red],  && \text{ or}
  \label{eq:backward-ired}
  \\
  \rho[\mu, i] & = \rho^{\star\!}[j_\blue]. && { }
  \label{eq:backward-iblue}
\end{alignat}

\subsection{General interpolation}
\label{subsec:general-interpolation}

One can perform the searches and interpolations described in
Sects.\,\ref{subsubsec:forward-transform}--\ref{subsubsec:backward-transform}
on the fly.
This does not require any additional storage, and can be used for a frequency
grid of arbitrary spacing, but the algorithm is slow.
We call this algorithm \textit{general interpolation}.

\subsubsection{General forward transform}
\label{subsubsec:general-forward}
\begin{algorithm}[!t]
  \SetInd{0.5em}{0.5em}
  \DontPrintSemicolon
  \LinesNotNumbered
  \SetCommentSty{textsl}
  \newcommand{\fwid}{\widthof{$ {\bigl[ \min(i + 1, i_\blue) \bigr]} $}}
  \KwData{Specific intensity $ I[x, y, z, \mu, i] $ at frequency $ q[i]$ and
    direction $ \vec{n}[\mu] $ in the inertial frame.}
  \KwResult{Incremented angle-averaged intensity $ J^{\star\!}[x, y, z, j] $ at
    corresponding frequencies $ q[j] $ in the comoving frame.}
  $ q_\mu \gets \bigl( \vec{n}[\mu] \cdot \vec{\varv}[x, y, z] \bigr) / \varv_\mathrm{B} $\;
  $ q^\star_\mathrm{L} \gets q\mathmakebox[\fwid][l]{\bigl[ \max(i - 1, i_\red) \bigr]} - q_\mu $\;
  $ q^\star_\mathrm{C} \gets q\mathmakebox[\fwid][l]{\bigl[i\bigr]} - q_\mu $\;
  $ q^\star_\mathrm{R} \gets q\bigl[ \min( i_\blue, i + 1 ) \bigr] - q_\mu $\;
  \eIf(\tcp*[f]{Linear interpolation on $ \bigl( q^\star_\mathrm{L}, q^\star_\mathrm{C} \bigr) $:})
    {$ q^\star_\mathrm{L} \neq q^\star_\mathrm{C} $}{
    $ \Delta \gets q^\star_\mathrm{C} - q^\star_\mathrm{L} $\;
    \For{$ j \gets j_\red $ \KwTo $ j_\blue $}{               
      \If{$ q^\star_\mathrm{L} < q[j] < q^\star_\mathrm{C} $}{  
        $ \xi^\prime \gets \big( q[j] - q^\star_\mathrm{L} \big) / \Delta $
          \tcp*[r]{The {r.h.s.}\ weight $ 1 - \varw_{i - 1} $.}
        $ J^{\star\!}[x, y, z, j] \overset{+}{\gets}
          \xi^\prime \dfrac{\omega_\mu}{4\pi} I[x, y, z, \mu, i] $\;
      }
    }
  }($ q^\star_\mathrm{L} = q^\star_\mathrm{C} $
    \tcp*[f]{Constant extrapolation on $ \bigl( {-}\infty, q^\star_\mathrm{C} \bigr) $:}){
    \For{$ j \gets j_\red $ \KwTo $ j_\blue $}{
      \If{$ q[j] < q^\star_\mathrm{C} $}{
        $ J^{\star\!}[x, y, z, j] \overset{+}{\gets} \dfrac{\omega_\mu}{4\pi} I[x, y, z, i] $\;
      }
    }
  }
  \eIf(\tcp*[f]{Linear interpolation on $ {\bigl[ q^\star_\mathrm{C}, q^\star_\mathrm{R} \bigr)} $:})
      {$ q^\star_\mathrm{C} \neq q^\star_\mathrm{R} $}{
    $ \Delta \gets q^\star_\mathrm{R} - q^\star_\mathrm{C} $\;
    \For{$ j \gets j_\red $ \KwTo $ j_\blue $}{
      \If{$ q^\star_\mathrm{C} \leq q[j] < q^\star_\mathrm{R} $}{
        $ \xi \gets \big( q^\star_\mathrm{R} - q[j] \big) / \Delta $
          \tcp*[r]{The {l.h.s.}\ weight $ \varw_i $.}
        $ J^{\star\!}[x, y, z, j] \overset{+}{\gets} \xi \dfrac{\omega_\mu}{4\pi} I[x, y, z, i] $\;
      }
    }
  }($ q^\star_\mathrm{C} = q^\star_\mathrm{R} $
    \tcp*[f]{Constant extrapolation on $ {\bigl[ q^\star_\mathrm{C}, {+}\infty \bigr)} $:}){
    \For{$ j \gets j_\red $ \KwTo $ j_\blue $}{
      \If{$ q^\star_\mathrm{C} \leq q[j] $}{
        $ J^{\star\!}[x, y, z, j] \overset{+}{\gets}
          \dfrac{\omega_\mu}{4\pi} I[x, y, z, i] $\;
      }
    }
  }
  \caption{General forward transform.  Comments in this algorithm and those
    listed below are given in slanted text after `//'.}
  \label{alg:general-forward}
\end{algorithm}
We search for all comoving knots $ j $ that lie on the intervals
$ {[i - 1, i)} $ and $ {[i, i + 1)} $, i.e., whose comoving frequencies
$ q^{\star\!} = q[j] + q_\mu $ are either between $ q[i - 1] $ and $ q[i] $, or
between $ q[i] $ and $ q[i + 1] $.

We take the current inertial knot $ i $ and its left neighbor
$ \max( i - 1, i_\red ) $ and right neighbor $ \min( i_\blue, i + 1 ) $.
We use $ \max() $ and $ \min() $ to keep the neighbors within the frequency grid
range $ i = i_\red, \dotsc, i_\blue $.
For these knots we compute their frequencies in the comoving frame, which we
denote by
\begin{alignat}{2}
  q^\star_\mathrm{L} & = q\bigl[ \max(i - 1,  i_\red) \bigr] && - q_\mu, \\
  q^\star_\mathrm{C} & = q\bigl[ i                    \bigr] && - q_\mu, \\
  q^\star_\mathrm{R} & = q\bigl[ \min(i_\blue, i + 1) \bigr] && - q_\mu.
\end{alignat}
Therefore, the comoving knots $ j $ related to the inertial knot $ i $ are
those, whose frequencies $ q[j] $ belong to the
$ {\bigl[ q^\star_\mathrm{L}, q^\star_\mathrm{C} \bigr)} $ or
$ {\bigl[ q^\star_\mathrm{C}, q^\star_\mathrm{R} \bigr)} $ intervals.

The knots $ j $ lie either on the left or on the right side of
$ q^\star_\mathrm{C} $, and they might fall outside the frequency coverage of
the line.
We thus have four different cases to test for and apply
Eqs.\,\eqref{eq:J*-lhs}--\eqref{eq:J*-rhs}.
Algorithm~\ref{alg:general-forward} shows the forward transform in detail.

\subsubsection{General backward transform}
%
\begin{algorithm}[!t]
  \SetInd{0.5em}{0.5em}
  \DontPrintSemicolon
  \LinesNotNumbered
  \SetCommentSty{textsl}
  \newcommand{\finvisible}{$ \phantom{ \rho[x, y, z, \mu, i] \gets \mathrm{interpolate}\big(\, } $}
  \KwData{Profile ratio $ \rho^{\star\!}[x, y, z, j] $ for all frequencies
    $ q[j]\colon j = j_\red, \dotsc, j_\blue $ in the comoving frame.}
  \KwResult{Profile ratio $ \rho[x, y, z, \mu, i] $ for given frequency $ q[i] $
    and direction $ \vec{n}[\mu] $ in the inertial frame.}
  $ q_\mu \gets \big( \vec{n}[\mu]\cdot \vec{\varv}[x, y, z] \big) / \varv_\mathrm{B} $\;
  $ \rho[x, y, z, \mu, i] \gets \mathrm{interpolate}\big(\, X = q[i] - q_\mu, $\;
  \finvisible$ \vec{x} = q[j_\red{:}j_\blue], $\;
  \finvisible$ \vec{y} = \rho^{\star\!}[x, y, z, j_\red{:}j_\blue] \,\big) $\;
  \caption{General backward transform.}
  \label{alg:general-backward}
\end{algorithm}
Algorithm~\ref{alg:general-backward} gives the backward transform, which is just
an interpolation.
It is performed by using Eq.\,\eqref{eq:interpolate} with
$ Y = \rho[x, y, z, \mu, i] $,
$ X = q[i] - q_\mu $,
$ \vec{x} = q[j_\red{:}j_\blue] $, and
$ \vec{y} = \rho^{\star\!}[x, y, z, j_\red{:}j_\blue] $.

\subsection{Precomputed interpolation}
\label{subsec:precomputed-interpolation}
The general algorithm given in Sect.\,\ref{subsec:general-interpolation} is slow
because for each $ x $, $ y $, $ z $, $ \mu $, and $ i $ we perform two linear
searches in the forward transform and a correlated bisection search in the
backward transform along the full frequency range
$ j = j_\red, \dotsc, j_\blue $.

To speed up the computations, one can precompute and store for each index tuple
$ (x, y, z, \mu, i) $ the comoving knots $ j_i^{\,(\alpha)} $ and the
{l.h.s}\ weights $ \varw_i^{(\alpha)} $ used in the forward transform and the
nearest comoving knot $ j_i $ and its {l.h.s.}\ weight $ \varw_i $ used in the
backward transform.

\subsubsection{Precomputation for the forward interpolation}
\label{subsubsec:precomputed-forward}
%
\begin{figure}
  \includegraphics[width=\columnwidth]{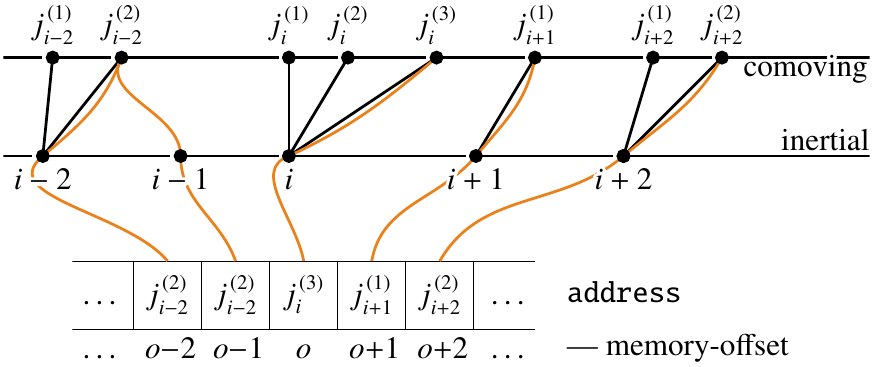}
  \caption{Illustrative contents of the \texttt{address} array for the forward
    precomputed interpolation.
    In the inertial frame (bottom black line), knots have indices $ i $.
    In the comoving frame (top black line), knots that lie on the inertial
    interval $ {[ i, i + 1 )} $, have indices
    $ j_i^{\,(\alpha)}\colon \alpha = 1,\dotsc, \alpha_\mathrm{max} $
    if at least one such knot exists.
    For each $ i $, corresponding memory-offset $ o $ is computed such that
    $ \mathtt{address}[o] $ points to $ j_i^{\,(\alpha_\mathrm{max})} $.
    Orange lines indicate
    $ i {\to} \mathtt{address}[o] {\to} j_i^{\,(\alpha_\mathrm{max})} $
    correspondence.
    Solid black lines connecting $ i $ to $ j_i^{\,(\alpha)} $,
    indicate {l.h.s.}\ weights $ \varw_i^{(\alpha)} $.
    There is no contribution from knot $ i - 1 $, because interval
    $ {[ i - 1, i )} $ does not contain any comoving knots.
    In this case, $ \mathtt{address}[o - 1] $ points to the knot $ j_{i - 2}^{\,(2)} $
    that lies on the previous interval.
    Note, that \texttt{address} keeps addresses but not the values of the
    shown comoving knots, those are stored in \texttt{forward\_j} instead. }
  \label{fig:precomputed-forward}
\end{figure}
\begin{algorithm}[]
  \SetInd{0.5em}{0.5em}
  \DontPrintSemicolon
  \LinesNotNumbered
  \SetCommentSty{textsl}
  \KwData{Line frequencies $ q[i_\red{:}i_\blue] $,
    directions $ \vec{n}[1{:}N_\mu] $,
    velocities $ \vec{\varv}[x_\mathrm{min}{:}x_\mathrm{max},
                             y_\mathrm{min}{:}y_\mathrm{max},
                             z_\mathrm{min}{:}z_\mathrm{max}] $.}
  \KwResult{Arrays \texttt{address}, \texttt{forward\_j}, and
    \texttt{forward\_w}.}
  $ \mathtt{address}[] \gets 0 $
    \tcp*[r]{Initialize array \texttt{address}}
  $ c \gets 0 $
    \tcp*[r]{and its counter.\makebox[38pt]{}}
  \For{$ x \gets x_\mathrm{min} $ \KwTo $ x_\mathrm{max} $}{
    \For{$ y \gets y_\mathrm{min} $ \KwTo $ y_\mathrm{max} $}{
      \For{$ z \gets z_\mathrm{min} $ \KwTo $ z_\mathrm{max} $}{
        \For{$ \mu \gets 1 $ \KwTo $ N_\mu $}{
          $ q_\mu
            \gets
            \big( \vec{n}[\mu]\cdot \vec{\varv}[x, y, z] \big) /
            \varv_\mathrm{B} $\;
          $ i \gets i_\red - 1 $
            \tcp*[r]{Start searching from the leftmost}
            \tcp*[r]{knot $ ({-}\infty) $ in the inertial frame.\makebox[6pt]{}}
          \For(\tcp*[f]{For each comoving knot $j$,})
            {$ j \gets j_\red $ \KwTo $ j_\blue $}{ 
            \tcp*[r]{find which inertial knot $ i $\makebox[7pt]{}}
            \tcp*[r]{contributes into it.\makebox[34pt]{}}
            $ q \gets q[j] + q_\mu $
              \tcp*[r]{Inertial frequency.}
            $i \gets \mathrm{locate}\big( q, q[i_\red{:}i_\blue], i \big)$
              \tcp*[r]{$ q[i] \leq q < q[i + 1] $}
            \uIf(\tcp*[f]{Linear interpolation})
              {$i_\red \leq i < i_\blue$}{
              $ \varw \gets \dfrac{ q[i + 1] - q\mspace{16mu} }
                                  { q[i + 1] - q[i]           } $
              \tcp*[r]{on $ {[ i, i + 1 )} $.\makebox[33pt]{}}
            }
            \uElseIf(\tcp*[f]{Constant extrapolation})
              {$ i = i_\red - 1 $}{
              $ \varw \gets 0 $
              \tcp*[r]{on $ {( {-}\infty, i_\red )} $.\makebox[37pt]{}}
            }
            \Else($ i = i_\blue $\tcp*[f]{Constant extrapolation}){
              $ \varw \gets 1 $
              \tcp*[r]{on $ {[ i_\blue, {+}\infty )} $.\makebox[35pt]{}}
            }
            $ o \gets o(x, y, z, \mu, i) $
              \tcp*[r]{Memory-offset by Eq.\,\eqref{eq:tuple}.}
            \tcp*[r]{Fill \texttt{address} between $ c $ and $ o $}
            \tcp*[r]{with the last stored address:\makebox[10pt]{}}
            $ \mathtt{address}[c + 1{:}o - 1] \gets \mathtt{address}[c] $\;

            \tcp*[r]{Update the current address:}
            $ \mathtt{address}[o] \gets \mathtt{address}[c] + 1 $\;
            $ c \gets o $
              \tcp*[r]{Set counter to the current address.}
            $ \mathtt{forward\_j}\left[\mathtt{address}[c]\right] \gets j $
              \tcp*[r]{Save comoving\makebox[3pt]{}}
            $ \mathtt{forward\_w}\left[\mathtt{address}[c]\right] \gets \varw $
              \tcp*[r]{index \& weight.}
          }
        }
      }
    }
  }
  \caption{Initialization of the precomputed forward interpolation.}
  \label{alg:precomputation-forward}
\end{algorithm}

In the forward transform, each inertial knot $ i $ contributes to zero, one
or more comoving knots $ j $.
Since the indices and the weights depend on $ x $, $ y $, $ z $, $ \mu $, $ i $,
and $ \alpha $, they require jagged 6D-arrays with the last variable-length
dimension for index $ \alpha $.
Neither Fortran nor C allows such a data structure in their latest language
standards.
A typical solution is to use 5D-arrays of pointers or linked lists.
Both solutions are inconvenient and memory-inefficient because they require at
least one 8-byte pointer (on the default 64-bit architecture) for each
$ \alpha $-series of $ j_i^{\,(\alpha)} $ and $ \varw_i^{(\alpha)} $.

It is more efficient to flatten these jagged arrays into two continous
1D-arrays, one for $ j_i^{\,(\alpha)} $ (\texttt{forward\_j}) and one for
$ \varw_i^{\,(\alpha)} $ (\texttt{forward\_w}), which use the same indexing
subscripts.
Each of them consecutively stores
$ N_\mathrm{X} N_\mathrm{Y} N_\mathrm{Z} N_\mu N_\nu $ values.

In that case, for each index tuple $ (x, y, z, \mu, i) $ we need to keep track
of the related knots $ j_i^{\,(\alpha)} $ and weights $ \varw_i^{(\alpha)} $.
We use the 1D-array \texttt{address} to do so.

For each $ i = i_\red - 1, \dots, i_\blue $, this array keeps the address of the
last comoving knot $ j_i^{\,(\alpha_\mathrm{max})} $ that lies on the inertial
interval $ {[i, i + 1)} $.
The extra knot $ i_\red - 1 $ corresponds to the comoving frequency
$ {-}\infty $ and is needed to treat extrapolation beyond the left end of the
frequency grid.
For convenience, one more element is added to the \texttt{address} array at the
beginning so that the array subscript $ o $ starts from $-1$.
Hence, \texttt{address} has
$ N_\mathrm{X} N_\mathrm{Y} N_\mathrm{Z} N_\mu (N_\nu + 1) + 1 $ elements.
The values stored in \texttt{address} start from 0 and are non-decreasing.

For each $ x $, $ y $, $ z $, and $ \mu $, we scan comoving knots
$ j = j_\red, \dotsc, j_\blue $ to find the inertial nearest-left neighbor knot
$ i $ such that $ q[i] \leq q[j] + q_\mu < q[i + 1] $.
Then we compute the {l.h.s.}\ weight $ \varw $, using interpolation for internal
knots, and constant extrapolation for external knots.

We set the subscript $ o $ to the memory-offset given by the index tuple
$ (x, y, z, \mu, i) $:
\begin{align} \label{eq:tuple}
  o( x, y, z, \mu, i )
    & = i - (i_\red - 1) + (N_\nu + 1) \cdot {} \nonumber\\
    & \qquad         ( \mu - 1 + N_\mu \cdot {} \nonumber\\
    & \qquad\quad     ( z - z_\mathrm{min} + N_\mathrm{Z} \cdot {} \nonumber\\
    & \qquad\qquad     ( y - y_\mathrm{min} + N_\mathrm{Y} \cdot {} \nonumber\\
    & \qquad\qquad\quad ( x - x_\mathrm{min} ) ) ) ) ).
\end{align}
It is important that the order of indices in the tuple follows the order of the
nested for-loops in
Algs.\,\ref{alg:precomputation-forward}--\ref{alg:precomputation-backward}
so that \texttt{address} is filled continuously.
Frequency is the fastest (continuous) dimension in memory.

At this point, counter $ c $ subscripts the last modified element of
\texttt{address} such that \texttt{address}[$ c $] points to the last comoving
knot saved in \texttt{forward\_j}.
Then we point \texttt{address}[$ o $] to the current comoving knot $ j $:
since $ j $ is stored contiguously in \texttt{index\_j}, we use an incremented
address of the last comoving knot:
\begin{equation}
  \mathtt{address}[o] = \mathtt{address}[c] + 1.
\end{equation}
If $ o = c $, then both the last and the current comoving knots lie on the same
interval $ {[i, i+1)} $ and \texttt{address}[$o = c$] is just incremented.
If $ o > c + 1 $, then there are inertial knots between the last and the current
comoving knots, which do not contribute to any comoving knots.
In this case, the elements of \texttt{address} between $ c $ and $ o $ are
filled with \texttt{address}[$ c $]:
\begin{equation}
  \mathtt{address}[c + 1{:}o - 1] = \mathtt{address}[c],
\end{equation}
so that they point to the last comoving knot.
Then we adjust the counter $ c $ to the current memory-offset $ o $
\begin{equation}
  c = o
\end{equation}
and store the current comoving knot and its weight:
\begin{align}
  \mathtt{forward\_j}[\mathtt{address}[c]] & = j, \\
  \mathtt{forward\_w}[\mathtt{address}[c]] & = \varw.
\end{align}
Now we proceed to the next comoving knot.
We summarize this complicated procedure in
Alg.\,\ref{alg:precomputation-forward} and illustrate it in
Fig.\,\ref{fig:precomputed-forward}.

\subsubsection{Precomputation for the backward interpolation}
%
\begin{algorithm}[!t]
  \SetInd{0.5em}{0.5em}
  \DontPrintSemicolon
  \LinesNotNumbered
  \SetCommentSty{textsl}
  \KwData{Line frequencies $ q[i_\red{:}i_\blue] $,
    directions $ \vec{n}[1{:}N_\mu] $,
    velocities $ \vec{\varv}[x_\mathrm{min}{:}x_\mathrm{max},
                             y_\mathrm{min}{:}y_\mathrm{max},
                             z_\mathrm{min}{:}z_\mathrm{max}] $.}
  \KwResult{Arrays \texttt{backward\_j} and \texttt{backward\_w}.}
  \newcommand{\wwid}{\widthof{$\varw$}}
  \For{$ x \gets x_\mathrm{min} $ \KwTo $ x_\mathrm{max} $}{
    \For{$ y \gets y_\mathrm{min} $ \KwTo $ y_\mathrm{max} $}{
      \For{$ z \gets z_\mathrm{min} $ \KwTo $ z_\mathrm{max} $}{
        \For{$ \mu \gets 1 $ \KwTo $ N_\mu $}{
          $ q_\mu
            \gets
            \big( \vec{n}[\mu]\cdot \vec{\varv}[x, y, z] \big) /
            \varv_\mathrm{B} $\;
          $ j \gets j_\red - 1 $
            \tcp*[r]{Start searching from the leftmost\makebox[5pt]{}}
            \tcp*[r]{knot $ ( {-}\infty ) $ in the comoving frame.}
          \For(\tcp*[f]{For each inertial knot $i$, find})
            {$ i \gets i_\red $ \KwTo $ i_\blue $}{ 
            \tcp*[r]{which comoving knot $j$\makebox[18pt]{}}
            \tcp*[r]{contributes into it.\makebox[39pt]{}}
            $ q^{\star\!} \gets q[i] - q_\mu $
              \tcp*[r]{Comoving frequency.}
            $ j \gets \mathrm{locate}\big( q^{\star\!}, q[j_\red{:}j_\blue], j \big) $
              \tcp*[r]{$ q[j]\,{\leq}\,q^\star{<}\,q[j{+}1] $}
            \uIf(){$ j = j_\red - 1 $}{
              $ \mathmakebox[\wwid][l]{j^\prime} \gets j_\red $
                \tcp*[r]{Constant extrapolation}
              $ \varw \gets 1 $
                \tcp*[r]{on $ {( {-}\infty, j_\red )} $.\makebox[37pt]{}}
            }
            \uElseIf{$ j = j_\blue $}{
              $ \mathmakebox[\wwid][l]{j^\prime} \gets j_\blue $
                \tcp*[r]{Constant extrapolation}
              $ \varw \gets 1 $
                \tcp*[r]{on $ {[ j_\blue, {+}\infty )} $.\makebox[34pt]{}}
            }
            \uElseIf{$ q^\star = q[j] $}{
              $ \mathmakebox[\wwid][l]{j^\prime} \gets j $
                \tcp*[r]{Exact match of $j$ to $i$.}
              $ \varw \gets 1$\;
            }
            \Else($ j_\red \leq j < j_\blue $){
              $ \mathmakebox[\wwid][l]{j^\prime} \gets j $
                \tcp*[r]{Linear interpolation}
              $ \varw \gets \dfrac{ q[j + 1] - q^\star\;\: }
                                  { q[j + 1] - q[j]        } $
                \tcp*[r]{on $ {[ j, j + 1 )} $.\makebox[31pt]{}}
            }
            $ \mathtt{backward\_j}[x, y, z, \mu, i] \gets j^\prime $
              \tcp*[l]{Store comoving}
            $ \mathtt{backward\_w}[x, y, z, \mu, i] \gets \varw $
              \tcp*[l]{index and weight.}
          }
        }
      }
    }
  }
  \caption{Initialization of the precomputed backward interpolation.}
  \label{alg:precomputation-backward}
\end{algorithm}
Precomputation for the backward transform is  similar to the one for the
forward interpolation and is explained in
Alg.\,\ref{alg:precomputation-backward}, but there are two differences.
First, each inertial knot $ i $ has only one related comoving knot $ j $, as
illustrated in Fig.\,\ref{fig:backward-transform}.
This strict correspondence allows us to use 5D-arrays to store the indices and
weights, no address array is needed.
Second, we scan the inertial knots to find the related comoving knots, instead
of the other way around.

\subsubsection{Precomputed forward transform}
%
\begin{algorithm}[!t]
  \SetInd{0.5em}{0.5em}
  \DontPrintSemicolon
  \LinesNotNumbered
  \SetCommentSty{textsl}
  \newcommand{\pfwid}{\widthof{$a_{i - 2}$}}
  \newcommand{\wwid}{\widthof{$\varw$}}
  \KwData{Specific intensity $ I[x, y, z, \mu, i] $ at frequency $ q[i] $ and
    direction $ \vec{n}[\mu] $ in the inertial frame.}
  \KwResult{Incremented angle-averaged intensity $ J^{\star\!}[x, y, z, j] $ at
    corresponding frequencies $ q[j] $ in the comoving frame.}
  $ o \gets o(x, y, z, \mu, i) $
    \tcp*[r]{Memory-offset by Eq.\,\eqref{eq:tuple}}
  $ \mathmakebox[\pfwid][l]{a_i} \gets \mathtt{address}[\mspace{14mu}o\mspace{15mu}] $
    \tcp*[r]{Addresses of comoving knots\makebox[8pt]{}}
  $ \mathmakebox[\pfwid][l]{a_{i - 1}} \gets \mathtt{address}[o - 1] $
    \tcp*[r]{for current $ i $, previous $ i - 1 $, and}
  $ \mathmakebox[\pfwid][l]{a_{i - 2}} \gets \mathtt{address}[o - 2] $
    \tcp*[r]{pre-previous $ i - 2 $ inertial knots.}
  \For(\tcp*[f]{Interpolate on $ {[ i - 1, i )} $:})
    {$ a \gets a_{i - 2} + 1 $ \KwTo $ a_{i - 1} $}{
    $ \mathmakebox[\wwid][l]{j} \gets \mathtt{forward\_j}[a] $\;
    $ \varw \gets \mathtt{forward\_w}[a] $\;
    $ J^{\star\!}[x, y, z, j] \overset{+}{\gets}
      (1 - \varw) \dfrac{\omega_\mu}{4\pi} I[x, y, z, \mu, i] $
      \tcp*[r]{Use {r.h.s.} weights.}
  }
  \For(\tcp*[f]{Interpolate on $ {[i, i + 1)} $:})
    {$ a \gets a_{i - 1} + 1 $ \KwTo $ a_i $}{
    $ \mathmakebox[\wwid][l]{j} \gets \mathtt{forward\_j}[a] $\;
    $ \varw \gets \mathtt{forward\_w}[a] $\;
    $ J^{\star\!}[x, y, z, j] \overset{+}{\gets}
      \varw \dfrac{\omega_\mu}{4\pi} I[x, y, z, \mu, i]$
      \tcp*[r]{Use {l.h.s.}\ weights.}
  }
  \caption{Precomputed forward transform.}
  \label{alg:precomputed-forward}
\end{algorithm}
With precomputed indices and weights, the forward transform can be done very
quickly following Alg.\,\ref{alg:precomputed-forward}.
At grid point $ (x, y, z) $, angle direction $ \vec{n}[\mu] $, and inertial
frequency $ q[i] $, we compute the memory-offset $ o $ by using
Eq.\,\eqref{eq:tuple}.

Now $ a_i = \mathtt{address}[o] $ is the address of the last comoving knot
$ j_i^{\,(\max\alpha)} $ that lies on the inertial interval $ {[ i, i + 1 )} $.
The previous knot $ i - 1 $ has memory-offset $ o - 1 $ so that
$ a_{i - 1} = \mathtt{address}[o - 1] $ is the address of the last comoving knot
$ j_{i - 1}^{\,(\max\alpha)} $ that lies on the inertial interval
$ {[i - 1, i)} $.

If $ a_{i - 1} < a_i $ then $ a_i - a_{i - 1} $ comoving knots lie on the
interval $ {[ i, i + 1 )} $, and their addresses are
$ a_{i - 1} + 1, \dotsc, a_i $.
Using these addresses, we extract indices the $ j_i^{\,(\alpha)} $ from
$ \mathtt{forward\_j} $ and the {l.h.s.}\ weights $ \varw_i^{(\alpha)} $ from
$ \mathtt{forward\_w} $ and perform the transform by using
Eq.\,\eqref{eq:J*-rhs}.

If $ a_{i - 1} = a_i $, then there are no comoving knots that lie on the
inertial interval $ {[ i, i + 1 )} $ and no interpolation is done.

Applying the same procedure to the previous $ i - 1 $ and pre-previous $ i - 2 $
knots, we obtain the corresponding comoving indices $ j_{i - 1}^{\,(\alpha)} $
and {r.h.s.}\ weights $ 1 - \varw_{i - 1}^{(\alpha)} $ for the inertial interval
$ {[ i - 1, i )} $ and perform the transform by using Eq.\,\eqref{eq:J*-lhs}.

It is necessary to use three knots (the current $ i $, the previous $ i - 1 $,
and the pre-previous $ i - 2 $) to compute the related indices and weights for
both the $ {[ i - 1, i )} $ and $ {[i, i + 1 )} $ intervals because we store
only the {l.h.s.}\ weights to save memory.
This is why the subscript of \texttt{address} starts at $-1$, so that the
algorithm handles the case $ i = i_\red $ without using an if-statement.

\subsubsection{Precomputed backward transform}
%
\begin{algorithm}[!t]
  \SetInd{0.5em}{0.5em}
  \DontPrintSemicolon
  \LinesNotNumbered
  \SetCommentSty{textsl}
  \newcommand{\wwid}{\widthof{$\varw$}}
  \KwData{Profile ratio $ \rho^{\star\!}[x, y, z, j] $ for all frequencies
    $ q[j]\colon j = j_\red, \dotsc, j_\blue $ in the comoving frame.}
  \KwResult{Profile ratio $ \rho[x, y, z, \mu, i] $ at given frequency $ q[i] $
    and direction $ \vec{n}[\mu] $ in the inertial frame.}
  $ \mathmakebox[\wwid][l]{j} \gets \mathtt{backward\_j}[x, y, z, \mu, i] $
    \tcp*[r]{Comoving index}
  $ \varw \gets \mathtt{backward\_w}[x, y, z, \mu, i] $
    \tcp*[r]{and weight.\makebox[19pt]{}}
  \eIf(\tcp*[f]{Linear interpolation on ${[j, j + 1)}$:}){$0 \leq \varw < 1$}{
    $ \rho[x, y, z, \mu, i] \gets \varw\,\rho^{\star\!}[x, y, z, j] +
        \left( 1 - \varw \right) \rho^{\star\!}[x, y, z, j + 1] $\;
  }($ \varw = 1 $
    \tcp*[f]{Either constant extrapolation if $j \notin {[ j_\red, j_\blue )}$,}){
    $ \rho[x, y, z, \mu, i] \gets \rho^{\star\!}[x, y, z, \mu, j] $
      \tcp*[r]{or $j$ exactly matches $i$.}
  }
  \caption{Precomputed backward transform.}
  \label{alg:precomputed-backward}
\end{algorithm}
With precomputed indices and weights the backward interpolation is performed as
in Alg.\,\ref{alg:precomputed-backward}.
We select the proper type of interpolation from
Eqs.\,\eqref{eq:backward-two-knots}--\eqref{eq:backward-iblue} depending on the
value of the {r.h.s.}\ weight $ \varw $.

\subsection{Optimization of memory storage}

The \texttt{address} array needs to store big numbers (typically, $ o < 10^9 $)
and therefore must be of 4-byte integer type, while values of
\texttt{forward\_j} and \texttt{backward\_j} usually fit in 2-byte integer type.
Values of \texttt{forward\_w} and \texttt{backward\_w} can be packed into 1-byte
integer type because interpolation weights are real numbers on the interval
$ {[0, 1]} $.
The limited precision of a one-byte representation is acceptable because linear
interpolation itself is not very accurate.
Hence, we need 10~bytes per spatial grid point, direction, and frequency.

\subsection{Equidistant interpolation}
\label{subsec:equidistant-interpolation}
%
\begin{figure}
  \includegraphics[width=\columnwidth]{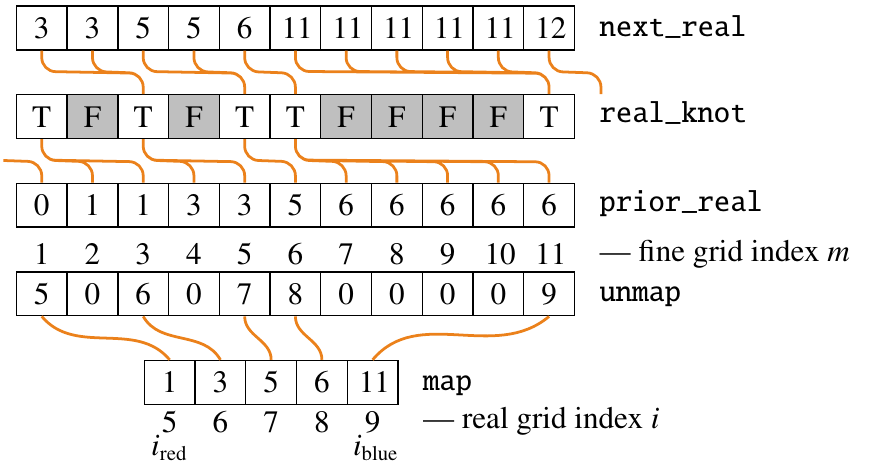}
  \caption{Illustrative contents of the five auxilliary arrays: \texttt{map},
    \texttt{unmap}, \texttt{real\_knot}, \texttt{prior\_real}, and
    \texttt{next\_real}.
    Orange lines between elements of \texttt{map} and \texttt{unmap} indicate
    that the arrays store each other's indices.
    Indices of \texttt{map} run on the real grid from $ i_\red = 5 $ to
    $ i_\blue =  9 $.
    Indices of \texttt{unmap} run on the fine grid from 1 to
    $ m_\mathrm{max} = 11 $.
    Logical array \texttt{real\_knot} contains true for real and false for
    virtual knots.
    Arrays \texttt{next\_real} and \texttt{prior\_real} contain fine grid
    indices of corresponding nearest real knots, illustrated by orange lines
    connected to the elements of \texttt{real\_knot}.
    The last element of \texttt{next\_real} points to $ m_\mathrm{max} + 1 $,
    while the first element of \texttt{prior\_real} points to 0.
    Note that the outmost knots $ i_\red \to m = 1 $ and
    $ i_\blue \to m = m_\mathrm{max} $ always have to be real.}
  \label{fig:equidistant-masks}
\end{figure}
To get rid of frequency dependence in the arrays that store precomputed indices
and weights, we modify the frequency grid so that it is equidistant.
For a given velocity resolution $ \delta\varv$ (typically, 1--2 km\,s$^{-1}$),
the new frequency grid resolution is
\begin{equation}
  \delta q = \frac{ \delta\varv }{ \varv_\mathrm{B} }.
\end{equation}
From $ q[i_\red] $ to $ q[i_\blue] $ we have
\begin{equation}
  m_\mathrm{max} = 1 + \frac{ q[i_\blue] - q[i_\red] }{ \delta q }
\end{equation}
knots separated by equal intervals of fine resolution $ \delta q $ and numbered
from 1 to $ m_\mathrm{max} $ having frequencies
\begin{equation}
  q^\prime[m] = q[i_\red] + ( m - 1 )\,\delta q.
\end{equation}
so that
$ q^\prime[1]              = q[i_\red ] $ and
$ q^\prime[m_\mathrm{max}] = q[i_\blue] $.
Note that indices $ m $ and $ i $ do not coincide.

Fine resolution is needed only in the line core but not in the line wings, so we
solve the radiative transfer equation only in selected knots called
\textit{real} and do not do this in the remaining knots called \textit{virtual}.

We call the ordered set of real knots the \textit{real grid} and we number them
in the same way as before using index $ i = i_\red, \dotsc, i_\blue $ for the
inertial real grid and using index $ j = j_\red, \dotsc, j_\blue $ for the
comoving real grid.

We call the ordered set of both real and virtual knots the \textit{fine grid}
and we number them using index $ m = 1, \dotsc, m_\mathrm{max} $ for the
inertial fine grid and using index $ n = 1, \dotsc, n_\mathrm{max} $ for the
comoving fine grid, with $ m_\mathrm{max} = n_\mathrm{max}$.

We make interpolation on the real grid fast by using extra maps, that track
positions of real and virtual knots and specify relations between them.
We store these maps as 1D arrays.
Example contents of these arrays are illustrated in
Fig.\,\ref{fig:equidistant-masks}.

The array \texttt{map} with size $ N_\nu $ is indexed from $ i_\red $ to
$ i_\blue $ and for the real grid index $ i $, \texttt{map}[$ i $] is the
related fine grid index $ m $.
The array \texttt{unmap} with size $ m_\mathrm{max} $ is indexed from 1 to
$ m_\mathrm{max} $, and for the fine grid index $ m $, \texttt{unmap}[$ m $]
either is the related real grid index $ i $ if $ m $ is a real knot, otherwise
it is 0.
In essence, \texttt{map} matches the real grid with the fine grid, while
\texttt{unmap} does the opposite.

The logical array \texttt{real\_knot} with size $ m_\mathrm{max} $ is indexed
from 1 to $ m_\mathrm{max} $ and specifies whether fine grid knots are real or
virtual.
The two arrays \texttt{prior\_real} and \texttt{next\_real} with size
$ m_\mathrm{max} $ are indexed from 1 to $ m_\mathrm{max} $.
They specify for each $ m $ in the fine grid, the nearest-left real or
nearest-right real knot.

\subsubsection{Precomputation for equidistant interpolation}
%
\begin{figure}
  \includegraphics[width=\columnwidth]{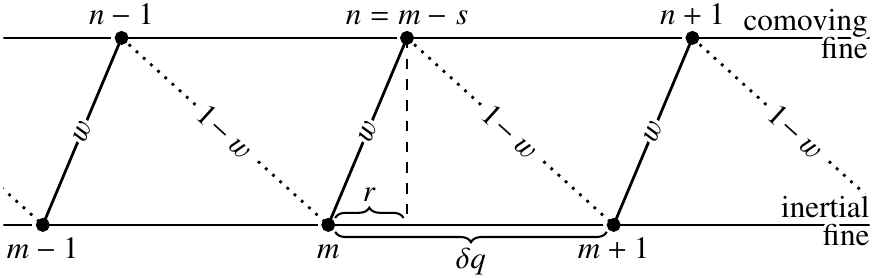}
  \caption{Linear interpolation on an infinite grid of knots separated by
    equidistant intervals $ \delta q $.
    Each knot $ n $ in the comoving frame (top line) is interpolated on the
    $ {[m, m + 1)} $ interval in the inertial frame (bottom line).
    The integer shift $ s = m - n $ defines how many $ \delta q $ intervals the
    comoving grid is Doppler-shifted with respect to the inertial grid.
    The {l.h.s.}\ $ \varw $ and {r.h.s.}\ $ 1 - \varw $ weights are
    indicated by sloped solid and dotted lines.
    The weight $ \varw $ is determined by the remainder
    $ r = q_\mu + q[n] - q[m] $, which is the frequency displacement between
    the inertial knot $ m $ and its related comoving knot $ n $.}
  \label{fig:equidistant-interpolation}
\end{figure}
\begin{algorithm}[!t]
  \SetInd{0.5em}{0.5em}
  \DontPrintSemicolon
  \LinesNotNumbered
  \SetCommentSty{textsl}
  \newcommand{\wwid}{\widthof{$\varw$}}
  \KwData{Line frequencies $ q[i_\red{:}i_\blue] $,
    directions $ \vec{n}[1{:}N_\mu] $,
    velocities $ \vec{\varv}[x_\mathrm{min}{:}x_\mathrm{max},
                             y_\mathrm{min}{:}y_\mathrm{max},
                             z_\mathrm{min}{:}z_\mathrm{max}] $.}
  \KwResult{Arrays \texttt{shift} and \texttt{weight}.}
  \newcommand{\weightwid}{\widthof{$\mathtt{weight}$}}
  \For{$ z \gets z_\mathrm{min} $ \KwTo $ z_\mathrm{max} $}{
    \For{$ y \gets y_\mathrm{min} $ \KwTo $ y_\mathrm{max} $}{
      \For{$ x \gets x_\mathrm{min} $ \KwTo $ x_\mathrm{max} $}{
        \For{$ \mu \gets 1 $ \KwTo $ N_\mu $}{
          $ q_\mu \gets
            \bigl( \vec{n}[\mu]\cdot \vec{\varv}[x, y, z] \bigr) /
            \varv_\mathrm{B} $\;
          $ s \gets \bigg\lfloor \dfrac{q_\mu}{\delta q} \bigg\rfloor $
            \tcp*[r]{Integer shift between the two grids.}
          $ r \gets q_\mu - s \cdot \delta q $
            \tcp*[r]{Fractional remainder.}
          $ \varw \gets 1 - \dfrac{r}{\delta q} $
            \tcp*[r]{The {l.h.s.}\ weight}
          $ \mathmakebox[\weightwid][l]{\mathtt{shift}}[x, y, z, \mu] \gets s $
            \tcp*[r]{Store shift and weight.}
          $ \mathtt{weight}[x, y, z, \mu] \gets \varw $\;
        }
      }
    }
  }
  \caption{Precomputation for equidistant interpolation.}
  \label{alg:precomputation-equidistant}
\end{algorithm}
At each grid point $ (x, y, z) $, for each direction $ \vec{n}[\mu] $, we
compute the Doppler shift $ q_\mu $.
The comoving knots $ n $ are Doppler-shifted along the inertial knots $ m $ by
$ s $ intervals of $ \delta q $ length and a fractional remainder $ r $ such
that
\begin{equation}
  q_\mu = s\cdot\delta q + r.
\end{equation}
If inertial knot $ m $ has a comoving nearest-right neighbor knot $ n $, then
$ m - n = s $.
The frequency displacement between these knots is $ q_\mu + q[n] - q[m] = r $.
The remainder $ r $ determines the interpolation weights for the comoving knot
$ n $ on the inertial interval $ {[m, m + 1)} $.
The {l.h.s.}\ weight is
\begin{equation}
  \varw = \frac{ \delta q - r }{ \delta q } = 1 - \frac{ r }{ \delta q }.
\end{equation}
There are three advantages of using equidistant frequency grids.
First, for a given index tuple $ (x, y, z, \mu) $, the values of $ s $, $ r $,
and $ \varw $ are independent of frequency as illustrated in
Fig.\,\ref{fig:equidistant-interpolation}.
Second, during the transforms, we can limit the searching range for comoving
knots by using the $ m = n + s $ relation.
Third, both the forward and backward transforms use the same values of $ s $ and
$ \varw $.
Therefore, we precompute and store $ s $ and $ \varw $ in two 4D-arrays
(\texttt{shift} and \texttt{weight}) following
Alg.\,\ref{alg:precomputation-equidistant}.

\subsubsection{Equidistant forward transform}
%
\begin{figure}
  \includegraphics[width=\columnwidth]{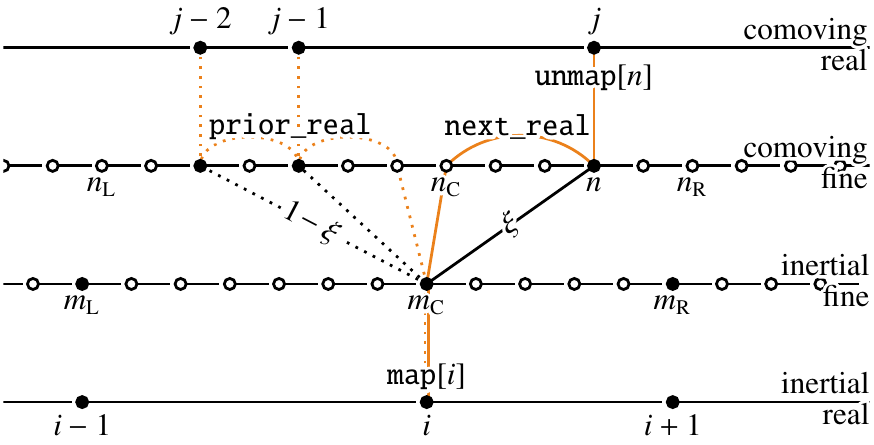}
  \caption{Equidistant forward interpolation from the inertial real grid onto
    the comoving real grid using the fine grids.
    Real knots $ (\bullet) $ are marked on the real and the fine grids; virtual
    knots $ (\circ) $ are marked on the fine grids.
    Knots $ i - 1 $, $ i $, and $ i + 1 $ are projected from the inertial real
    onto inertial fine grid using the \texttt{map} array ($ m_\mathrm{L} $,
    $ m_\mathrm{C} $, $ m_\mathrm{R} $) and corrected for the shift $ s $ to
    obtain the corresponding nearest-right neighbors $ n_\mathrm{L} $,
    $ n_\mathrm{C} $, and $ n_\mathrm{R} $ in the comoving fine grid.
    In the comoving fine grid, two scans with running index $ n $ are done to
    find real knots on the intervals $ {[ m_\mathrm{L}, m_\mathrm{C} ]} $ and
    $ {[ m_\mathrm{C}, m_\mathrm{R} ]} $: from $ n_\mathrm{C} $ to
    $ n_\mathrm{R} - 1 $ using the \texttt{next\_real} array (solid orange
    line), and from $ n_\mathrm{C} - 1 $ to $ n_\mathrm{L} $ using the
    \texttt{prior\_real} array (dotted orange line).
    When a real knot is found, we find the corresponding real comoving grid
    index $ j $ using the \texttt{unmap} array.
    The desired {l.h.s.}\ $ \xi $ (sloped black solid line) or
    {r.h.s.}\ $ 1 - \xi $ (sloped black dotted line) weights are computed from
    $ n $, $ n_\mathrm{L} $, $ n_\mathrm{C} $, $ n_\mathrm{R} $, and $ \varw $.}
  \label{fig:equidistant-forward}
\end{figure}
\begin{algorithm}[!t]
  \SetInd{0.5em}{0.5em}
  \DontPrintSemicolon
  \LinesNotNumbered
  \SetCommentSty{textsl}
  \newcommand{\weightwid}{\widthof{$\mathtt{weight}$}}
  \newcommand{\fwid}{\widthof{$[\min(i_\blue, i + 1)]$}}
  \newcommand{\wwid}{\widthof{$\varw$}}
  \KwData{Specific intensity $ I[x, y, z, \mu, i] $ at frequency $ q[i] $ and
    direction $ \vec{n}[\mu] $ in the inertial frame.}
  \KwResult{Incremented angle-averaged intensity $ J^{\star\!}[x, y, z, j] $ at
    corresponding frequencies $ q[j] $ in the comoving frame.}
  $ n_\mathrm{max} \gets \mathrm{size}( \mathtt{map} ) $
    \tcp*[f]{Fine grid length.}\;
  $ \mathmakebox[\wwid][l]{s} \gets \mathmakebox[\weightwid][l]{\mathtt{shift}}[x, y, z, \mu] $\;
  $ \varw \gets \mathtt{weight}[x, y, z, \mu] $\;
  $ n_\mathrm{L} \gets \mathtt{map}\mathmakebox[\fwid][l]{\bigl[\max(i - 1, i_\red)\bigr]} - s $
    \tcp*[h]{Nearest-right neighbors of}\;
  $ n_\mathrm{C} \gets \mathtt{map}\mathmakebox[\fwid][l]{\bigl[i\bigr]} - s $
    \tcp*[h]{$i-1$, $i$, and $i+1$ on the fine}\;
  $ n_\mathrm{R} \gets \mathtt{map}\bigl[\min(i_\blue, i + 1)\bigr] - s $
    \tcp*[h]{grid in the comoving frame.}\;
  \eIf(\tcp*[f]{Linear interpolation on $ {[i - 1, i)} $:}){$ n_\mathrm{L} \neq n_\mathrm{C} $}{
    $ n \gets \min\bigl( n_\mathrm{C} - 1, n_\mathrm{max} \bigr) $\;
    \If(){$ 1 \leq n $}{
      \lIf(){\KwSty{not} $ \mathtt{real\_knot}[n] $}{$ n \gets \mathtt{prior\_real}[n] $}
      \While(){$ \max\bigl( 1, n_\mathrm{L} \bigr) \leq n $}{
        $ \xi^\prime \gets \dfrac{ n - n_\mathrm{L} + 1 - \varw }
                                 { n_\mathrm{C} - n_\mathrm{L}  } $
          \tcp*[f]{The {r.h.s.}\ weight.}\;
        $ J^{\star\!}\bigl[x, y, z, \mathtt{unmap}[n]\bigr] \overset{+}{\gets}
          \xi^\prime \dfrac{\omega_\mu}{4\pi} I[x, y, z, \mu, i] $\;
        $ n \gets \mathtt{prior\_real}[n] $\;
      }
    }
  }($ n_\mathrm{L} = n_\mathrm{C} $\tcp*[f]{Constant extrapolation on $ {({-}\infty, i_\red)} $:}){
    $ n \gets \min\bigl( n_\mathrm{C} - 1, n_\mathrm{max} \bigr) $\;
    \If(){$ 1 \leq n $}{
      \lIf(){\KwSty{not} $ \mathtt{real\_knot}[n] $}{$ n \gets \mathtt{prior\_real}[n] $}
      \While(){$ 1 \leq n $}{
        $ J^{\star\!}\bigl[x, y, z, \mathtt{unmap}[n]\bigr] \overset{+}{\gets}
          \dfrac{\omega_\mu}{4\pi} I[x, y, z, \mu, i] $\;
        $ n \gets \mathtt{prior\_real}[n] $\;
      }
    }
  }
  \eIf(\tcp*[f]{Linear interpolation on $ {[i, i+1)} $:}){$ n_\mathrm{C} \neq n_\mathrm{R} $}{
    $ n \gets \max\bigl( 1, n_\mathrm{C} \bigr) $\;
    \If(){$ n \leq n_\mathrm{max} $}{
      \lIf(){\KwSty{not} $ \mathtt{real\_knot}[n] $}{$ n \gets \mathtt{next\_real}[n] $}
      \While(){$ n \leq \min\bigl( n_\mathrm{R} - 1, n_\mathrm{max} \bigr) $}{
        $ \xi \gets \dfrac{ n_\mathrm{R} - n - (1 - \varw) }
                          { n_\mathrm{R} - n_\mathrm{C}    } $
          \tcp*[f]{The {l.h.s.}\ weight.}\;
        $ J^{\star\!}\bigl[x, y, z, \mathtt{unmap}[n]\bigr] \overset{+}{\gets}
          \xi \dfrac{\omega_\mu}{4\pi} I[x, y, z, \mu, i] $\;
        $ n \gets \mathtt{next\_real}[n] $\;
      }
    }
  }($ n_\mathrm{C} = n_\mathrm{R} $\tcp*[f]{Constant extrapolation on $ {[i_\blue, {+}\infty)} $:}){
    $ n \gets \max\bigl(1, n_\mathrm{C} \bigr) $\;
    \If(){$ n \leq n_\mathrm{max} $}{
      \lIf(){\KwSty{not} $ \mathtt{real\_knot}[n] $}{$ n \gets \mathtt{next\_real}[n] $}
      \While(){$ n \leq n_\mathrm{max} $}{
        $ J^{\star\!}\bigl[x, y, z, \mathtt{unmap}[n]\bigr] \overset{+}{\gets}
          \dfrac{\omega_\mu}{4\pi} I[x, y, z, \mu, i] $\;
        $ n \gets \mathtt{next\_real}[n] $\;
      }
    }
  }
  \caption{Equidistant forward transform.}
  \label{alg:forward-equidistant}
\end{algorithm}
The forward transform for the equidistant grid is given by
Alg.\,\ref{alg:forward-equidistant}.
Figure~\ref{fig:equidistant-forward} provides an illustration.

Interpolation is performed on the fine grid.
We project the current knot $ i $ and its left $ i - 1 $ and right $ i + 1 $
neighbors from the inertial real grid onto the inertial fine grid:
\begin{align}
  m_\mathrm{L} & = \mathtt{map}\bigl[\max(i - 1, i_\red) \bigr], \\
  m_\mathrm{C} & = \mathtt{map}\bigl[i                   \bigr], \\
  m_\mathrm{R} & = \mathtt{map}\bigl[\min(i_\blue, i + 1)\bigr].
\end{align}
We use $ \max(i - 1, i_\red) $ and $ \min(i_\blue, i + 1) $ to avoid going
beyond the profile ends $ i = i_\red $ and $ i = i_\blue $.

Indices of the related nearest-right neighbors in the comoving fine grid are
computed from the shift $ s $:
\begin{alignat}{2}
  & n_\mathrm{L} = \mathtt{map}\bigl[\max(i - 1, i_\red) \bigr] && {} - s, \\
  & n_\mathrm{C} = \mathtt{map}\bigl[i                   \bigr] && {} - s, \\
  & n_\mathrm{R} = \mathtt{map}\bigl[\min(i_\blue, i + 1)\bigr] && {} - s.
\end{alignat}
To interpolate on the $ {[i, i + 1)} $ interval, we scan the comoving fine grid
knots $ n_\mathrm{C}, \dotsc, n_\mathrm{R} - 1 $ that lie on the
$ {[ m_\mathrm{C}, m_\mathrm{R} ]} $ interval.

Moving from $ n = n_\mathrm{C} $ to $ n_\mathrm{R} - 1$, we look for real knots.
If $ n $ is a virtual knot, we jump to the next real knot using the
\texttt{next\_real} array.
If $ n $ is a real knot, we find the corresponding index $ j $ in the comoving
real grid using the \texttt{unmap} array.
Then the desired {l.h.s.}\ weight $ \xi $ is computed from $ n $,
$ n_\mathrm{C} $, $ n_\mathrm{R} $, and the fine grid weight $ \varw $:
\begin{equation}
  \xi = \dfrac{ n_\mathrm{R} - n - (1 - \varw) }
              { n_\mathrm{R} - n_\mathrm{C}    }.
\end{equation}

To interpolate on the $ {[i - 1, i)} $ interval, we scan the comoving fine grid
knots $ n_\mathrm{L}, \dotsc, n_\mathrm{C} - 1 $ that lie on the
$ {[ m_\mathrm{L}, m_\mathrm{C} ]} $ interval.
Moving from $ n = n_\mathrm{C} - 1 $ to $ n_\mathrm{L} $ we look for real knots.
If $ n $ is a virtual knot, we jump to the prior real knot using the
\texttt{prior\_real} array.
If $ n $ is a real knot, we find the corresponding index $ j $ in the comoving
real grid using the \texttt{unmap} array.
The desired {r.h.s.}\ weight $ 1 - \xi $ is then computed as
\begin{equation}
 \xi^\prime
  \equiv 1 - \xi
  = \dfrac{ n - n_\mathrm{L} + 1 - \varw }
          { n_\mathrm{C} - n_\mathrm{L}  }.
\end{equation}

Extrapolation at the $ i = i_\red $ or $ i = i_\blue $ ends is done likewise.

In all four cases, we truncate the running index $ n $ by taking $ \max(1, n) $
or $ \min(n, n_\mathrm{max}) $ to keep it within the fine grid range
$ 1, \dotsc, n_\mathrm{max} $.

\subsubsection{Equidistant backward transform}
%
\begin{figure}
  \includegraphics[width=\columnwidth]{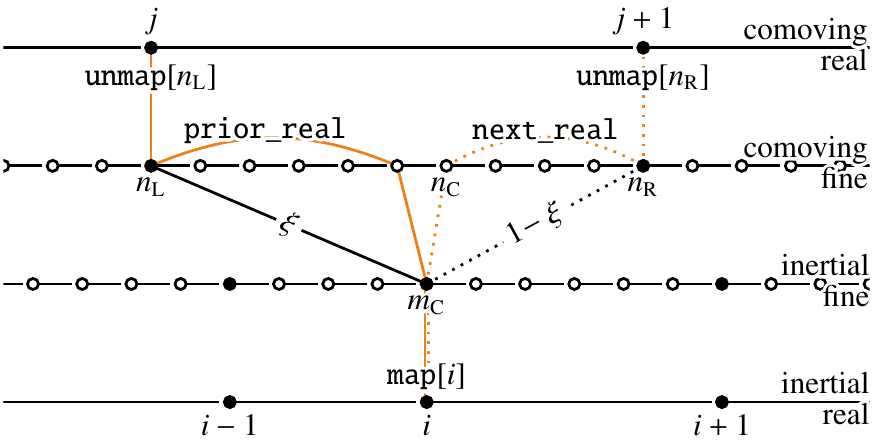}
  \caption{Equidistant backward interpolation from the comoving real grid onto
    the inertial real grid using the two fine grids.
    Real knots $ (\bullet) $ are marked on the real and the fine grids;
    virtual knots $ (\circ) $ are marked on the fine grid.
    The inertial real knot $ i $ is projected onto the inertial fine grid using
    the \texttt{map} array ($ m_\mathrm{C} $), and corrected for the shift $ s $
    to obtain the related nearest-right neighbor $ n_\mathrm{C} $ in the
    comoving fine grid.
    If $ n_\mathrm{C} < 1 $ or $ n_\mathrm{max} \leq n_\mathrm{C} $, then
    constant extrapolation is done.
    If $ n_\mathrm{C} $ is an internal knot, then we set
    $ n_\mathrm{L} = n_\mathrm{C} - 1 $ and $ n_\mathrm{R} = n_\mathrm{C} $ and
    check whether they both are real.
    If not, we find the nearest real knots using \texttt{prior\_real} (indicated
    by the solid orange curve) or \texttt{next\_real} (indicated by the orange
    dotted curve).
    Then $ n_\mathrm{L} $ and $ n_\mathrm{R} $ are projected onto the comoving
    real grid to get the corresponding knots $ j $ and $ j + 1 $.
    The desired {l.h.s.}\ $ \xi $ (sloped solid black line) and
    {r.h.s.}\ $ 1 - \xi $ (sloped dotted black line) weights are computed using
    $ n_\mathrm{L} $, $ n_\mathrm{C} $, $ n_\mathrm{R} $, and $\varw$.}
  \label{fig:equidistant-backward}
\end{figure}
\begin{algorithm}[!t]
  \SetInd{0.5em}{0.5em}
  \DontPrintSemicolon
  \LinesNotNumbered
  \SetCommentSty{textsl}
  \newcommand{\weightwid}{\widthof{$\mathtt{weight}$}}
  \newcommand{\wwid}{\widthof{$\varw$}}
  \KwData{Profile ratio $ \rho^{\star\!}[x, y, z, j] $ for all frequencies
    $ q[j]\colon j = j_\red, \dotsc, j_\blue $ in the comoving frame.}
  \KwResult{Profile ratio $ \rho[x, y, z, \mu, i] $ at given frequency $ q[i] $
    and direction $ \vec{n}[\mu] $ in the inertial frame.}
  $ n_\mathrm{max} \gets \mathrm{size}( \mathtt{map} ) $
    \tcp*[f]{Fine grid length.}\;
  $ \mathmakebox[\wwid][l]{s} \gets \mathmakebox[\weightwid][l]{\mathtt{shift}}[x, y, z, \mu] $\;
  $ \varw \gets \mathtt{weight}[x, y, z, \mu] $\;
  $ n_\mathrm{C} \gets \mathtt{map}[i] - s $
    \tcp*[r]{Nearest-right neighbor of $i$ on the}
    \tcp*[r]{fine grid in the comoving frame.\makebox[6pt]{}}
  \uIf(\tcp*[f]{Linear interpolation on $ {[j, j+1)} $:}){$ 1 < n_\mathrm{C} \leq n_\mathrm{max} $}{
    $ n_\mathrm{L} \gets n_\mathrm{C} - 1 $
      \tcp*[r]{Nearest-left and right neighbors}
    $ n_\mathrm{R} \gets n_\mathrm{C}     $
      \tcp*[r]{in the comoving fine grid.\makebox[24pt]{}}
    \lIf(){\KwSty{not} $ \mathtt{real\_knot}[n_\mathrm{L}] $}
      {$ n_\mathrm{L} \gets \mathtt{prior\_real}[n_\mathrm{L}] $}
    \lIf(){\KwSty{not} $ \mathtt{real\_knot}[n_\mathrm{R}] $}
      {$ n_\mathrm{R} \gets \mathtt{next\_real\ \,}[n_\mathrm{R}] $}
    $ \xi \gets \dfrac{ n_\mathrm{R} - n_\mathrm{C} + 1 - \varw }
                      { n_\mathrm{R} - n_\mathrm{L}             } $
      \tcp*[f]{The {l.h.s.}\ weight.}\;
    \tcp*[r]{Interpolate on the comoving grid:}
    $ \rho[x, y, z, \mu, i] \gets \xi\, \rho^{\star\!}\bigl[x, y, z, \mathtt{unmap}[n_\mathrm{L}]\bigr] $\;
    $ \mspace{57mu}     {} + (1 - \xi)\,\rho^{\star\!}\bigl[x, y, z, \mathtt{unmap}[n_\mathrm{R}]\bigr] $\;
  }
  \uElseIf(\tcp*[f]{Constant extrapolation on $ {[ j_\blue, {+}\infty )} $:}){$ n_\mathrm{max} < n_\mathrm{C} $}{
    $ \rho[x, y, z, \mu, i] \gets \rho^{\star\!}[x, y, z, j_\blue] $\;
  }
  \Else($ n_\mathrm{C} \leq 1 $\tcp*[f]{Constant extrapolation on $ {( {-}\infty, j_\red )} $:}){
    $ \rho[x, y, z, \mu, i] \gets \rho^{\star\!}[x, y, z, j_\red] $\;
  }
  \caption{Equidistant backward transform.}
  \label{alg:backward-equidistant}
\end{algorithm}
The backward interpolation employs the same ideas as the equidistant forward
interpolation.
It is given in Alg.~\ref{alg:backward-equidistant} and illustrated in
Fig.\,\ref{fig:equidistant-backward}.

The current knot $ i $ is projected from the inertial real onto the inertial
fine grid using the \texttt{map} array to get $ m_\mathrm{C} $.
Then we compute its nearest-right neighbor $ n_\mathrm{C} = m_\mathrm{C} - s $
in the comoving grid.

We test whether $ n_\mathrm{C} $ is beyond the fine grid ends 1 or
$ n_\mathrm{max} $.
If so, then we extrapolate from the comoving real grid ends $ j_\red $ or
$ j_\blue $.

If not, then knot $ i $ lies between knots $ n_\mathrm{C} - 1 $ and
$ n_\mathrm{C} $.
We set $ n_\mathrm{L} = n_\mathrm{C} - 1 $ and $ n_\mathrm{R} = n_\mathrm{C} $.
If $ n_\mathrm{L} $ or $ n_\mathrm{R} $ are virtual, we use \texttt{prior\_real}
or \texttt{next\_real} to set them to the nearest real knots.

Then we project $ n_\mathrm{L} $ and $ n_\mathrm{R} $ onto the comoving real
grid using the \texttt{unmap} array to get the comoving real knots $ j $ and
$ j + 1 $.
The desired {l.h.s.}\ weight is computed using $ n_\mathrm{L} $,
$ n_\mathrm{C} $, $ n_\mathrm{R} $, and $ \varw $:
\begin{equation}
  \xi = \dfrac{ n_\mathrm{R} - n_\mathrm{C} + 1 - \varw }
              { n_\mathrm{R} - n_\mathrm{L}             }.
\end{equation}

\end{appendix}

\end{document}